\documentclass[aps,pre,reprint,superscriptaddress]{revtex4-1}

\usepackage{subfigure}
\usepackage{t1enc,latexsym,amsfonts,amssymb,amsmath}
\usepackage{bm}
\usepackage{graphics}
\usepackage{graphicx}
\usepackage{wasysym}
\usepackage{setspace}
\usepackage{url}

\usepackage[english]{babel}
\usepackage[utf8x]{inputenc}
\usepackage[T1]{fontenc}

\usepackage[usenames,dvipsnames]{xcolor}

\usepackage{color}
\definecolor{darkblue}{rgb}{0,0,0.6}
\definecolor{darkred}{rgb}{0.6,0,0}

\usepackage{hyperref}
\hypersetup{
bookmarksopen=true,
pdftitle="On the relevance of disorder in athermal amorphous materials under shear",
pdfauthor="Elisabeth Agoritsas", 
pdfsubject="", 
pdftoolbar=false, 
pdfstartview={FitH},		
pdfmenubar=true,			
pdfhighlight=/O,			
colorlinks=true,			
urlcolor=darkblue,
citecolor=darkblue,		
linkcolor=MidnightBlue,	
}

\newcommand{\argc}[1]{\left[#1\right]}
\newcommand{\arga}[1]{\left\lbrace #1\right\rbrace }
\newcommand{\argp}[1]{\left(#1\right)}
\newcommand{\valabs}[1]{\vert #1\vert}
\newcommand{\moy}[1]{\left\langle  #1 \right\rangle }

\begin{document}

\title{
On the relevance of disorder in athermal amorphous materials under shear
}

\author{Elisabeth Agoritsas}
\email[]{Elisabeth.Agoritsas@ujf-grenoble.fr}
\affiliation{Univ. Grenoble Alpes, LIPHY, F-38000 Grenoble, France}
\affiliation{CNRS, LIPHY, F-38000 Grenoble, France}
\author{Eric Bertin}
\affiliation{Univ. Grenoble Alpes, LIPHY, F-38000 Grenoble, France}
\affiliation{CNRS, LIPHY, F-38000 Grenoble, France}
\author{Kirsten Martens}
\affiliation{Univ. Grenoble Alpes, LIPHY, F-38000 Grenoble, France}
\affiliation{CNRS, LIPHY, F-38000 Grenoble, France}
\author{Jean-Louis Barrat}
\affiliation{Univ. Grenoble Alpes, LIPHY, F-38000 Grenoble, France}
\affiliation{CNRS, LIPHY, F-38000 Grenoble, France}

\date{\today}


\begin{abstract}

We show that, at least at a mean-field level, the effect of structural disorder in sheared amorphous media is very dissimilar depending on the thermal or athermal nature of their underlying dynamics.
We first introduce a toy model, including explicitly two types of noise (thermal \textit{versus} athermal).
Within this interpretation framework, we argue that mean-field athermal dynamics can be accounted for by the so-called H{\'e}braud-Lequeux (HL) model, in which the mechanical noise stems explicitly from the plastic activity in the sheared medium.
Then, we show that the inclusion of structural disorder, by means of a distribution of yield energy barriers, has no qualitative effect in the HL model, while such a disorder is known to be one of the key ingredients leading kinematically to a finite macroscopic yield stress in other mean-field descriptions, such as the Soft-Glassy-Rheology model.
We conclude that the statistical mechanisms at play in the emergence of a macroscopic yield stress, and a complex stationary dynamics at low shear rate, are different in thermal and athermal amorphous systems.

\end{abstract}


\maketitle


\section{Introduction}
\label{section-intro}

Understanding the nature of the plastic response of amorphous media to externally applied forces, and their resulting mechanical and rheological properties, is a challenging issue in current material science research.
Interestingly, materials that seem at first sight very different, like amorphous solids (\textit{e.g.}~metallic glasses, polymer glasses, granular materials) or yield stress fluids (\textit{e.g.}~gels, foams, dense emulsions),
share qualitatively a very similar yielding behavior in their plastic response to shear. 
Such a resemblance between the dynamics of hard and soft materials has already been highlighted in the field of crystal studies, where L.~Bragg and J.~F.~Nye used monodisperse bubble rafts in compression experiments, in order to mimic the dynamical properties of crystalline structures~\cite{bragg_nye_1947_ProcRSocLondA190_474}.

For dense, structurally disordered materials, it has been known for a long time that the yielding process starts at a well-defined material-dependent stress threshold. 
The ensuing flowing regimes are traditionally characterized by constitutive laws that describe plasticity as a homogeneous steady flow.
It was A.~Argon who first introduced the idea of localized shear events on a microscopic level as the physical mechanism underlying plasticity \cite{argon_1979_ActaMetallurgica27_47, argon_kuo_1979_MaterialsScienceAndEngineering39_101}, similar to the existence of defects in crystalline materials.
This idea of localized events is also at the basis of the Princen theory of foams, that features the so-called ``T1'' events, small rearrangements of bubbles in a disordered foam, that collectively generate the flow \cite{princen_1983_JColloidInterfaceScience91_160}. 
These pioneering ideas inspired many works both in numerics and in experiments.
Taking advantage of the modern developments in particle tracking techniques and of the boost in computer power, the idea of localized plastic events has been widely validated as a microscopic scenario.
Important experimental verifications have been given \textit{e.g.} in the field of colloids \cite{schall_weitz_spaepen_2007_Science318_1895}, gels \cite{manneville_2004_EurPhyJApplPhysP28_361} and granular materials \cite{amon_2012_PhysRevLett108_135502}, as well as in simulations of 
the response to shear in glasses at the particle scale \cite{maloney_lemaitre_2006_PhysRevE74_016118, tanguy_leonforte_JLB_2006_EPJE20_355}.

The local yielding picture has also led to the development of a variety of  models at a mesoscopic scale, with currently an increased research activity in order to connect qualitatively and quantitatively their predictions, both to simulations starting from a microscopic modeling, and to the observed macroscopic mechanical and rheological properties. For a recent review on this topic see ref.~\cite{rodney_2011_ModellingSimulMatterSciEng19_083001}.
The most important challenge for these mesoscopic approaches is to find the correct way to model the local yielding dynamics and to implement the effect of the long-range elastic response of the surrounding medium to the locally plastic zones.
Numerical studies on mesoscopic models with a spatial resolution do actually implement these interactions explicitly, using the Eshelby theory of elastic response to a local deformation \cite{baret_vandembroucq_2002_PhysRevLett89_195506,picard_2005_PhysRevE71_010501, homer_schuh_2009_ActaMaterialia57_2823}.
The resulting mechanical noise is in this way triggered by the yielding dynamics itself and is thus, by construction, self-consistent.
When constructing mean-field descriptions, it is then crucial to develop a mesoscopic picture integrating accurately into the evolution equations this mechanical noise and the associated activation processes.
This issue is even more important for the so-called athermal systems, where the stress relaxation due to thermal noise can be neglected, and new plastic events are triggered solely by the macroscopic shear and the resulting mechanical noise.

At a phenomenological level, there are several ways to describe the mechanical noise.
One of the first proposals was put forward in the Soft-Glassy-Rheology (SGR) model \cite{sollich_1997_PhysRevLett78_2020, sollich_1998_PhysRevE58_738}, by assuming that the mechanical noise acts as an effective activation temperature $x$, controlling an Arrhenius-like rate of plastic events.
Combined with structural disorder, this description yields a broad distribution of relaxation time scales for a sufficiently low effective temperature.
It leads in particular to a complex fluid behavior, to the emergence at sufficiently small $x$ of a Herschel-Bulkley type of rheological curve \cite{herschel_bulkley_1926_kolloid-zeitschrift39_291}. In other words, its macroscopic stress $\sigma_M$ as a function of the applied shear rate ${\dot{\gamma}}$ follows a power law with a threshold:
${\sigma_M (\dot{\gamma}) \stackrel{(\dot{\gamma} \to 0)}{=} \sigma_Y^{\text{SGR}} + A^{\text{SGR}} \, \dot{\gamma}^{(1-x)}}$, its exponent depending explicitly on the effective temperature.
This approach has recently been brought into question, at least in the scope of strictly athermal dynamics \cite{nicolas_martens_barrat_2014_EurPhysLett107_44003}, by pointing out that the subtle interplay between the different time scales at play in the sheared dynamics can actually jeopardize the exponential, Arrhenius-like rate of plastic events, at the core of the SGR model.

Another way to describe the mechanical noise, as proposed in the H{\'e}braud-Lequeux (HL) model \cite{hebraud_lequeux_1998_PhysRevLett81_2934}, is to model it more explicitly as a diffusion of the local stress with a diffusion coefficient proportional to the plastic activity, \textit{i.e.}, the rate at which plastic events occur.
As we shall see below, this modeling of the mechanical noise differs from the SGR approach not only because the noise amplitude is a dynamical variable, but also because the noise does not act on the same physical observables in the SGR and HL models.
This second model predicts in particular three different scaling regimes for the rheological law ${\sigma_M(\dot{\gamma})}$ controlled by a single coupling parameter, including a Herschel-Bulkley behavior of fixed exponent $1/2$.
It is important to note another crucial difference between these models: the SGR model includes the structural disorder as a key ingredient, without which no complex rheological behavior can emerge, while the original version of the HL model does not include any structural disorder.

In this paper, we study a generalization of the HL model including a distribution of yield energy barriers, and we show that its predictions for the rheological law at low shear rate are qualitatively robust with respect to the introduction of structural disorder.
The paper is organized as follows.
We first present in sect.~\ref{section-thermal-vs-athermal} a toy model that we use as an interpretation framework in order to distinguish between the thermal and athermal types of noise in the description of sheared amorphous media.
Then we recall in sect.~\ref{section-HL-model} the definition and main properties of the original HL model, before studying in sect.~\ref{section-disord-HL} its generalized version in the presence of structural disorder.
We conclude that the physical mechanisms that lead to a non-linear macroscopic response to shear are different in thermal and athermal sheared amorphous systems. Our mean-field mesoscopic analysis suggests that the impact of structural disorder strongly depends on the nature of noise (mechanical \textit{versus} thermal) in the local stress dynamics.

\section{Mechanical noise \textit{versus} thermal dynamics}
\label{section-thermal-vs-athermal}

In order to understand the distinction between the thermal or athermal nature of noise, it is useful to consider the following simple description, distinguishing two different local degrees of freedom on which a noise might act.
In a coarse-grained description, we decompose the system into small boxes, for which we can define the local  strain $\ell$ (that we assume to be scalar for simplicity) and the local stress $\sigma$ at each time $t$, under an external shear rate~${\dot{\gamma}(t)}$.
We emphasize the different interpretations of these two variables:
$\ell$ characterizes the local configuration of the system within the box,
while $\sigma$ characterizes the forces exerted by neighboring boxes, in line with the continuum mechanics interpretation of stress.

If the local element behaves as a linear Hookean spring, $\sigma$ and $\ell$ are linearly related.  However, as plasticity results from irreversible rearrangements of particles at a microscopic scale, a complete mesoscopic description should also include,  in addition to the evolution of the stress, the dynamics of an the 'internal' degree of freedom  $\ell$.
In what follows, we first define a toy model for the coupled dynamics of $\ell$ and $\sigma$, and we use it as an interpretation framework in order to infer and discuss the assumptions made on the underlying dynamics of the strain $\ell$ for a given effective dynamics of the stress $\sigma$.

\begin{center}
\begin{figure*}[!ht]
\includegraphics[width=2 \columnwidth]{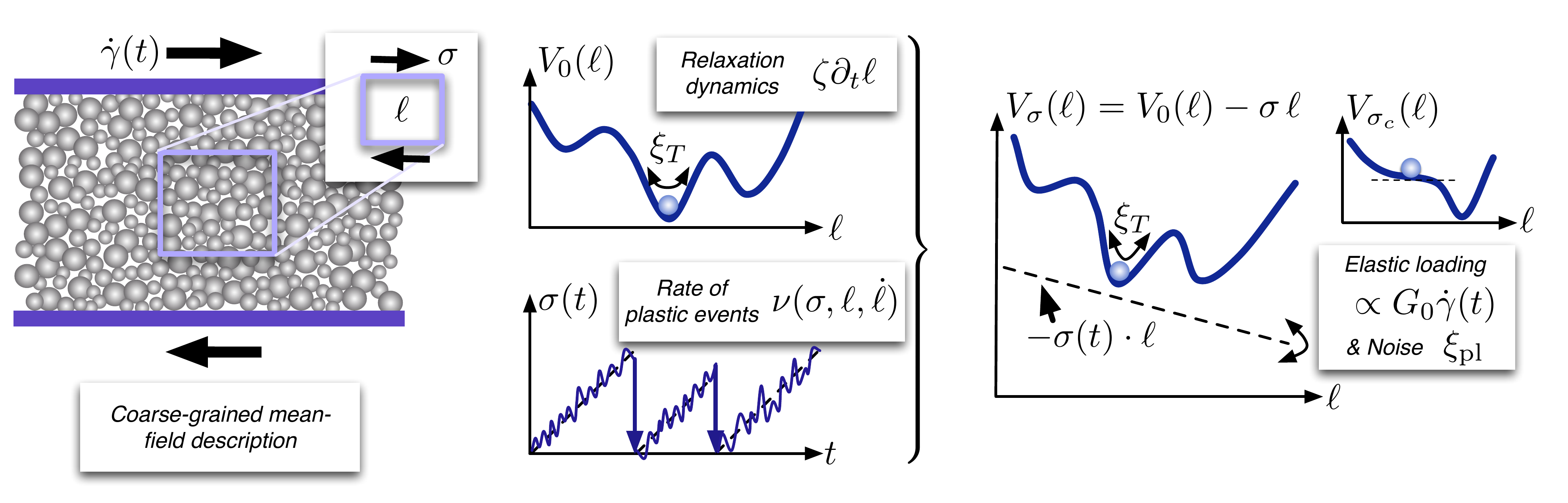}
 \caption{
 Summary of the mean-field picture presented in sect.~\ref{section-thermal-vs-athermal}, centered on two coarse-grained degrees of freedom:
 the local deformation $\ell$ and the local stress $\sigma$ (its shear component),
 whose coupled dynamics is governed by eqs.~\eqref{eq-evol-ell}-\eqref{eq-variance-xiplast}.
 \textit{Left:}~Starting from a microscopic description of an amorphous material, subjected to an external shear rate ${\dot{\gamma}(t)}$, we distinguish for a given mesoscopic box the local configuration $\ell$ of the particles inside the box, from the local stress $\sigma$ resulting from the particles outside the box.
 \textit{Center:}~In this simplified picture, if these two degrees of freedom were considered  separately, $\ell$ would relax in the configuration potential ${V_0(\ell)}$ accounting for the local structural disorder, whereas $\sigma$ would fluctuate around a sawtooth behavior controlled by a given rate ${\nu (\sigma,\ell,\dot{\ell})}$ of plastic events.
 \textit{Right:}~Their dynamics are actually coupled, tilting the bare potential ${V_0}$ with the fluctuating stress, and reciprocally relating the local yield stress $\sigma_c$ to the closest inflection point of ${V_0}$;
 the thermal noise $\xi_T$ and the mechanical noise ${\xi_{\text{pl}}}$ are defined as the noise of the individual Langevin dynamics of the two degrees of freedom $\ell$ and $\sigma$.
 }
 \label{fig:toymodel}
\end{figure*}
\end{center}

In the absence of stress, we first assume that the evolution of the coarse-grained internal variable $\ell$ is described as a dynamics in a potential ${V_0(\ell)}$, characterizing the local potential energy landscape of the possible configurations within the box.
For simplicity, we also assume this dynamics to be overdamped.
Secondly, we further assume that in the presence of a local stress $\sigma$, the potential ${V_0(\ell)}$ is modified by a linear contribution proportional to $\sigma$, becoming thus ${V_{\sigma}(\ell) = V_0(\ell) - \sigma \ell}$. In other words the stress tilts the potential energy landscape.
Thirdly, we assume that the local strain is subjected to a Gaussian white noise ${\xi_{T}(t)}$ --~ interpreted as a \emph{thermal} noise as it acts on the local configurational degree of freedom $\ell$~-- of zero mean and fixed variance, satisfying 
\begin{equation}
\label{eq-variance-xiT}
 \moy{\xi_{T}(t) \, \xi_{T}(t')}
 = 2 T \zeta \, \delta(t-t')
\end{equation}
with $T$ the `temperature' of the system, $\zeta$ the damping coefficient controlling its relaxation and ${\moy{\dots}}$ denoting the statistical average over the noise. 
This term describes the thermal contribution to the forces exerted by the surroundings, due \textit{e.g.} to acoustic-like vibrations. It can be significant for a small system at non-zero temperature, even in the absence of a macroscopic driving, but would be strictly zero in an athermal system.
The evolution equation for $\ell$ in this tilted potential then reads:
\begin{equation}
\label{eq-evol-ell}
 \zeta \partial_t \ell
  = - \frac{dV_0(\ell)}{d \ell} + \sigma (t) + \xi_{T}(t)
\end{equation}
In this picture, a fluctuation of $\ell$ within a local minimum of ${V_{\sigma}(\ell)}$ corresponds to an elastic local displacement of particles, whereas a jump above an energy barrier towards a new local minimum corresponds to a plastic local rearrangement, which can be triggered either by the `thermal' noise $\xi_T$ or by the fluctuating local stress ${\sigma (t)}$.
The dynamics of the local stress ${\sigma (t)}$ is assumed to be driven by the externally imposed shear rate which increase the stress, and by the yielding events which relax the stress. The dynamics is also affected by distant plastic rearrangements, that we model through a white noise $\xi_{\text{pl}}(t)$ acting on $\sigma$. We thus postulate the following dynamics,
\begin{equation}
\partial_t \sigma = G_0 \dot{\gamma}(t) + F(\sigma,\ell,\dot\ell) + \xi_{\text{pl}}(t)
\end{equation}
where $G_0$ is the shear elastic modulus, ${\dot{\ell}\equiv \partial_t \ell}$, and $F(\sigma,\ell,\dot\ell)$ is a coupling term between stress and deformation, leading to stress relaxation during a yielding event. The form of this coupling term is not known, although one might guess that it may crucially depend on the deformation rate $\dot\ell$, which becomes significantly larger after $\ell$ has crossed the energy barrier.
For simplicity, ${\xi_{\text{pl}}(t)}$ is assumed to be a Gaussian white noise, with a dynamical amplitude ${D_{\text{pl}}(t)}$ that may (slowly) depend on time:
\begin{equation}
\label{eq-variance-xiplast}
 \moy{\xi_{\text{pl}}(t) \, \xi_{\text{pl}}(t')}
 = D_{\text{pl}}(t) \, \delta(t-t')
\end{equation}
In the following, we further simplify the dynamics, and replace both the explicit dynamics of $\ell$ given in eq.~(\ref{eq-evol-ell}) by an effective stochastic rate $\nu(\sigma)$ with which the local stress $\sigma$ instantaneously relaxes to zero.
We thus end up with an effective hybrid stochastic evolution equation for $\sigma$, of the form
\begin{eqnarray}
 \label{eq-continuous-evol-sigma}
 && \partial_t \sigma = G_0 \dot{\gamma}(t) + \xi_{\text{pl}}(t) \\
 \label{eq-jump-sigma}
 && \sigma \mapsto 0 \; \text{with rate} \; \nu(\sigma,\ell,\dot{\ell})
\end{eqnarray}
The first term on the right hand side of eq.~\eqref{eq-continuous-evol-sigma} describes the continuous elastic load of the local stress due to the shear rate ${\dot{\gamma}(t)}$ ($G_0$ is the elastic shear modulus), while the second term ${\xi_{\text{pl}}(t)}$ is a noise accounting for the effect of distant plastic rearrangements, and as such can be interpreted as a \emph{mechanical} noise.
Combining eqs.~\eqref{eq-evol-ell} and~\eqref{eq-jump-sigma}, it is already clear (see also discussion below) that this noise has a cumulative effect on the dynamics of $\ell$, very different from the uncorrelated thermal noise $\xi_T$.
In order to fix the diffusion coefficient $D_{\text{pl}}(t)$, a closure relation has to be found, involving for instance the rate of plastic events in the system.
For example, in the HL model \cite{hebraud_lequeux_1998_PhysRevLett81_2934}, a simple closure relation is provided by the physically reasonable assumption that the diffusion coefficient is proportional to the average rate of plastic events.
This whole physical picture is summarized in fig.~\ref{fig:toymodel}.

If we make a change of variables and use, instead of the strain $\ell$, the local energy barrier  $E$ (\textit{i.e.} the distance in energy to the yield point, which depends on $\sigma$ and $\ell$),
the dynamics given by the eqs.~\eqref{eq-continuous-evol-sigma}-\eqref{eq-jump-sigma}-\eqref{eq-variance-xiplast} leads, eventually, to the following evolution equation for the joint distribution of the local stress $\sigma$ and local yield energy barrier $E$~:
\begin{equation}
\label{eq-dist-PsigmaE-generically}
\begin{split}
 \partial_t \widetilde{\mathcal{P}}(E,\sigma ,t)
 =	&	- G_0 \dot{\gamma}(t) \, \partial_\sigma \widetilde{\mathcal{P}}
 		+ D_{\text{pl}}(t) \, \partial_{\sigma}^2 \widetilde{\mathcal{P}} \\
 	& 	- \nu (\sigma , E) \, \widetilde{\mathcal{P}}
 		+ \Gamma (t) \, \delta (\sigma) \, \rho(E)
\end{split}
\end{equation}
with the plastic activity ${\Gamma (t)}$ defined as the averaged plastic rate ${\moy{\nu(\sigma,E)}_{\widetilde{\mathcal{P}}}=\int dE \int d\sigma \,\nu(\sigma,E) \, \widetilde{\mathcal{P}}(E,\sigma ,t)}$,
and ${\rho(E)}$ the probability distribution of energy barriers.
Alternatively, $E$ could be replaced by the local yield stress $\sigma_c$ or any other relevant feature of the configurational potential ${V_0(\ell)}$.
Such an evolution equation, with this specific structure, is precisely the starting point of the SGR~\cite{sollich_1997_PhysRevLett78_2020, sollich_1998_PhysRevE58_738} and HL~\cite{hebraud_lequeux_1998_PhysRevLett81_2934} models, two prototypal mean-field models for sheared amorphous systems.
In our toy model picture, the assumptions made on the underlying dynamics of $\ell$ thus translates into a given set of effective parameters $\arga{D_{\text{pl}}(t),\nu (\sigma , E),\rho(E)}$.

In the above formulation, the fact that two different types of noise may be present appears clearly.
On the one hand, the noise ${\xi_{T}(t)}$ in eq.~\eqref{eq-evol-ell} may be interpreted as a thermal noise, as it act on the local configurational degree of freedom $\ell$.
On the other hand, the noise ${\xi_{\text{pl}}(t)}$ expresses the effect of the mechanical noise, originating from distant plastic events.
This picture is of course a simplified local mean-field approximation, because
\textit{(i)}~the local deformation $\ell$ is only an effective coarse-grained variable, which does not describe all the positions of the particles in the box considered,
\textit{(ii)}~the relative proximity of distant plastic events is encoded in non-trivial correlations of the noise ${\xi_{\text{pl}}}$, which are completely neglected via the Gaussian white noise simplification,
and \textit{(iii)}~the local stress does not necessarily fully relax to zero after a plastic rearrangement, nor does it relax instantaneously.
Nevertheless, the important point is that, in our picture, both noises act on different degrees of freedom, and thus do not play an equivalent role. The noise $\xi_{T}(t)$ acts on the configurational variable $\ell$, which experiences a return force making hard to overcome the energy barrier (for a fixed energy landscape), leading to long and broadly-distributed escape times through the Arrhenius relation assumed in the SGR model.
On the contrary, the noise $\xi_{\text{pl}}(t)$ acts on the local stress $\sigma$ without any return force, so that reaching the value $\sigma_c$ at which the potential energy landscape changes of minima is comparatively easier, and takes a much shorter time. An alternative formulation \cite{nicolas_martens_barrat_2014_EurPhysLett107_44003} is to consider the fluctuations of $\sigma$ as a mechanical noise acting on the deformation $\ell$. However, in this view, the correlation of the noise does not decay to zero on a short time, but rather increases due to the persistent deformation induced by plastic events (or equivalently, due to the absence of recoil force acting on $\sigma$).
An important physical consequence of this property is related to the influence of disorder, as we shall see in Section~\ref{section-disord-HL}.

Within the present framework, the situation without mechanical noise (${\xi_{\text{pl}}(t)=0}$) corresponds to a purely `thermal' dynamics of the local strain $\ell$.
In this case, the local stress has a sawtooth behavior without fluctuations, whose stress drops can be identified as fast relaxations of the strain $\ell$ into a new local minimum of the potential energy landscape.
If we approximate a given well of the unstressed potential ${V_0(\ell)}$ by a harmonic potential centered on $\ell_0$, we have in presence of a local stress $\sigma$ that ${V_{\sigma}(\ell) \approx \frac{1}{2} k (\ell-\ell_0)^2 -\sigma (\ell-\ell_0)}$ and the local strain $\ell_{\sigma}$ corresponding to its minimum is given by ${\ell_{\sigma}-\ell_0 = \sigma/k}$.
If we assume furthermore that plasticity is dominated by plastic events triggered by the thermal noise $\xi_T(t)$ (and not by the local stress), then the dynamics of $\ell$ given by eq.~\eqref{eq-evol-ell} can be implicitly reduced to an Arrhenius escape rate above an energy barrier $\Delta E(\sigma)$.
%
The effective plastic rate in the stress evolution equation~\eqref{eq-jump-sigma} can then be written as
${\nu(\Delta E(\sigma)) \propto \exp \argc{-\Delta E(\sigma)/T}}$.
Since the stress diffusion coefficient has been assumed to be ${D_{\text{pl}}(t) =0}$, this Arrhenius-like activation description of plastic events is qualitatively similar to the SGR model~\cite{sollich_1997_PhysRevLett78_2020,sollich_1998_PhysRevE58_738},
in which the structural disorder encoded in ${\rho(E)}$ has been shown to play a crucial role on the rheological properties.

In contrast, the situation described by the HL model corresponds to a purely relaxational dynamics, without thermal noise, so that $\ell$ simply relaxes to the local minimum of the potential $V_{\sigma}(\ell)$.
In this picture, a threshold $\sigma_c$ is defined as the maximum value of $\valabs{\frac{dV_0(\ell)}{d \ell}}$ of the current barrier to overcome, and as such it determines the minimum value of local stress allowing for a change of potential well --a plastic rearrangement within the mesoscopic region-- in the athermal case, \textit{i.e.}, ${\xi_T(t) =0}$).
The rate $\nu$ is simply taken as zero if the local stress $\sigma$ is below the threshold $\sigma_c$  and takes a constant value ${1/\tau}$ if the local stress exceeds this threshold.
Moreover, in the original HL model \cite{hebraud_lequeux_1998_PhysRevLett81_2934}, $\sigma_c$ can take only a single typical value, whereas we will consider more generally a distribution of threshold values ${\rho (\sigma_c)}$.
Note that in the HL model, the degree of freedom $\ell$ is not explicitly described either, but including it allows for a better understanding of the origin of the local rearrangements in terms of the evolution of the local potential energy landscape.
We can for instance infer, at least formally, the distribution ${\rho (\sigma_c)}$ from the distribution of the disordered potential ${\overline{\mathcal{P}} \argc{V_0 (\ell)}}$ itself.
To sum up, in the HL model, only the mechanical noise $\xi_{\text{pl}}(t)$ is taken into account and there is no `thermal' noise on the configurational variable $\ell$,
so in our interpretation framework the HL description corresponds an athermal dynamics of a sheared amorphous system, whereas an Arrhenius-activated plasticity corresponds to a thermal dynamics.
The main goal of this paper is to investigate the role of disorder, encoded in the distribution of barrier heights $E$ or yield stress $\sigma_c$, on the rheological properties in the case of a pure athermal dynamics as described by the HL model.

Let us conclude this section with a word of caution.
First, in the simplified framework presented above and summarized in fig.~\ref{fig:toymodel}, it appears rather natural to identify the temperature $T$ with the `real' temperature, \textit{i.e.}, to set it to zero in athermal systems.
Nevertheless, most interpretations of the SGR model introduce the notion of an effective temperature associated with the mechanical noise, which has the same role, but may be unrelated to the physical temperature.
Some other models, such as the Shear-Transformation-Zone (STZ) model \cite{falk_langer_2011_AnnuRevCondensMatterPhys2_353},
also introduce the notion of an effective temperature, which is however not used, in general, to compute an activation rate but rather as an internal variable that characterizes the state of the material.
Secondly, in the view presented above, the mechanical noise is only described by the fluctuating term in the stress evolution, and is thus totally absent from the equation of the strain evolution (except for its coupling to the fluctuating stress $\sigma$). Whether or not a remnant of the mechanical noise should also be considered at this level, due to the very strong simplification that replaces the complex energy landscape of a few tens of particles with a single degree of freedom $\ell$, remains an open question.

\section{The H{\'e}braud-Lequeux model}
\label{section-HL-model}


Although the HL model \cite{hebraud_lequeux_1998_PhysRevLett81_2934} has been previously studied in great detail in the  mathematical literature
\cite{phdthesis_YousraGati2004,
cances_catto_gati_2006_SIAMJMathAnal37_60,
cances_catto_gati_lebris_2006_MultiscaleModelSimul4_1041,
phdthesis_JulienOlivier2011,
olivier_2010_ZAngewMathPhys61_445,
olivier_renardy_2011_SIAMJApplMath71_1144,
olivier_2012_SciChinaMath55_435},
no concise explicit account of the derivation of its rheological law is available so far  in the physics literature.
Before generalizing the HL model by including a distribution of threshold stresses (see Sect.~\ref{section-disord-HL}), we find  useful to provide the main steps of the derivation for the standard HL model.
In this section, we thus recall
first the definitions of the HL model,
second the main steps of the derivation at fixed shear rate ${\dot{\gamma}}$ (and further in the limit ${\dot{\gamma} \to 0}$) of its stationary solution,
third its corresponding predictions for the macroscopic stress ${\sigma_M (\dot{\gamma})}$,
and fourth its connection with the Kinetic-Elasto-Plastic (KEP) model \cite{bocquet_PhysRevLett103_036001},
since all these points will prove useful for the study of its disordered generalization in the next section.
We detail in particular the scalings of the macroscopic stress, with their associated prefactors, in the different limits of interest,
expressing them in a way that will be systematically generalized in sect.~\ref{section-disord-HL}.

\subsection{Definition of the HL model}
\label{section-HL-model-defmodel}

Introduced closely after the SGR model, the original HL model \cite{hebraud_lequeux_1998_PhysRevLett81_2934} is defined by the following evolution equation for the probability distribution  function (PDF) ${\mathcal{P}(\sigma,t)}$ of the local stress $\sigma$ at time $t$, under an external shear rate ${\dot{\gamma}(t)}$:
\begin{equation}
\label{eq-dist-Psigma-HL}
\begin{split}
 \partial_t \mathcal{P}(\sigma ,t)
 =	&	- G_0 \dot{\gamma}(t) \, \partial_\sigma \mathcal{P}
 		+ D_{\text{HL}}(t) \, \partial_{\sigma}^2 \mathcal{P} \\
 	& 	- \nu_{\text{HL}} (\sigma , \sigma_c) \, \mathcal{P}
 		+ \Gamma (t) \, \delta (\sigma)
\end{split}
\end{equation}
It corresponds to the hybrid stochastic dynamics defined by eq.~\eqref{eq-continuous-evol-sigma}-\eqref{eq-variance-xiplast} for the mean-field local stress,
with ${D_{\text{pl}}(t)=D_{\text{HL}}(t)}$, and the following specific rate $\nu_{\text{HL}}$ and plastic activity ${\Gamma(t)}$:
\begin{eqnarray}
 \label{eq-nu-HL}
 && \nu_{\text{HL}} (\sigma , \sigma_c)
 \equiv \frac{1}{\tau} \theta (\valabs{\sigma} - \sigma_c) \\
 \label{eq-Gamma-nu-HL}
 && \Gamma(t)
 = \moy{\nu_{\text{HL}} (\sigma , \sigma_c)}_{\mathcal{P}(\sigma ,t)}
 = \frac{1}{\tau} \int_{\valabs{\sigma'}> \sigma_c} \!\!\!\!\!\!\!\! d \sigma' \, \mathcal{P}(\sigma' ,t)
\end{eqnarray}
where $\theta$ is the Heaviside function and $\delta$ the Dirac distribution.
The choice \eqref{eq-nu-HL} assumes that there is a single typical value of the threshold stress $\sigma_c$ and that the rate of plastic event can be approximated by a fixed value $1/\tau$ in any overstressed region.
As for the plastic activity, it is defined as the mean rate of plastic events --the first equality in \eqref{eq-Gamma-nu-HL} is imposed in general by the normalization ${\int_{-\infty}^{\infty} d\sigma \, \mathcal{P}(\sigma ,t)=1}$ of the PDF-- and is then quantified by the proportion of overstressed regions in the system.
The Dirac distribution ${\delta (\sigma)}$ in \eqref{eq-dist-Psigma-HL} corresponds to the full relaxation condition stated in \eqref{eq-jump-sigma}, and this condition could be replaced more generally by a distribution ${\Delta (\sigma)}$ of the stress after relaxation.
Finally, and this is the key ingredient of the model, the diffusion coefficient \eqref{eq-variance-xiplast} is assumed to be proportional to the plastic activity, hence the following linear closure relation:
\begin{equation}
\label{eq-closure-DHL}
 D_{\text{HL}}(t) = \alpha \, \Gamma(t)
\end{equation}
where $\alpha >0$ is for the time being an ad hoc parameter of the model.
Since the plastic activity depends on the PDF, this relation implies that the evolution equation~\eqref{eq-dist-Psigma-HL} is nonlinear in ${\mathcal{P}(\sigma,t)}$, the nonlinearity being encoded in the diffusion coefficient and leading to the non-trivial features of the HL model.

\subsection{Stationary solution at fixed shear rate}
\label{section-HL-model-stationary-solution}

We focus exclusively on the case of constant shear rate ${\dot{\gamma}}$, which has also been studied in the mathematical literature \cite{phdthesis_YousraGati2004,phdthesis_JulienOlivier2011,olivier_2010_ZAngewMathPhys61_445,olivier_renardy_2011_SIAMJApplMath71_1144}, proving in particular the existence and uniqueness of its stationary solution at a constant shear rate, in the specific limit ${\dot{\gamma} \to 0}$.

Assuming that the stationary PDF ${\mathcal{P}_{\text{st}}(\sigma)}$ exists, its corresponding plastic activity ${\Gamma_{\text{st}}}$ and stationary diffusion coefficient $D_{\text{HL}}$ are well-defined,
hence the determination of the stationary solution of the HL model proceeds generically in the following way.
\textit{(i)}~One does not take into account the closure relation eq.~\eqref{eq-closure-DHL}, and considers ${D_{\text{HL}}}$ as a fixed diffusion constant $D$.
\textit{(ii)}~The stationary solution of eq.~\eqref{eq-dist-Psigma-HL} can then be determined, and in this specific case one has to solve a second-order differential equation with constant coefficients on each of the intervals $(-\infty,-\sigma_c)$, $(-\sigma_c,0)$, $(0, \sigma_c)$ and $(\sigma_c,+\infty)$, connecting them using the continuity of ${\mathcal{P}_{\text{st}}(\sigma)}$ and of its derivative at ${\sigma=\pm \sigma_c}$.
\textit{(iii)}~${\mathcal{P}_{\text{st}}(\sigma)}$ has actually an overall multiplicative constant, that can be identified with the (as yet unknown) plastic activity $\Gamma_{\text{st}}$ defined by \eqref{eq-Gamma-nu-HL}; using the normalization condition ${\int_{-\infty}^{\infty} d\sigma \, \mathcal{P}_{\text{st}}(\sigma)=1}$, one ends up with an equation of the form:
\begin{equation}
\label{eq-Gamma-HL-D-factorf}
 \Gamma_{\text{st}} \tau
 = \frac{D \tau}{\tilde{f}_{\sigma_c} \argp{\sqrt{D \tau}, \frac{G_0 \dot{\gamma} \tau}{D \tau}}}
\end{equation}
where ${\tilde{f}_{\sigma_c}  (x,y)}$ is a known function (cf.~Appendix~\ref{A-appendix-factorf-sigmac}).
\textit{(iv)}~At this stage, the closure relation \eqref{eq-closure-DHL} can at last be taken into account, yielding either ${D=0}$, or a finite diffusion coefficient according to the condition:
\begin{equation}
\label{eq-implicit-for-D}
 \tilde{f}_{\sigma_c} \argp{\sqrt{D \tau}, \frac{G_0 \dot{\gamma} \tau}{D \tau}}
 = \alpha
\end{equation}
from which ${D=D_{\text{HL}}}$ can be determined uniquely as a function of the shear rate ${\dot{\gamma}}$ and of the coupling parameter $\alpha$.

\begin{center}
\begin{figure}[!htb]
\includegraphics[width=\columnwidth]{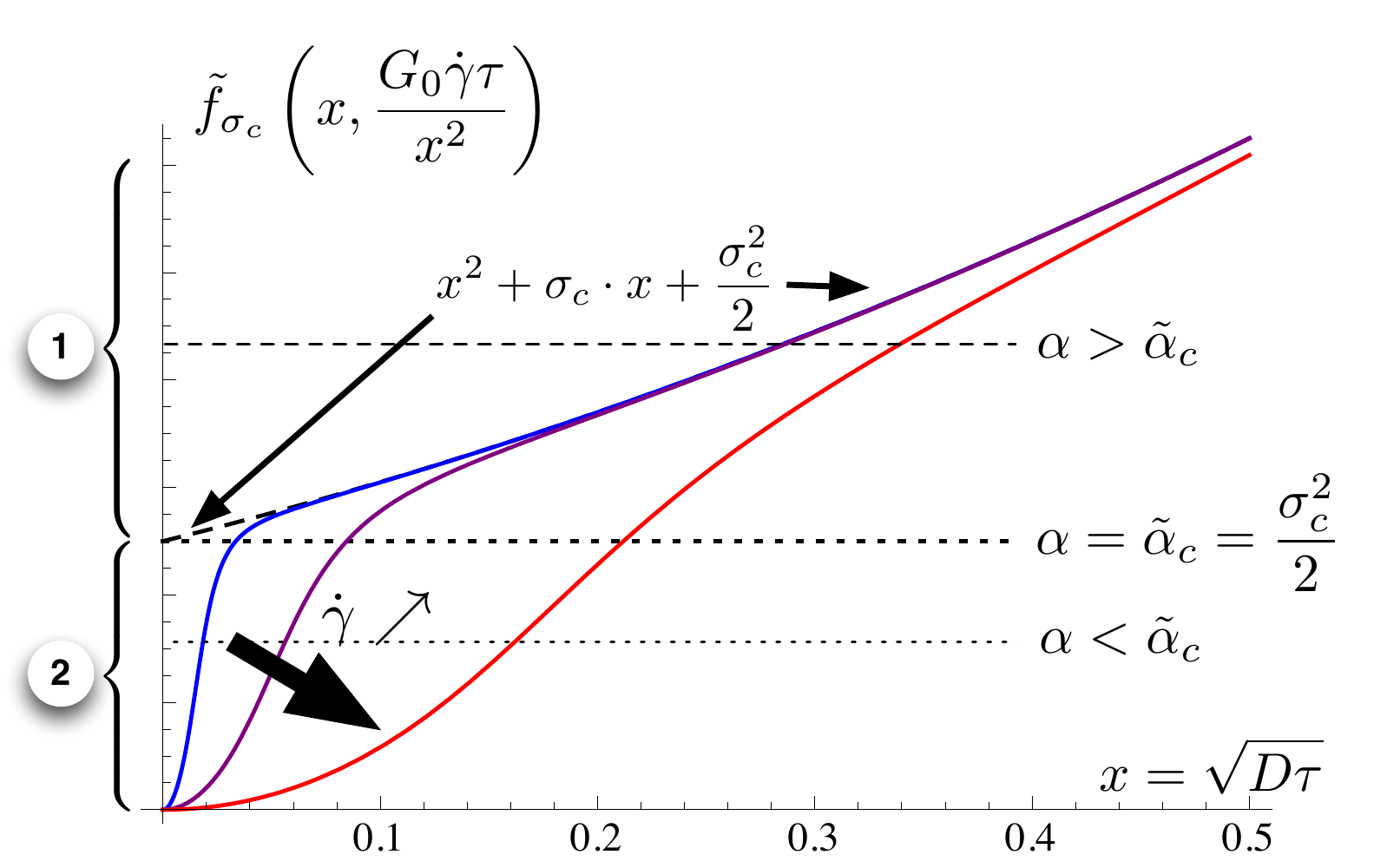}
 \caption{
 Schematic plot of the function ${\tilde{f}_{\sigma_c} \argp{\sqrt{D \tau}, \frac{G_0 \dot{\gamma} \tau}{D \tau}}}$ defined in eq.~\eqref{eq-Gamma-HL-D-factorf},
in the presence of a constant external shear rate ${\dot{\gamma} \geq 0}$.
 According to eq.~\eqref{eq-implicit-for-D}, this plot provides a geometrical interpretation of the solution ${x=\sqrt{D_{\text{HL}} \, \tau}}$ imposed by the closure relation \eqref{eq-closure-DHL}, as the intersection between the function ${\tilde{f}_{\sigma_c}}$ and the horizontal line $\alpha$.
 The black dashed parabola is the limiting case at ${\dot{\gamma}=0}$ of eq.~\eqref{eq-factorf-HL-noshear}, whose minimum defines the critical value $\tilde{\alpha}_c$.
 The three curves with increasing ${\dot{\gamma}}$ correspond to the more general case of eq.~\eqref{eq-factorf-HL-withshear}.
 The different cases for ${\lim_{\dot{\gamma} \to 0} D_{\text{HL}}(\dot{\gamma}, \alpha)}$ listed in eqs.~\eqref{eq-DHL-withshear-smallgammadot-1}-\eqref{eq-DHL-withshear-smallgammadot-2} correspond to the intersections with ${\tilde{f}_{\sigma_c}}$ which occur above $\tilde{\alpha}_c$ (area 1), below $\tilde{\alpha}_c$ (area 2) or exactly at $\tilde{\alpha}_c$.
 }
 \label{fig:factorf-standardHL}
\end{figure}
\end{center}

Note first that this procedure is quite generic, in the sense that another choice for the rate ${\nu (\sigma,\sigma_c)}$ in eq.~\eqref{eq-nu-HL} will only modify the functional form of ${\tilde{f}_{\sigma_c}}$,
and secondly we are free to choose a different closure relation than \eqref{eq-closure-DHL} starting from eq.~\eqref{eq-Gamma-HL-D-factorf}.
The last equation \eqref{eq-implicit-for-D} has actually a straightforward geometrical interpretation, illustrated in fig.~\ref{fig:factorf-standardHL}.
The two arguments of the function ${\tilde{f}_{\sigma_c}}$, respectively ${x=\sqrt{D\tau}}$ and ${y = G_0 \dot{\gamma} \tau/x^2}$, are natural choices given the structure of the HL equations~\eqref{eq-dist-Psigma-HL}-\eqref{eq-nu-HL},
that we thus keep on purpose in all our results.

\subsection{Stationary solution in the absence of shear rate}
\label{section-HL-model-stationary-noshearrate}

In absence of shear rate ($\dot{\gamma}=0$), the stationary PDF is symmetric so it predicts no macroscopic stress:
${\sigma_M \equiv \int_{-\infty}^{\infty} d \sigma \, \sigma \, \mathcal{P}_{\text{st}}(\sigma) =0}$.
This case is nevertheless a benchmark of the model, as it exhibits two regimes for the diffusion coefficient depending on the value of the coupling parameter $\alpha$ in the relation \eqref{eq-closure-DHL}.
Indeed, the HL model at ${\dot{\gamma}=0}$ yields:
\begin{equation}
\label{eq-factorf-HL-noshear}
 x \equiv \sqrt{D \tau}
 , \quad
 \tilde{f}_{\sigma_c} \argp{x,0}
 = x^2 + \sigma_c \, x + \frac{\sigma_c^2}{2}
\end{equation}
Defining the critical value ${\tilde{\alpha}_c = \tilde{f}_{\sigma_c}(0,0)=\sigma_c^2/2}$ as the minimum of this function, there is no solution of eq.~\eqref{eq-implicit-for-D} at ${\alpha < \tilde{\alpha}_c}$ (\textit{i.e.}, no intersection of ${\tilde{f}_{\sigma_c}}$ with ${\alpha}$ in fig.~\ref{fig:factorf-standardHL}) and hence ${D_{\text{HL}}(\alpha < \tilde{\alpha}_c)=0}$.
On the contrary, at ${\alpha \geq \tilde{\alpha}_c}$ eq.~\eqref{eq-implicit-for-D} always has a solution:
\begin{equation}
 \sqrt{D_{\text{HL}} \tau} =
 \left\lbrace \begin{array}{ll}
	\frac12 \sigma_c \argp{\sqrt{\frac{4 (\alpha - \tilde{\alpha}_c)}{\sigma_c^2} + 1} -1}
	& (\alpha \geq \tilde{\alpha}_c) \\ \\
	\frac{\alpha-\tilde{\alpha}_c}{\sigma_c} + \mathcal{O} \argp{(\alpha-\tilde{\alpha}_c)^2}
	& (\alpha \gtrsim \tilde{\alpha}_c)
 \end{array} \right.
\label{eq-solution-D-without-gammadot}
\end{equation}
Physically, $\tilde{\alpha}_c$ is thus the lower threshold for the coupling of the diffusion coefficient with the overstressed regions in eq.~\eqref{eq-closure-DHL},
such that a self-sustainable plastic diffusion can be reached even in the absence of an external shear rate.
However, because of the dissipative nature of plastic processes, such a stationary regime cannot be realized without some energy input to the system, and it is thus an artifact of the model for $\dot{\gamma}=0$. In the next section we will study the situation where such an external energy input will be provided via the macroscopically applied steady shear.

\subsection{Stationary solution at low shear rate}
\label{section-HL-model-stationary-low-shearrate}

In the presence of a finite constant shear rate, the stationary PDF is asymmetric and predicts a finite macroscopic stress ${\sigma_M(\dot{\gamma},\alpha)}$.
We have at ${\dot{\gamma}>0}$:
\begin{equation}
\label{eq-factorf-HL-withshear}
 \tilde{f}_{\sigma_c} \argp{x, y}
 \equiv x^2 + \frac{\sigma_c}{y} \, \frac{1 + \argp{\sqrt{1+\frac{4}{x^2y^2}}+\frac{2}{\sigma_c y}} \, \tanh \argp{\frac{\sigma_c y}{2}}}{\tanh \argp{\frac{\sigma_c y}{2}} + \sqrt{1+\frac{4}{x^2 y^2}}}
\end{equation}
with ${x = \sqrt{D \tau}}$ and ${y = G_0 \dot{\gamma} \tau/x^2}$
and, as illustrated in fig.~\ref{fig:factorf-standardHL} for three values of ${\dot{\gamma}>0}$, there is always a solution of eq.~\eqref{eq-implicit-for-D} for the stationary diffusion coefficient ${D_{\text{HL}}}$, which can be determined numerically by combining eqs.~\eqref{eq-factorf-HL-withshear} and \eqref{eq-implicit-for-D}.
Note that, dimensionally, we have ${\argc{x}=\argc{1/y}=\argc{\sigma_c}}$ and ${\argc{ \tilde{f}_{\sigma_c}}=\argc{\alpha}=\argc{D\tau}=\argc{\sigma_c^2}}$.

In the limit of vanishing shear rate ${\dot{\gamma} \to 0}$, the lowest-order scalings of ${D_{\text{HL}}(\dot{\gamma})}$ can be written down explicitly (see Appendix~\ref{A-appendix-perturb-diff-coeff}).
Distinguishing the two regimes ${\alpha \gtrless \tilde{\alpha}_c}$ and their limiting case ${\alpha=\tilde{\alpha}_c}$, one thus obtains:
\begin{equation}
\label{eq-DHL-withshear-smallgammadot-1}
\left\lbrace \begin{array}{cl}
 \alpha > \tilde{\alpha}_c :
 & D_{\text{HL}} \tau
 	\approx D_{\text{HL}}(\dot{\gamma}=0) \tau
 	\quad \text{(see eq.~\eqref{eq-solution-D-without-gammadot})}
 \\ \\
 \alpha = \tilde{\alpha}_c :
 &  D_{\text{HL}} \tau	
	\approx \argp{\frac{\sigma_c^{3}}{24}}^{2/5} \argp{G_0 \dot{\gamma} \tau}^{4/5}	
 \\ \\
 \alpha < \tilde{\alpha}_c :
 &  D_{\text{HL}} \tau
 	\approx C \, G_0 \dot{\gamma} \tau
\end{array} \right.
\end{equation}
where $C$ satisfies the following implicit equation
\begin{equation}
C \sigma_c \tanh \argp{\frac{\sigma_c}{2 C}}= \alpha
\end{equation}
In addition, we get the specific perturbative expansions for $\alpha$ close to or much smaller than $\tilde{\alpha}_c$:
\begin{equation}
\label{eq-DHL-withshear-smallgammadot-2}
\begin{split}
 \alpha \gtrsim \tilde{\alpha}_c :
 & \quad D_{\text{HL}} \tau
 	\stackrel{\eqref{eq-solution-D-without-gammadot}}{\approx} \argp{\frac{\alpha-\tilde{\alpha}_c}{\sigma_c}}^2
 \\
 \alpha \lesssim \tilde{\alpha}_c :
 & \quad D_{\text{HL}} \tau
 	\approx \frac{1}{2 \sqrt{6}} \frac{\sigma_c^2}{\sqrt{\tilde{\alpha}_c - \alpha}} \, G_0 \dot{\gamma} \tau
 \\
 \alpha \ll \tilde{\alpha}_c :
 & \quad D_{\text{HL}} \tau
 	\approx \frac{\alpha}{\sigma_c} \, G_0 \dot{\gamma} \tau
\end{split}
\end{equation}
Note that the perturbative expansion eq.~\eqref{eq-DHL-withshear-smallgammadot-2} involves two limits that cannot be exchanged: the limit $\dot{\gamma} \to 0$ should be taken first, before the limit 
$\alpha \to \tilde{\alpha}_c$. Considering the limit $\alpha \to \tilde{\alpha}_c$ at fixed $\dot{\gamma} >0$ in eq.~\eqref{eq-DHL-withshear-smallgammadot-2} leads to inconsistencies, as $D_{\text{HL}}$ would then diverge when $\alpha \to \tilde{\alpha}_c^{-}$.
This divergence is non-physical, since ${\Gamma_{\text{st}} \tau = \frac{D_{\text{HL}} \tau}{\alpha} \leq 1}$ is bounded by $1$ as a probability,
and it actually appears as soon as the assumption of a `low' $\dot{\gamma}$ breaks down,
restricting the validity range of the previous perturbative expansion of ${D_{\text{HL}} (\alpha \lesssim \tilde{\alpha}_c)}$ to shear rates at least lower than ${\dot{\gamma}_*(\alpha) = \frac{\sqrt{24 (\tilde{\alpha}_c - \alpha)}}{G_0 \tau \, \sigma_c^2 \alpha}}$. This upper bound goes to $0$ in the limit ${\alpha \to \tilde{\alpha}_c^-}$.
As for the third limit $\alpha \ll \tilde{\alpha}_c$, we recover as expected by self-consistency that, if we remove the coupling between the diffusion and the plastic activity, by setting the coupling parameter ${\alpha}$ close to zero, the diffusion coefficient should vanish as well.

Once the diffusion coefficient ${D_{\text{HL}}}$ is known, the stationary distribution ${\mathcal{P}_{\text{st}}(\sigma)}$ is fully determined, and its average value ${\sigma_M \equiv \int_{-\infty}^{\infty} d \sigma \, \sigma \, \mathcal{P}_{\text{st}}(\sigma)}$ can be computed, yielding eventually an analytical prediction for the macroscopic stress in the limit ${\dot{\gamma} \to 0}$:
\begin{equation}
\label{eq-sigmaM-withshear-smallgammadot-1}
\begin{split}
 \alpha > \tilde{\alpha}_c :
 & \quad \sigma_M \approx
 \argc{1+
	\argp{\frac{\sigma_c^3}{6 x_0} + \frac{\sigma_c^4}{24 x_0^2}} \frac{1}{ \tilde{f}_{\sigma_c} \argp{x_0,0}}
	} \, G_0 \dot{\gamma} \tau
\\
 \alpha = \tilde{\alpha}_c :
 & \quad \sigma_M \approx
 \frac{1}{2^{4/5} \times 3^{3/5}} \, \sigma_c^{4/5} \, \argp{G_0 \dot{\gamma} \tau}^{1/5}
\\
 \alpha < \tilde{\alpha}_c :
 & \quad \sigma_M \approx
 \tilde{\sigma}_Y(\sigma_c) + \widetilde{A}(\sigma_c) \cdot \argp{G_0 \dot{\gamma} \tau}^{1/2}
\end{split}
\end{equation}
with ${x_0 = \sqrt{D_{\text{HL}}(\dot{\gamma}=0) \tau}}$ and ${\tilde{f}_{\sigma_c}}$ respectively given by eqs.~\eqref{eq-solution-D-without-gammadot} and~\eqref{eq-factorf-HL-noshear}. In the last case $\alpha < \tilde{\alpha}_c$, the prediction of a threshold stress and of a power-law dependence on the shear rate, corresponds to a Herschel-Bulkley law of exponent $1/2$.
The corresponding macroscopic stress is given by:
\begin{equation}
 \label{eq-HL-law-sigmaY-1}
 \tilde{\sigma}_Y(\sigma_c)
 = C \argc{\frac{\sigma_c^2 /2}{C \sigma_c \, \tanh \argp{\frac{\sigma_c}{2 C}}} -1 }
\end{equation}
and the prefactor of the power law by:
\begin{equation}
\begin{split}
 \label{eq-HL-law-prefactorA-1}
 \widetilde{A}(\sigma_c)
 =& \frac{3 \sqrt{C}}{2} \coth \argp{\frac{\sigma_c}{2 C}} \\
 & \quad \quad + \frac{C^{3/2}}{\sigma_c} \argp{1+ \frac{\cosh\argp{\frac{\sigma_c}{C}}-1}{1- \frac{C}{\sigma_c} \sinh \argp{\frac{\sigma_c}{C}}}}
\end{split}
\end{equation}
with the factor $C$ determined by eq.~\eqref{eq-DHL-withshear-smallgammadot-1} (see Appendix~\ref{A-appendix-perturb-diff-coeff}).
We can finally give the perturbative expansions for the macroscopic stress corresponding to the limits given in eq.~\eqref{eq-DHL-withshear-smallgammadot-2},
first in the limit of small ${x_0}$,
then expanding the hyperbolic tangent depending on whether ${C}$ diverges or tends to zero:
\begin{equation}
\label{eq-sigmaM-withshear-smallgammadot-2}
\begin{split}
 \alpha \gtrsim \tilde{\alpha}_c :
 & \quad \frac{\sigma_M}{G_0 \dot{\gamma} \tau}
 	\stackrel{\eqref{eq-factorf-HL-noshear}}{\approx}
 		\frac{\sigma_c^2}{12 x_0^2}
 	\stackrel{\eqref{eq-solution-D-without-gammadot}}{\approx}
 		\frac{\sigma_c^4}{12} \argp{\alpha -\tilde{\alpha}_c}^{-2}
 	 \\ \\
 \alpha \lesssim \tilde{\alpha}_c :
 & \quad \tilde{\sigma}_Y
	\stackrel{\eqref{eq-HL-law-sigmaY-1}}{\approx}
 		\frac{\sigma_c^2}{12 C}
 	\stackrel{\eqref{eq-DHL-withshear-smallgammadot-2}}{\approx}
 		\frac{1}{\sqrt{6}} \argp{\tilde{\alpha}_c - \alpha}^{1/2}
 	\\
 & \quad	\widetilde{A}
 	\stackrel{\eqref{eq-HL-law-prefactorA-1}}{\approx}
 		\frac{C^{3/2}}{\sigma_c}
 	 \stackrel{\eqref{eq-DHL-withshear-smallgammadot-2}}{\approx}
		\frac{\sigma_c^2}{2^{3/2} \times 6^{3/4}} \argp{\tilde{\alpha}_c - \alpha}^{-3/4}
 \\ \\
 \alpha \ll \tilde{\alpha}_c :
 & \quad \tilde{\sigma}_Y
   \stackrel{\eqref{eq-HL-law-sigmaY-1}}{\approx}
  		\frac{\sigma_c}{2} - C
	\stackrel{\eqref{eq-DHL-withshear-smallgammadot-2}}{\approx}
 			\frac{\sigma_c}{2} - \frac{\alpha}{\sigma_c}
 	\\
 & \quad \widetilde{A}
 	\stackrel{\eqref{eq-HL-law-prefactorA-1}}{\approx}
 		\frac{\sqrt{C}}{2}
	\stackrel{\eqref{eq-DHL-withshear-smallgammadot-2}}{\approx}
		\frac12 \argp{\frac{\alpha}{\sigma_c}}^{1/2}
\end{split}
\end{equation}
Note that although ${\tilde{\sigma}_Y}$ tends to zero in the limit ${\alpha \to \tilde{\alpha}_c^-}$ and is thus physically well-behaved (predicting the disappearance of the macroscopic yield stress close to ${\tilde{\alpha}_c}$),
the validity range of its expression has been fixed previously in the perturbative expansion of the diffusion coefficient, with the upper bound ${\dot{\gamma}_*(\alpha) \sim  (\tilde{\alpha}_c - \alpha)^{1/2}}$.
A similar validity range can be defined in the limit ${\alpha \to \tilde{\alpha}_c^+}$,
when ${x_0^2}$ becomes comparable to ${G_0 \dot{\gamma} \tau}$,
with another upper bound ${\dot{\gamma}_{**}(\alpha) = \frac{12 (\tilde{\alpha}_c - \alpha)^2}{G_0 \tau \, \sigma_c^4}}$ squeezing this regime in the vicinity of ${\tilde{\alpha}_c}$.
As for the third limit ${\alpha \ll \tilde{\alpha}_c}$, we find that if we  progressively remove the diffusion term (\textit{i.e.}, ${\xi_{\text{pl}}(t) \equiv 0}$ in eq.~\eqref{eq-continuous-evol-sigma} of our toy model), we eventually destroy the Herschel-Bulkley behavior and the macroscopic yield stress tends to ${\sigma_c/2}$, which is the arithmetic average between the local yield stress $\sigma_c$ and the local stress after a plastic event ($0$ in the full relaxation assumption).

One of the strengths of the HL model is that it can predict three different scaling regimes for the stationary macroscopic stress ${\sigma_M (\dot{\gamma}) \sim \dot{\gamma}^{\delta}}$ at low shear rate, depending on the intensity of the plastic events feedback on the distribution of local stress, a feedback quantified by ${\alpha = D_{\text{HL}}/\Gamma_{\text{st}}}$.
Note that, on the one hand, a complete study of the existence and uniqueness of the stationary solution in the limit ${\dot{\gamma}} \to 0$ can be found in refs.~\cite{phdthesis_JulienOlivier2011,olivier_2010_ZAngewMathPhys61_445,olivier_renardy_2011_SIAMJApplMath71_1144}, in a dimensionless formulation.
On the other hand, the only three scalings given explicitly in the original paper by H\'ebraud and Lequeux~\cite{hebraud_lequeux_1998_PhysRevLett81_2934} are partial expressions for ${\sigma_M (\dot{\gamma})}$ at
${\alpha \gtrsim \tilde{\alpha}_c}$, ${\alpha \lesssim \tilde{\alpha}_c}$ and ${\alpha = \tilde{\alpha}_c}$.
However, the explicit dependence of all the predictions with respect to the fixed value of threshold stress $\sigma_c$ is crucial for the disordered generalization of the model in sect.~\ref{section-HL-model-stationary-low-shearrate-disord}.

\subsection{Connection with the KEP model}
\label{section-HL-model-stationary-KEP}

The original HL model \cite{hebraud_lequeux_1998_PhysRevLett81_2934} defined by eqs.~\eqref{eq-dist-Psigma-HL}-\eqref{eq-closure-DHL} thus proposes a simplified mean-field mesoscopic scenario that can reproduce different rheological laws for ${\sigma_M(\dot{\gamma})}$ depending on the value of $\alpha$.
However, it does not provide any justification for either the diffusion contribution ${D_{\text{HL}}(t) \, \partial_{\sigma}^2 \mathcal{P}}$ in the evolution equation \eqref{eq-dist-Psigma-HL}, or for the linear closure relation \eqref{eq-closure-DHL} relating the diffusion coefficient to the plastic activity with the control parameter ${\alpha >0}$.
The KEP model~\cite{bocquet_PhysRevLett103_036001} precisely provides such a justification, and as such we briefly recall its assumptions in order to complete the framework presented in this section.

Providing a derivation of the HL model from a spatial mesoscopic picture, the KEP model actually allows for the description of spatial inhomogeneities and the spatial propagation of stress after a plastic rearrangement.
It thus starts from an evolution equation ${\partial_t \mathcal{P}_i(\sigma ,t)}$ similar to eqs.~\eqref{eq-dist-Psigma-HL}-\eqref{eq-nu-HL}, but instead of the diffusion term, it includes an explicit spatial (discrete) dependence $i$ and a nonlocal operator (denoted `${\mathcal{L}(\mathcal{P},\mathcal{P})}$') coupling distant regions through an elastic propagator.
Note however that the local distributions ${\mathcal{P}_i(\sigma ,t)}$ are assumed to be independent in this operator, allowing for the factorization of their joint distributions (\textit{i.e.}, ${\mathcal{P}_{ij}=\mathcal{P}_i \, \mathcal{P}_j}$).

The KEP model eventually transforms into an effective HL equation, with an effective local shear rate ${\dot{\gamma}_i (t)}$ and a diffusion coefficient ${D_i(t)}$, using the following set of assumptions:
\textit{(i)}~A plastic rearrangement occurs at a site $j$ soon after its local stress ${\sigma'}$ has exceeded the local yield stress, so ${\valabs{\sigma'} \approx \sigma_c}$;
\textit{(ii)}~The stress is then fully relaxed at site $j$ (${\sigma' \to 0}$), so that the stress fluctuation that propagates from site $j$ to site $i$ is ${\delta \sigma_i^{(j)} \approx - G_{ij} \sigma' \approx - G_{ij} \sigma_c}$, with ${G_{ij}}$ the microscopic elastic propagator;
\textit{(iii)}~This stress ${\delta \sigma_i^{(j)}}$ acts as a perturbation that can trigger a plastic event only if the site $i$ is already on the verge of yielding, \textit{i.e.}, ${\valabs{\frac{\delta \sigma_i^{(j)}}{\sigma_c}} \approx \valabs{G_{ij}} \ll 1}$.
In other words, assuming either that a plastic rearrangement occurs sufficiently far away, or that the amplitude of the elastic propagator is small enough, then it can be treated as a noise acting on the mean-field local stress, in the spirit of eq.~\eqref{eq-continuous-evol-sigma}-\eqref{eq-jump-sigma}.
The KEP model then relates the effective diffusion coefficient $D$ to the plastic activity ${\Gamma}$, both at a given position ${\mathbf{r}}$, according to:
\begin{eqnarray}
\label{eq-closure-DHL-versionKEP-1}
 && D(\mathbf{r},t)
  = \tilde{m}(\sigma_c) \, \partial_{\mathbf{r}}^2 \Gamma (\mathbf{r},t) + \tilde{\alpha}(\sigma_c) \, \Gamma (\mathbf{r},t) \\
 \label{eq-closure-DHL-versionKEP-2} 
 && \left\lbrace \begin{array}{ccl}
 \tilde{m}(\sigma_c)
 &=& b^2 \, \sigma_c^2 \, G_{\text{nn}}^2 \\
  \tilde{\alpha}(\sigma_c)
 &=& \sigma_c^2 \, \sum_{i (\neq j)} G_{ij}^2
 \end{array} \right.
\end{eqnarray}
with $b$ the discrete lattice parameter, and $G_{\text{nn}}$ the nearest-neighbor propagator, that is the elastic propagator between neighboring blocks.
In the homogeneous and stationary case, the linear closure relation \eqref{eq-closure-DHL} is recovered, and the coupling parameter $\alpha$ can be identified with ${\tilde{\alpha} (\sigma_c)}$ given in eq.~\eqref{eq-closure-DHL-versionKEP-2}.
So the diffusion term in the HL model is justified and properly related to the physical quantities of the systems, namely the typical value of the yield stress $\sigma_c$ and the microscopic propagator ${G_{ij}}$.

The structural disorder is not included explicitly in the HL or KEP models, since the local potential energy landscape --~denoted $V_0(\ell)$ in sect.~\ref{section-thermal-vs-athermal}~-- is summarized into a single value of the yield stress $\sigma_c$.
In the next section, we will consider the generalization both of these constructions, by including a distribution of yield stress values ${\rho (\sigma_c)}$.

\section{H{\'e}braud-Lequeux model with structural disorder}
\label{section-disord-HL}



The assumption of a single value for the yield stress $\sigma_c$ is of course restrictive, and a natural generalization of the HL and KEP models relies on the existence of an a priori distribution of such threshold stresses ${\rho(\sigma_c)}$.
In the physical picture presented in sect.~\ref{section-thermal-vs-athermal}, it can formally be derived from the distribution of the local potential energy landscape ${V_0(\ell)}$, more specifically from the distribution of its inflection points, as illustrated in fig.~\ref{fig:toymodel} (right inset).
Our motivation for studying a generalization of the HL model that includes a structural disorder explicitly, via a distribution of yield stresses, is in particular to examine the robustness of the HL predictions for the macroscopic stress at vanishing constant shear rate ${\dot{\gamma}\to 0}$:
\begin{equation}
\label{eq-recap-HL-predictions}
\left\lbrace \begin{array}{cll}
 \alpha > \alpha_c :
& D \sim \dot{\gamma}^0
& \Rightarrow \sigma_M \sim \dot{\gamma}
\\
 \alpha = \alpha_c :
& D \sim \dot{\gamma}^{4/5}
& \Rightarrow \sigma_M \sim \dot{\gamma}^{1/5}
\\
 \alpha < \alpha_c :
& D \sim \dot{\gamma}
& \Rightarrow \sigma_M = \sigma_Y+A \, \dot{\gamma}^{1/2}
\end{array} \right.
\end{equation}
The questions we wish to address are the following.
Do we still predict three regimes controlled by the parameter $\alpha$, with these specific exponents?
If we predict qualitatively the same behaviors, what are the corresponding prefactors?
How to define the `critical' value $\alpha_c$?
And can we still derive such a disordered HL model from a KEP-like construction?

\subsection{Definition of the disordered HL model}
\label{section-HL-model-defmodel-disord}

Our disordered HL model is defined by the following evolution equation for the joint PDF ${\widetilde{\mathcal{P}}(\sigma_c,\sigma ,t)}$ of having a local stress $\sigma$ and a local yield stress $\sigma_c$ at time $t$, under an external shear rate ${\dot{\gamma}(t)}$:
\begin{equation}
\label{eq-dist-Psigma-HL-disord}
\begin{split}
 \partial_t \widetilde{\mathcal{P}}(\sigma_c,\sigma ,t)
 =	&	- G_0 \dot{\gamma}(t) \, \partial_\sigma \widetilde{\mathcal{P}}
 		+ \widetilde{D}(\sigma_c,t) \, \partial_{\sigma}^2 \widetilde{\mathcal{P}} \\
 	& 	- \nu_{\text{HL}} (\sigma , \sigma_c) \, \widetilde{\mathcal{P}}
 		+ \Gamma (t) \, \delta (\sigma) \, \rho(\sigma_c)
\end{split}
\end{equation}
with the following specific rate $\nu_{\text{HL}}$ and the plastic activities ${\widetilde{\Gamma}(\sigma_c,t)}$ and ${\Gamma(t)}$:
\begin{eqnarray}
 \label{eq-nu-HL-disord}
 && \nu_{\text{HL}} (\sigma , \sigma_c)
 \stackrel{\eqref{eq-nu-HL}}{=} \frac{1}{\tau} \theta (\valabs{\sigma} - \sigma_c) \\
 \label{eq-Gamma-nu-HL-disord-global}
 && \Gamma(t)
 = \moy{\nu_{\text{HL}} (\sigma , \sigma_c)}_{\widetilde{\mathcal{P}}(\sigma_c,\sigma ,t)}
 = \int_0^{\infty} \!\!\!\! d \sigma_c \, \widetilde{\Gamma}(\sigma_c,t) \\
  \label{eq-Gamma-nu-HL-disord-partial}
 && \widetilde{\Gamma}(\sigma_c,t)
 \equiv \frac{1}{\tau} \int_{\valabs{\sigma'}> \sigma_c} \!\!\!\!\!\!\!\!\!\! d \sigma' \, \widetilde{\mathcal{P}}(\sigma_c,\sigma' ,t)
\end{eqnarray}
where $\theta$ is the Heaviside function and $\delta$ the Dirac distribution.
The evolution equation \eqref{eq-dist-Psigma-HL-disord} differs from eq.~\eqref{eq-dist-Psigma-HL} only in its last term, which includes the a priori distribution ${\rho (\sigma_c)}$.
It corresponds to the hybrid stochastic dynamics defined by eq.~\eqref{eq-continuous-evol-sigma}-\eqref{eq-variance-xiplast} for the mean-field local stress, but randomly selecting according to ${\rho (\sigma_c)}$ a new yield stress value $\sigma_c$, after each plastic stress release in eq.~\eqref{eq-jump-sigma} (in the same spirit as in the SGR model \cite{sollich_1997_PhysRevLett78_2020,sollich_1998_PhysRevE58_738} and its precursor the Bouchaud trap model \cite{bouchaud_1995_JPhysIFrance5_1521}).
The choice \eqref{eq-nu-HL-disord} assumes that the rate of plastic events can be approximated by a fixed value $1/\tau$ in a locally overstressed region (where ${\valabs{\sigma} > \sigma_c}$).

We can now distinguish the global plastic activity ${\Gamma(t)}$ from its partial counterpart ${\widetilde{\Gamma}(\sigma_c,t)}$, defined respectively as the mean rate of plastic events for all the system, and the rate restricted to a given value of yield stress.
Similarly, we can distinguish the joint PDF ${\widetilde{\mathcal{P}}(\sigma_c,\sigma ,t)}$ from the PDFs of local stress and local yield stress, respectively:
\begin{eqnarray}
\label{eq-PDF-sigma-HLdisord}
 \mathcal{P}(\sigma,t)
 &=& \int_0^{\infty} \!\!\!\! d \sigma_c \, \widetilde{\mathcal{P}}(\sigma_c,\sigma ,t) \\
\label{eq-PDF-sigmac-HLdisord}
 \tilde{\rho} (\sigma_c,t)
 &=& \int_{-\infty}^{\infty} \!\!\!\! d \sigma \, \widetilde{\mathcal{P}}(\sigma_c,\sigma ,t)
\end{eqnarray}
The normalization ${\int_{-\infty}^{\infty} d\sigma \, \mathcal{P}(\sigma ,t)=1}$ imposes the first equality in eq.~\eqref{eq-Gamma-nu-HL-disord-global}.
By integrating eq.~\eqref{eq-dist-Psigma-HL-disord} over $\sigma$ under the assumption that both $\widetilde{\mathcal{P}}(\sigma_c,\sigma ,t)$ and $\partial_{\sigma}\widetilde{\mathcal{P}}(\sigma_c,\sigma ,t)$ vanish when $|\sigma| \to \infty$, we obtain the additional relation,
\begin{equation}
 \label{eq-rhotilde-evolution}
 \partial_t \tilde{\rho} (\sigma_c,t)
 = \Gamma(t) \, \rho (\sigma_c) - \widetilde{\Gamma}(\sigma_c,t)
\end{equation}
This equation describes how the distribution of local yield stress evolves in the system, depending on how the sample has been prepared (the initial condition ${\tilde{\rho} (\sigma_c,0)}$) and how fast plastic events refresh the potential energy landscape.
This picture simplifies only in the stationary state at fixed shear rate, where
\begin{equation}
\label{eq-rhotilde-stationary}
\partial_t \tilde{\rho}_{\text{st}} (\sigma_c) =0
 \; \Rightarrow \; \widetilde{\Gamma}_{\text{st}}(\sigma_c) = \Gamma_{\text{st}} \, \rho (\sigma_c)
\end{equation}
The partial activity $\widetilde{\Gamma}_{\text{st}}(\sigma_c)$ is then proportional to the a priori distribution ${\rho (\sigma_c)}$
(which in general differs from the dynamical steady-state distribution ${\tilde{\rho}_{\text{st}} (\sigma_c)}$).

As for the diffusion coefficient, it can generically be denoted ${\widetilde{D}(\sigma_c,t)}$, allowing for a different diffusion coefficient for each value of $\sigma_c$. Such a dependence on $\sigma_c$ would mean that the mechanical diffusion stemming from the plastic events would be different depending on the yield barrier to overcome, \textit{i.e.}, modifying the proportion of active sites depending on the barrier height.
This case cannot be completely ruled out physically, and it will be briefly discussed in Appendix~\ref{A-appendix-diffusion-coeff-depending-on-sigmac}.
Nevertheless, a generalization of the KEP construction --~that we will present at the end of this section~-- rather suggests that the mechanical diffusion coefficient should be assumed to be the same for all regions of the system, disregarding the local yield stress value, in the sense that it should gather in one parameter the collective feedback of all the overstressed regions that might yield.
In that case, since a plastic rearrangement with a higher barrier will release a larger stress in the rest of the system, we will assume that the diffusion coefficient ${\widetilde{D}(\sigma_c,t)}$ will simply be replaced by the generalized linear closure relation:
\begin{equation}
\label{eq-closure-DHL-disord}
 D_{\text{HL}}(t)
 = \int_0^{\infty} \!\!\!\! d \sigma_c \, \tilde{\alpha}{(\sigma_c)} \, \widetilde{\Gamma}(\sigma_c,t)
\end{equation}
where ${\tilde{\alpha}{(\sigma_c)}}$ is an ad hoc set of parameters of the model.
In the stationary case, we recover the same closure relation \eqref{eq-closure-DHL} as in the original HL model, by using eq.~\eqref{eq-rhotilde-stationary}:
\begin{eqnarray}
\label{eq-closure-DHL-disord-stat1}
 && D_{\text{HL}}
 = \alpha_{\text{eff}} \, \Gamma_{\text{st}} \\
\label{eq-closure-DHL-disord-stat2}
 && \alpha_{\text{eff}}
  \equiv \int_0^{\infty} \!\!\!\! d \sigma_c \, \tilde{\alpha}{(\sigma_c)} \, \rho (\sigma_c) 
  = \moy{ \tilde{\alpha}{(\sigma_c)}}
\end{eqnarray}
Note the introduction of the notation ${\moy{\mathcal{O}(\sigma_c)}}$ for the disorder average over ${\rho(\sigma_c)}$ of an arbitrary observable~$\mathcal{O}(\sigma_c)$.

The expressions for our disordered HL model will be given for a generic distribution ${\rho (\sigma_c)}$, and the perturbative expansions for the diffusion coefficient and the macroscopic stress will be valid as long as the moments of this distributions are finite (which is always the case in physical systems).
In particular, the choice ${\rho_{\delta}(\sigma_c)=\delta(\sigma_c - \sigma_c^\star)}$ corresponds to the standard HL model.
However, whenever needed for explicit computations regarding the graphs, we will consider specifically an exponential distribution for the threshold energy ${E=\frac{1}{2}\sigma_c^2}$, similarly to the structural disorder included in the SGR model~\cite{sollich_1997_PhysRevLett78_2020, sollich_1998_PhysRevE58_738}. In terms of the distribution of $\sigma_c$, this corresponds to
\begin{equation}
\label{eq-def-rho-SGR}
 \rho_0(\sigma_c) = 2 \sigma_c \, \exp \argp{-\sigma_c^2}
\end{equation}
with the mean value ${\moy{\sigma_c} = \sqrt{\pi}/2}$ and the second moment ${\moy{\sigma_c^2} = 1}$.
Its higher moments can be computed straightforwardly, and their rescalings with respect to the mean value ${\moy{\sigma_c^k}/\moy{\sigma_c}^k}$ yield only constant factors of order $1$.
%

\subsection{Stationary solution at fixed shear rate}
\label{section-HL-model-stationary-solution-disord}

We focus exclusively on the stationary solution at constant shear rate ${\dot{\gamma}}$, as it is the generalization of the HL predictions recalled in sect.~\ref{section-HL-model}. Thanks to eq.~\eqref{eq-rhotilde-stationary} the partial plastic activity can be straightforwardly related to its global counterpart with the distribution ${\rho (\sigma_c)}$.
Assuming that the stationary PDF ${\widetilde{\mathcal{P}}_{\text{st}}(\sigma_c,\sigma)}$ exists, the corresponding global plastic activity $\Gamma_{\text{st}}$ and stationary diffusion coefficient ${D}$ are well-defined. Hence the determination of the stationary solution of our disordered HL model proceeds in the same way as in the original HL model (see sect.~\ref{section-HL-model}). The corresponding explicit expressions are given in Appendix~\ref{A-appendix-explicit-expressions}, along with some technical hints regarding their derivation.

For a given $\sigma_c$, the stationary joint PDF ${\widetilde{\mathcal{P}}_{\text{st}}(\sigma_c,\sigma)}$ can be written as 
$\widetilde{\mathcal{P}}_{\text{st}}(\sigma_c,\sigma)=\mathcal{P}_{\sigma_c}(\sigma)\, \rho(\sigma_c)$, which defines $\mathcal{P}_{\sigma_c}(\sigma)$.
Using eq.~(\ref{eq-dist-Psigma-HL-disord}) in the stationary case, we find that the distribution $\mathcal{P}_{\sigma_c}(\sigma)$ obeys, for a fixed $\sigma_c$, an equation of the same form as that of the standard HL model, namely eq.~(\ref{eq-dist-Psigma-HL}). However, the distribution $\mathcal{P}_{\sigma_c}(\sigma)$ is not normalized to $1$, but instead it satisfies, using eq.~(\ref{eq-PDF-sigmac-HLdisord})
\begin{equation}
\int_{-\infty}^{\infty} \!\!\!\! d\sigma \, \mathcal{P}_{\sigma_c}(\sigma) = \frac{\tilde{\rho}_{\text{st}}(\sigma_c)}{\rho(\sigma_c)}
\end{equation}
Then, using the same solution procedure as in Sect.~\ref{section-HL-model-stationary-solution}, we end up with the relation
\begin{equation}
 \label{eq-rhotilde-stationary-disord-0}
 \tilde{\rho}_{\text{st}} (\sigma_c) = \Gamma_{\text{st}} \tau \, \rho(\sigma_c) \, \frac{\tilde{f}_{\sigma_c} \argp{\sqrt{D \tau}, \frac{G_0 \dot{\gamma} \tau}{D \tau}}}{D\tau}
\end{equation}
where ${\tilde{f}_{\sigma_c}}$ is exactly the same function as in eq.~\eqref{eq-Gamma-HL-D-factorf},
for instance the parabola \eqref{eq-factorf-HL-noshear} in the absence of shear rate and the function \eqref{eq-factorf-HL-withshear} in the presence of a constant shear rate.
We emphasize the key role that will be played by this function ${\tilde{f}_{\sigma_c}}$ in the present study of our disordered HL model.
Integrating  this last expression over $\sigma_c$, we obtain the counterpart of eq.~\eqref{eq-Gamma-HL-D-factorf} for the global plastic activity,
\begin{equation}
 \label{eq-rhotilde-stationary-disord-1}
 \Gamma_{\text{st}} \tau \int_0^{\infty} \!\!\!\! d \sigma_c\, \rho(\sigma_c) \, \frac{\tilde{f}_{\sigma_c} \argp{\sqrt{D \tau}, \frac{G_0 \dot{\gamma} \tau}{D \tau}}}{D\tau} = 1.
\end{equation}
Since we have restricted ourselves to the case where the stationary diffusion coefficient takes a fixed value $D$ independent of $\sigma_c$, as in eq.~\eqref{eq-closure-DHL-disord},
the previous relation simplifies to
\begin{equation}
 \label{eq-rhotilde-stationary-disord-2}
\Gamma_{\text{st}} \tau \, \frac{f_{\text{eff}} \argp{\sqrt{D \tau}, \frac{G_0 \dot{\gamma} \tau}{D\tau}}}{D \tau} = 1
\end{equation}
defining the following effective function,
\begin{equation}
 \label{eq-rhotilde-stationary-disord-3}
f_{\text{eff}}(x,y) \equiv \langle \tilde{f}_{\sigma_c}(x,y) \rangle
\end{equation}
Eq.~\eqref{eq-rhotilde-stationary-disord-1} can be used to compute ${\Gamma_{\text{st}}}$, at least numerically if not analytically, for any choice of ${\rho(\sigma_c)}$ and $D$.
Combined with the closure relation \eqref{eq-closure-DHL-disord-stat1}, it provides us with the generalized counterpart of eq.~\eqref{eq-implicit-for-D}:
\begin{equation}
\label{eq-implicit-for-D-disord}
 f_{\text{eff}} \argp{\sqrt{D \tau}, \frac{G_0 \dot{\gamma} \tau}{D\tau}}
 = \alpha_{\text{eff}}
\end{equation}
from which ${D=D_{\text{HL}}}$ can be determined uniquely as a function of the shear rate $\dot{\gamma}$ and of the effective coupling parameter ${\alpha_{\text{eff}}}$.

We emphasize that all this procedure is again quite generic --~with respect to the choice of the rate ${\nu (\sigma,\sigma_c)}$ and of the closure relation~-- and that it has the same geometrical interpretation as the one illustrated in fig.~\ref{fig:factorf-standardHL}.
The more generic case of a diffusion coefficient that would depend on $\sigma_c$, as initially included in eq.~\eqref{eq-dist-Psigma-HL-disord} for the evolution ${\partial_t \widetilde{\mathcal{P}}(\sigma_c,\sigma ,t)}$, is discussed in Appendix~\ref{A-appendix-diffusion-coeff-depending-on-sigmac}.

Moreover, in the stationary case and with a diffusion coefficient independent of $\sigma_c$, we have direct access to the distribution of local yield stress values, by combining eqs.~\eqref{eq-rhotilde-stationary}, \eqref{eq-rhotilde-stationary-disord-0} and \eqref{eq-rhotilde-stationary-disord-3}:
\begin{equation}
\label{eq-rhotilde-stationary-disord-final-generic}
 \tilde{\rho}_{\text{st}} (\sigma_c)
 = \rho(\sigma_c) \, \frac{\tilde{f}_{\sigma_c} \argp{\sqrt{D \tau}, \frac{G_0 \dot{\gamma} \tau}{D \tau}}}{\moy{\tilde{f}_{\sigma_c} \argp{\sqrt{D \tau}, \frac{G_0 \dot{\gamma} \tau}{D \tau}}}} \\
\end{equation}
confirming that the distributions $\tilde{\rho}_{\text{st}} (\sigma_c)$ and $\rho(\sigma_c)$ differ in general, except if $\rho(\sigma_c)$ reduces to a Dirac distribution.
This generic expression further simplifies with the closure relation \eqref{eq-implicit-for-D-disord}, replacing the denominator with ${\alpha_{\text{eff}}}$ and fixing ${D_{\text{HL}}=D(\dot{\gamma},\alpha_{\text{eff}})}$.

\begin{center}
\begin{figure}[!htb]
 \subfigure{\includegraphics[width=\columnwidth]{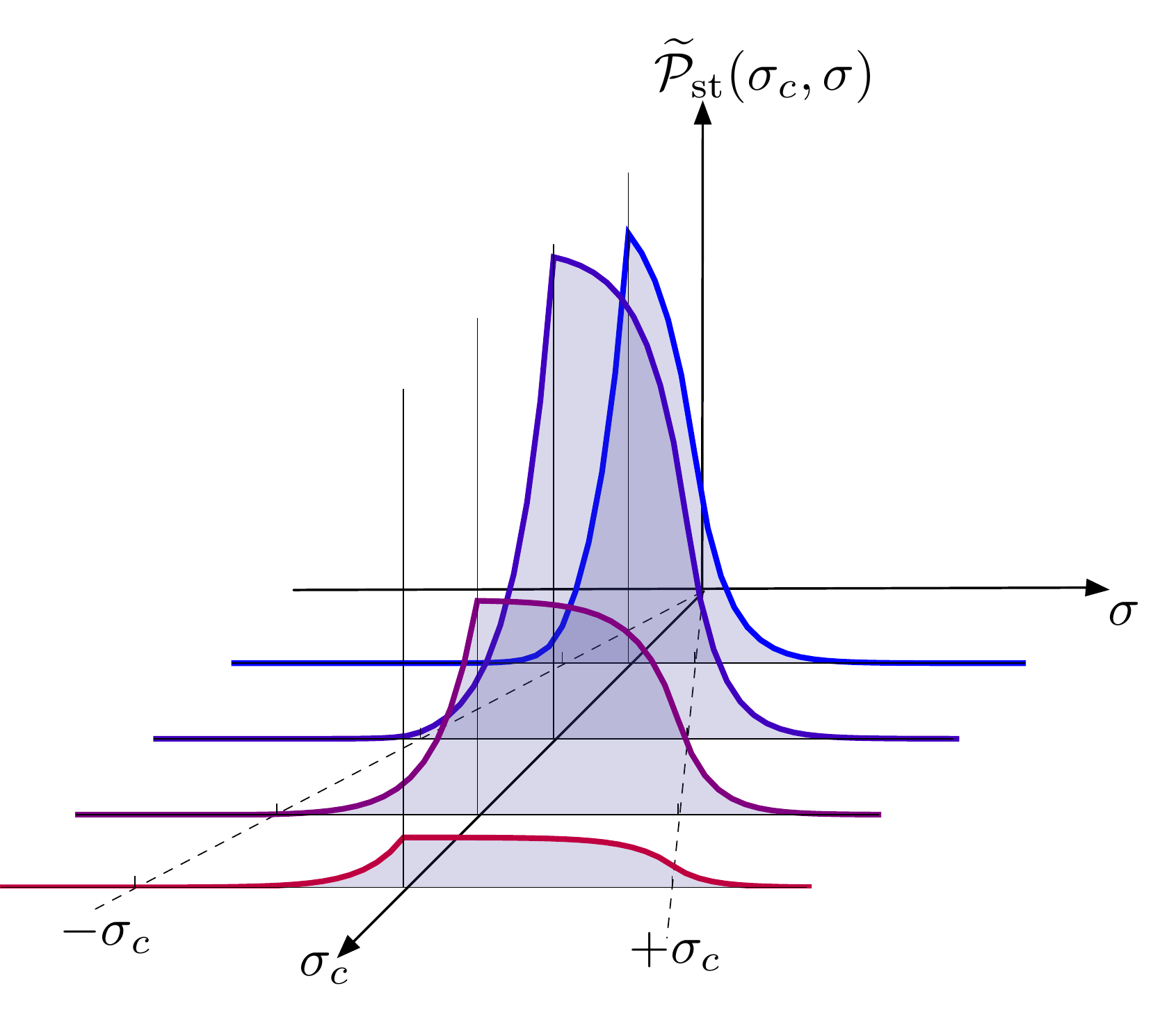}}
 \subfigure{\includegraphics[width=\columnwidth]{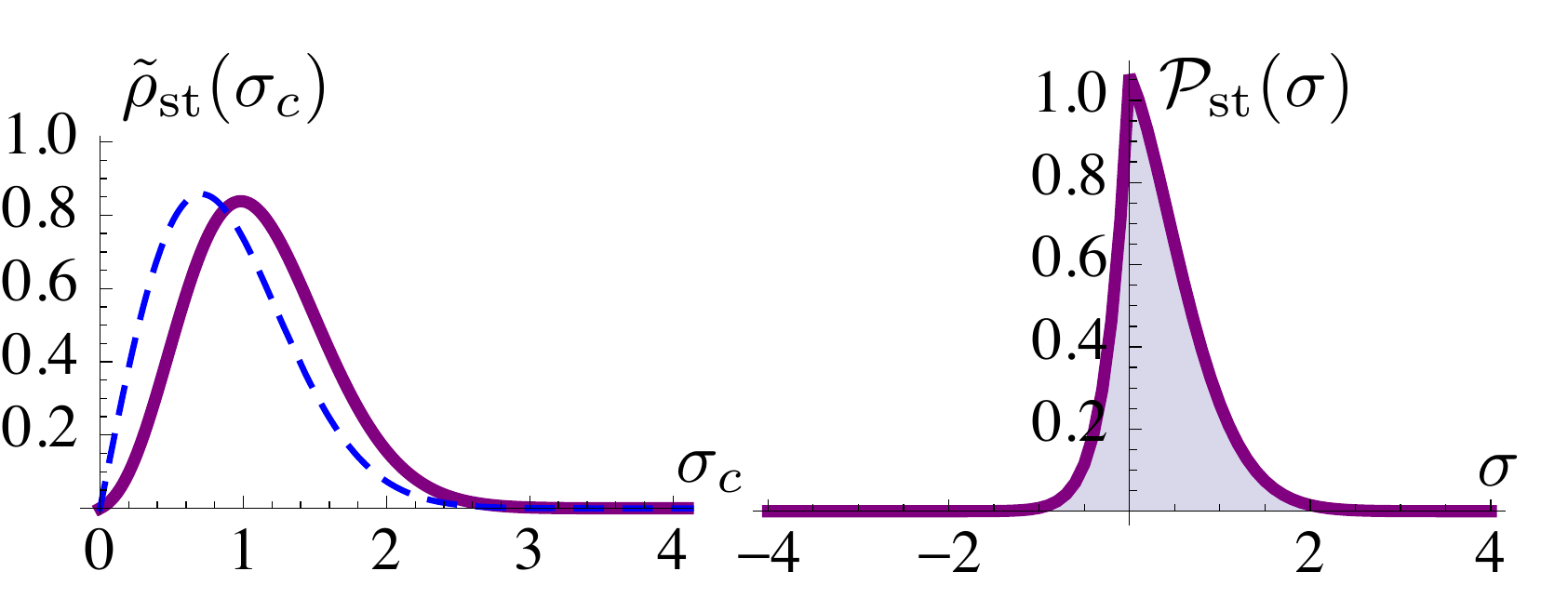}}
 \caption{
 \textit{Top:}~Stationary joint distribution ${\widetilde{\mathcal{P}}(\sigma_c,\sigma)}$ at ${\alpha_{\text{eff}}=0.3 < \alpha_c}$ and ${G_0 \dot{\gamma}\tau = 0.1}$, for four fixed values ${\sigma_c \in \arga{0.5,1,1.5,2}}$.
 Each slice of the distribution has the same functional form at each $\sigma_c$, with an asymmetry due to the finite shear rate ${\dot{\gamma}}$, and is normalized by definition to ${\tilde{\rho}_{\text{st}} (\sigma_c)}$.
 \textit{Bottom left:}~Corresponding dynamical distribution of local yield stress ${\tilde{\rho}_{\text{st}} (\sigma_c)}$, as defined in eq.~\eqref{eq-PDF-sigmac-HLdisord}, with the underlying a priori distribution ${\rho_0 (\sigma_c)}$ of eq.~\eqref{eq-def-rho-SGR} plotted as a blue dashed line.
 \textit{Bottom right:}~Corresponding distribution of stress ${\mathcal{P}_{\text{st}}(\sigma)}$, as defined in eq.~\eqref{eq-PDF-sigma-HLdisord}; its precise shape depends on ${\rho_0 (\sigma_c)}$.
 %
 }
 \label{fig:PDF-3D}
\end{figure}
\end{center}

In fig.~\ref{fig:PDF-3D}, we have illustrated the stationary joint distribution ${\widetilde{\mathcal{P}}_{\text{st}}(\sigma_c,\sigma)}$ for different values $\sigma_c$ (which have the same functional form in the stationary case)
and the corresponding complete distributions ${\tilde{\rho}_{\text{st}}(\sigma_c)}$ and ${\mathcal{P}_{\text{st}}(\sigma)}$ for the specific distribution ${\rho_0(\sigma_c)}$ given in eq.~\eqref{eq-def-rho-SGR}.
%
%
The behavior of the distribution ${\tilde{\rho}_{\text{st}} (\sigma_c)}$ depending on ${\alpha}$ and the shear rate ${\dot{\gamma}}$  will be examined in sect.~\ref{section-HL-model-stationary-rho-tilde-disord}.
But for now we will consider the diffusion coefficient and the mean stress in the stationary case, following the same structure as for the standard HL model in sect.~\ref{section-HL-model-stationary-noshearrate}-\ref{section-HL-model-stationary-low-shearrate}.

\subsection{Stationary solution in the absence of shear rate}
\label{section-HL-model-stationary-noshearrate-disord}

The unsheared case (${\dot{\gamma}=0}$) can be used as a benchmark for the comparison between the disordered HL model and its original counterpart.

The stationary joint PDF is symmetric with respect to $\sigma$, and it thus predicts as expected no macroscopic stress.
But more importantly, applying the definition of ${f_{\text{eff}}}$ on eq.~\eqref{eq-factorf-HL-noshear}, we have:
\begin{eqnarray}
\label{eq-factorf-HL-noshear-disord}
 && x \equiv \sqrt{D \tau}
 , \quad
 f_{\text{eff}}  \argp{x,0}
 = x^2 + \moy{\sigma_c} \, x + \frac12 \moy{\sigma_c^2} \\
\label{eq-def-alphac-HL-noshear}
 && \alpha_c \equiv f_{\text{eff}} (0,0)=\frac12 \moy{\sigma_c^2}
\end{eqnarray}
where all the factors ${\sim \sigma^k}$ in eq.~\eqref{eq-factorf-HL-noshear} have been transformed into moments ${\moy{\sigma^k}}$.
With the updated definition \eqref{eq-def-alphac-HL-noshear} of the `critical' value $\alpha_c$, we recover the same two regimes for the diffusion coefficient as before, respectively $D_{\text{HL}} =0$ at $\alpha_{\text{eff}}< \alpha_c$
and, at ${\alpha_{\text{eff}} \geq \alpha_c}$:
\begin{equation}
 \sqrt{D_{\text{HL}} \tau} =
 \left\lbrace \begin{array}{ll}

	\frac12 \moy{\sigma_c} \argp{\sqrt{\frac{4 (\alpha_{\text{eff}} - \alpha_c)}{\moy{\sigma_c}^2} + 1} -1}
	& (\alpha_{\text{eff}} \geq \alpha_c) \\ \\
	\frac{\alpha_{\text{eff}}-\alpha_c}{\moy{\sigma_c}} + \mathcal{O} \argp{(\alpha_{\text{eff}}-\alpha_c)^2}
	& (\alpha_{\text{eff}} \gtrsim \alpha_c)
 \end{array} \right.
\label{eq-solution-D-without-gammadot-disord}
\end{equation}

So we can conclude that, at least in the case ${\dot{\gamma}=0}$, the physical interpretation of ${\alpha_c}$ --~as the lower threshold for a possible self-sustained plastic diffusion~-- is robust to the addition of structural disorder.
Both the function ${\tilde{f}_{\sigma_c}}$ \eqref{eq-factorf-HL-noshear} and the predicted diffusion coefficient ${D_{\text{HL}}}$ \eqref{eq-solution-D-without-gammadot} can actually be straightforwardly generalized by a proper averaging over the random values of yield stress $\sigma_c$, resulting only in slight quantitative modifications of the predictions.

\subsection{Stationary solution at low shear rate}
\label{section-HL-model-stationary-low-shearrate-disord}

In the presence of a finite constant shear rate, the stationary joint PDF is asymmetric at each fixed $\sigma_c$, and it thus predicts a finite macroscopic stress ${\sigma_M (\dot{\gamma},\alpha_{\text{eff}})}$.
From eq.~\eqref{eq-rhotilde-stationary-disord-3}, we can compute at ${\dot{\gamma}>0}$ the function ${f_{\text{eff}}}$ for an arbitrary distribution ${\rho(\sigma_c)}$, at least numerically, from the known expressions for ${\tilde{f}_{\sigma_c}}$ defined in eq.~\eqref{eq-factorf-HL-withshear}.

\begin{center}
\begin{figure}[!htb]
\includegraphics[width=\columnwidth]{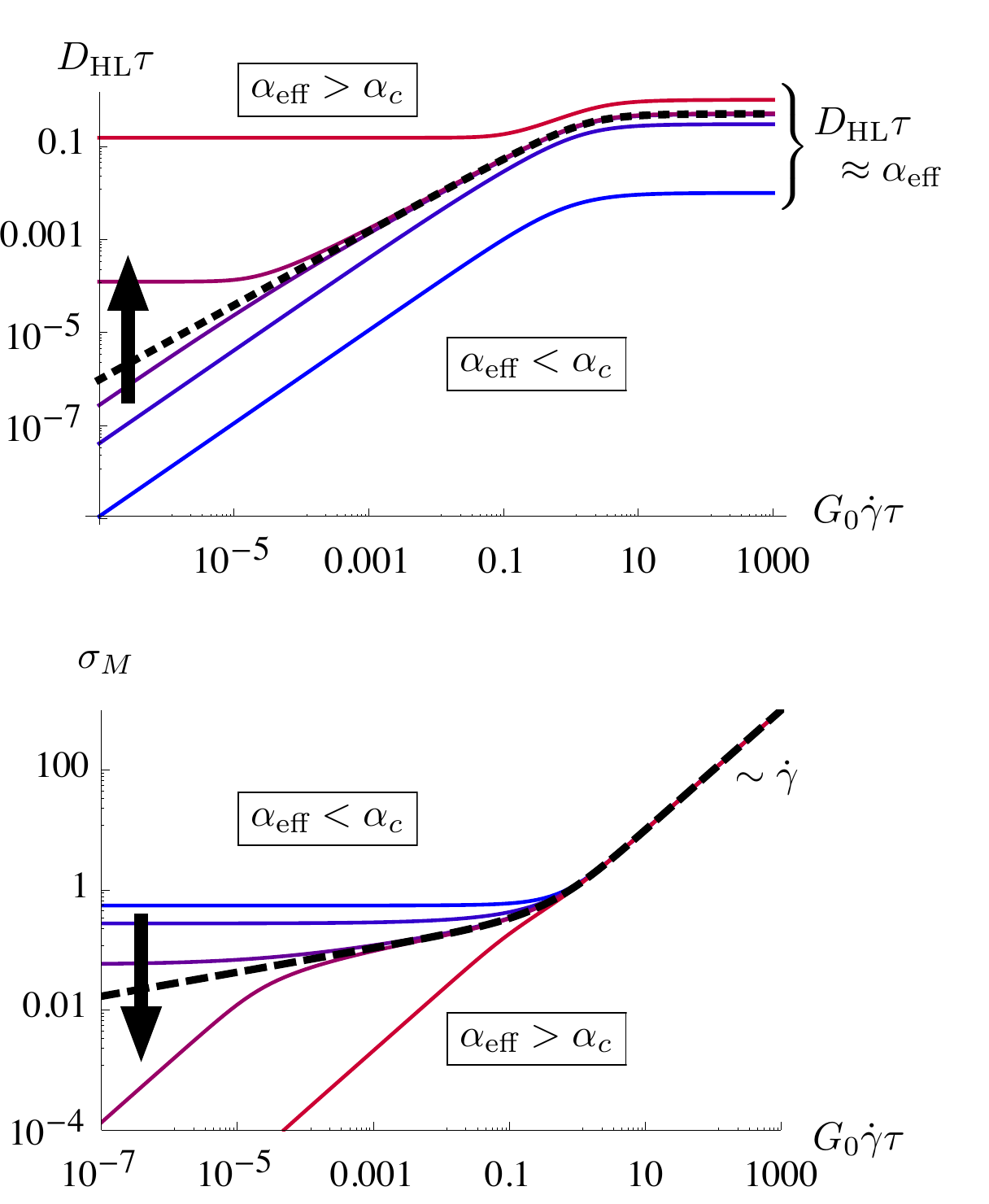}
 \caption{
 (Color online.)
 \textit{Top:}~Stationary diffusion coefficient as a function of the shear rate, for increasing values of ${\alpha_{\text{eff}} \in \arga{0.01,0.3,0.49,0.5,0.51,1}}$ (blue to red), as indicated by the black arrow. The dashed black curve corresponds to ${\alpha_{\text{eff}}=\alpha_c=0.5}$. At large shear rates we have ${D_{\text{HL}}\tau \approx \alpha_{\text{eff}}}$.
 \textit{Bottom:}~Corresponding macroscopic stress as a function of the shear rate, for the same increasing values of ${\alpha_{\text{eff}}}$ (blue to red), as indicated again by the black arrow. At large shear rates ${\sigma_M \sim \dot{\gamma}}$ so that the predicted behavior is Newtonian, whereas at low shear rates we recover the three regimes in $\alpha_{\text{eff}}$ that we have predicted analytically for ${\sigma_M (\dot{\gamma},\alpha_{\text{eff}})}$ in eqs.~\eqref{eq-sigmaM-withshear-smallgammadot-disord-1}-\eqref{eq-sigmaM-withshear-smallgammadot-disord-2}.
 This graph is reminiscent, as expected, of fig.~2 in the original HL paper~\cite{hebraud_lequeux_1998_PhysRevLett81_2934}, although there are quantitative discrepancies due to the structural disorder included via ${\rho_0(\sigma_c)}$ of eq.~\eqref{eq-def-rho-SGR}.
 }
 \label{fig:diffcoeff-6alpha}
\end{figure}
\end{center}

For the disordered HL model \eqref{eq-dist-Psigma-HL-disord}-\eqref{eq-Gamma-nu-HL-disord-partial}, we need the following definition of `critical' values for the coupling parameter:
\begin{equation}
 \tilde{\alpha}_c (\sigma_c)
 = \frac{1}{2} \sigma_c^2
 \, , \quad
 \alpha_c = \moy{\tilde{\alpha}_c (\sigma_c)}= \frac12 \moy{\sigma_c^2}
\end{equation}
Combining it with the stationary closure relation \eqref{eq-closure-DHL-disord-stat1}, we eventually obtain, in the limit of vanishing shear rate ${\dot{\gamma}}$, the following lowest-order scaling of ${D_{\text{HL}}(\dot{\gamma})}$, as generalizations of eqs.~\eqref{eq-DHL-withshear-smallgammadot-1}-\eqref{eq-DHL-withshear-smallgammadot-2}-\eqref{eq-sigmaM-withshear-smallgammadot-1} (see also Appendix~\ref{A-appendix-perturb-diff-coeff}):
\begin{equation}
\label{eq-DHL-withshear-smallgammadot-disord-1}
\left\lbrace \begin{array}{cl}
 \alpha_{\text{eff}} > \alpha_c :
 & D_{\text{HL}} \tau
 	\approx D_{\text{HL}}(\dot{\gamma}=0) \tau
 	\quad \text{(see eq.~\eqref{eq-solution-D-without-gammadot-disord})} 
 \\ \\
 \alpha_{\text{eff}} = \alpha_c :
 &  D_{\text{HL}} \tau
 	\approx \widetilde{C} \, \argp{G_0 \dot{\gamma} \tau}^{4/5}
 	\, ,  \; \;\, \widetilde{C} = \argc{\frac{\moy{\sigma_c^4}}{24 \moy{\sigma_c}}}^{2/5}
	 \\ \\
 \alpha_{\text{eff}} < \alpha_c :
 &  D_{\text{HL}} \tau
 	\approx C \, G_0 \dot{\gamma} \tau \quad ,
 \\ \\
 & \quad \, \text{with} \;\;\, \moy{C \sigma_c \tanh \argp{\frac{\sigma_c}{2 C}}}=\alpha_{\text{eff}} .
\end{array} \right.
\end{equation}
This last equation, which implicitly determines the prefactor $C$ at ${\alpha_{\text{eff}} < \alpha_c}$, can be rewritten as:
\begin{equation}
\label{eq-DHL-belowalphac-smallgammadot-disord-1}
 \moy{\tilde{\alpha}_c (\sigma_c) \argc{\frac{\tanh \argp{\frac{\sigma_c}{2 C}}}{\frac{\sigma_c}{2 C}}-\frac{\alpha_{\text{eff}}}{\tilde{\alpha}_c (\sigma_c)}}}
 = 0
\end{equation}
where we see the competition between terms ${\tanh (u) /u}$ and ${\alpha_{\text{eff}}/\tilde{\alpha}_c (\sigma_c)}$, depending on the value of $\sigma_c$.
For a generic distribution ${\rho (\sigma_c)}$ and an arbitrary value of ${\alpha_{\text{eff}}}$, the prefactor ${C(\alpha_{\text{eff}})}$ has to be determined numerically.
The different behaviors of the diffusion coefficient are illustrated in fig.~\ref{fig:diffcoeff-6alpha} (\textit{top}): at low shear rates it scales according to eqs.~\eqref{eq-DHL-withshear-smallgammadot-disord-1}, whereas at large shear rates we have ${D_{\text{HL}}\tau =\alpha_{\text{eff}} \, \Gamma_{\text{st}} \tau \approx \alpha_{\text{eff}}}$, the system being essentially overstressed everywhere.

We can nevertheless obtain analytically the following specific perturbative expansions, which are the counterparts of eqs.~\eqref{eq-DHL-withshear-smallgammadot-2}:
\begin{equation}
\label{eq-DHL-withshear-smallgammadot-disord-2}
\begin{split}
 \alpha_{\text{eff}} \gtrsim \alpha_c :
 & \quad D_{\text{HL}} \tau
 	\stackrel{\eqref{eq-solution-D-without-gammadot-disord}}{\approx} 
 	\argp{\frac{\alpha_{\text{eff}}-\alpha_c}{\moy{\sigma_c}}}^2
 	 \\
 \alpha_{\text{eff}} \lesssim \alpha_c :
 & \quad D_{\text{HL}} \tau
 	\approx \argc{\frac{\moy{\sigma_c^4}}{24 \argp{\alpha_c -\alpha_{\text{eff}}}}}^{1/2} \!\!\!\!\!\! \, G_0 \dot{\gamma} \tau
 \\
 \alpha_{\text{eff}} \ll \alpha_c :
 & \quad D_{\text{HL}} \tau
 	\approx \frac{\alpha_{\text{eff}}}{\moy{\sigma_c}} \, G_0 \dot{\gamma} \tau
\end{split}
\end{equation}
Note that these two last expressions are obtained by neglecting the contribution of some values of $\sigma_c$ in the total average of eq.~\eqref{eq-DHL-belowalphac-smallgammadot-disord-1}.
As discussed in Appendix~\ref{A-appendix-perturb-diff-coeff}, we can define for a given ${\alpha_{\text{eff}}}$ a typical value ${\sigma_c^*=2 C}$.
Close to $\alpha_c$, ${C \to \infty}$ and we can safely neglect the contributions of ${\sigma_c > \sigma_c^*}$,
whereas when ${\alpha_{\text{eff}}}$ is close to zero, ${C \to 0}$ and we can neglect the contributions  of ${\sigma_c < \sigma_c^*}$.
%
So, in the vicinity of ${\alpha_c = \moy{\tilde{\alpha}_c}}$, we recover in particular the same expressions as in eq.~\eqref{eq-DHL-withshear-smallgammadot-2} and there is an increasing quantitative correction of the prefactors the further the coupling parameter $\alpha_{\text{eff}}$ moves away from $\alpha_c$.
These approximations are of course valid only if we are sufficiently close to $\alpha_c$ or to $0$, but also if there is a `reasonable' cutoff in the distribution ${\rho (\sigma_c)}$ for very large or very small $\sigma_c$, respectively, as is expected to be the case physically.

Once the diffusion coefficient ${D_{\text{HL}}}$ is known, we can at last compute the corresponding macroscopic stress ${\sigma_M}$ using the explicit expressions eqs.~\eqref{eq-sigmaM-over-appendix}-\eqref{eq-sigmaM-under-appendix}-\eqref{eq-sigmaM-appendix} given in Appendix~\ref{A-appendix-explicit-expressions}, as illustrated in fig.~\eqref{fig:diffcoeff-6alpha}.
The overstressed regions (${\valabs{\sigma} > \sigma_c}$) contribute only linearly in $\dot{\gamma}$ to the macroscopic stress, and hence it is only the asymmetry of the PDF of the understressed regions (${\valabs{\sigma} < \sigma_c}$) that can modify the dominant scaling in $\dot{\gamma}$, at least in the vanishing shear rate limit that we want to consider.
We can derive analytical predictions in the limit ${\dot{\gamma} \to 0}$ for the three regimes in ${\alpha_{\text{eff}}}$, starting from the case at ${\alpha_{\text{eff}} > \alpha_c}$:
\begin{equation}
\label{eq-sigmaM-withshear-smallgammadot-disord-1}
\begin{split}
 \sigma_M
 \approx
 & \argc{1 + \frac{4 x_0 \, \moy{\sigma_c^3} + \moy{\sigma_c^4} }{24 x_0^2 \, \moy{\tilde{f}_{\sigma_c} \argp{x_0,0}}}}
 \, G_0 \dot{\gamma} \tau \\
 \stackrel{\eqref{eq-factorf-HL-noshear}}{=}
 & \argc{1 + \frac{4 x_0 \, \moy{\sigma_c^3} + \moy{\sigma_c^4} }{24 x_0^2 \, \argp{ x_0^2 + x_0 \moy{\sigma_c} + \moy{\sigma_c^2}/2}}}
 \, G_0 \dot{\gamma} \tau
\end{split}
\end{equation}
with ${x_0 = \sqrt{D_{\text{HL}}(\dot{\gamma}=0) \tau}}$ given by eq.~\eqref{eq-solution-D-without-gammadot-disord},
and in the two other regimes in ${\alpha_{\text{eff}}}$:
\begin{equation}
\begin{split}
 \alpha_{\text{eff}} = \alpha_c :
& \quad \sigma_M \approx \frac{1}{2^{4/5} \times 3^{3/5}} \, \frac{\moy{\sigma_c^4}^{3/5} \moy{\sigma_c}^{2/5}}{\moy{\sigma_c^2}} \! \, \! \argp{G_0 \dot{\gamma} \tau}^{1/5}
\\
 \alpha_{\text{eff}} < \alpha_c :
 & \quad \sigma_M \approx \sigma_Y+A \, \argp{G_0 \dot{\gamma} \tau}^{1/2}
\end{split}
\end{equation}
As for the predicted Herschel-Bulkley behavior at ${\alpha_{\text{eff}} < \alpha_c}$, the macroscopic yield stress ${\sigma_Y}$ and the prefactor $A$ are \emph{not} simply given by averaging eqs.~\eqref{eq-HL-law-sigmaY-1}-\eqref{eq-HL-law-prefactorA-1} over the distribution ${\rho (\sigma_c)}$ of threshold stress values.
In fact, taking care of the different averages according to eq.~\eqref{eq-sigmaM-properaverages} (and detailed furthermore in Appendix~\ref{A-appendix-explicit-expressions}),
we obtain for the macroscopic yield stress:
\begin{equation}
 \label{eq-HL-law-sigmaY-disord-1}
 \sigma_Y
 = C \argc{\frac{\moy{\sigma_c^2} /2}{C \moy{\sigma_c \, \tanh \argp{\frac{\sigma_c}{2 C}}}} -1 }
\end{equation}
with $C$ defined by eq.~\eqref{eq-DHL-belowalphac-smallgammadot-disord-1}.
As for the prefactor $A$, it is more subtly obtained by a Taylor expansion of ${D_{\text{HL}}}$ to the next order (whose derivation is sketched in Appendix~\ref{A-appendix-perturb-diff-coeff}).
Its complete expression, rather cumbersome, is given in eq.~\eqref{eq-HL-law-prefactorA-disord-1} and it simplifies in the usual limiting cases of interest ${\alpha_{\text{eff}} \lesssim \alpha_c}$ and ${\alpha_{\text{eff}} \ll \alpha_c}$, as presented thereafter.
The complete behavior of ${\sigma_Y}$ and $A$ as functions of $\alpha_{\text{eff}}$ are illustrated in fig.~\ref{fig:sigmaY-A-totale} (see Appendix~\ref{A-appendix-prefactorA-disordHL} for the complete expressions).

We can finally give the perturbative expansions for the macroscopic stress corresponding to the limits given in eq.~\eqref{eq-DHL-withshear-smallgammadot-disord-2},
first in the limit of small ${x_0}$,
then expanding the hyperbolic tangent depending on whether ${C}$ diverges (${\alpha_{\text{eff}} \gtrsim \alpha_c}$, see eq.~\eqref{eq-HL-law-prefactorA-disord-1-justbelowalphac}) or tends to zero (${\alpha_{\text{eff}} \ll \alpha_c}$, see eq.~\eqref{eq-HL-law-prefactorA-disord-1-wellbelowalphac}):
\begin{equation}
\label{eq-sigmaM-withshear-smallgammadot-disord-2}
\begin{split}
 \alpha_{\text{eff}} \gtrsim \alpha_c :
 & \quad \frac{\sigma_M}{G_0 \dot{\gamma} \tau}
	\stackrel{\eqref{eq-sigmaM-withshear-smallgammadot-disord-1}}{\approx}
		\frac{\moy{\sigma_c^4}/\moy{\sigma_c^2}}{12 x_0^2}
	\stackrel{\eqref{eq-solution-D-without-gammadot-disord}}{\approx}
 		\frac{\moy{\sigma_c^4} \moy{\sigma_c}^2 \!\! / \! \moy{\sigma_c^2}}{12 \argp{\alpha_{\text{eff}} -\alpha_c}^2}
 	 \\
 \alpha_{\text{eff}} \lesssim \alpha_c :
 & \quad \sigma_Y
 	\approx \frac{\argp{\alpha_c - \alpha_{\text{eff}}}^{1/2}}{\sqrt{6}} \frac{\moy{\sigma_c^4}^{1/2}}{\moy{\sigma_c^2}}
 	\\
 & \quad A
	\approx \frac{\argp{\alpha_c - \alpha_{\text{eff}}}^{-3/4}}{2^{3/2} \times 6^{3/4}} \frac{\moy{\sigma_c^4}^{3/4} \moy{\sigma_c} }{\moy{\sigma_c^2}}
	\\
 \alpha_{\text{eff}} \ll \alpha_c :
 & \quad \sigma_Y
	\approx \frac{\alpha_c - \alpha_{\text{eff}}}{\moy{\sigma_c}}
 	\\
 & \quad A
	\approx \argp{1- \frac{\moy{\sigma_c^2}}{2 \moy{\sigma_c}^2}} \argp{\frac{\alpha_{\text{eff}}}{\moy{\sigma_c}}}^{1/2}
\end{split}
\end{equation}
Illustrated in fig.~\ref{fig:sigmaY-A-totale}, these behaviors are qualitatively similar to the predictions of the HL model without disorder. They actually differ only quantitatively from their counterpart expressions~\eqref{eq-sigmaM-withshear-smallgammadot-2} since they involve the moments of the distribution ${\rho (\sigma_c)}$ instead of simple powers of ${\sigma_c}$. 
Consequently, we can transpose the physical discussion of their counterpart expressions~\eqref{eq-sigmaM-withshear-smallgammadot-2} to our disordered HL model, taking into account the quantitative corrections due to the distribution ${\rho(\sigma_c)}$:
\textit{(i)}~${\sigma_Y}$ tends to zero in the limit ${\alpha_{\text{eff}} \to \alpha_c^-}$ and is thus physically well-behaved (predicting the disappareance of macroscopic yield stress and hence of the Herschel-Bulkley behavior at low shear rate);
\textit{(ii)}~nevertheless, the validity range of its expression has been fixed previously in the perturbative expansion of the diffusion coefficient, with the upper bound ${\dot{\gamma}_*(\alpha_{\text{eff}}) \sim  (\alpha_c - \alpha_{\text{eff}})^{1/2}}$;
\textit{(iii)}~finally, a similar validity range can be defined in the limit ${\alpha \to \alpha_c^+}$,
when ${x_0^2}$ becomes comparable to ${G_0 \dot{\gamma} \tau}$,
with another upper bound
${\dot{\gamma}_{**}(\alpha_{\text{eff}}) = \frac{12 (\alpha_c - \alpha_{\text{eff}})^2 \moy{\sigma_c^2}}{G_0 \tau \, \moy{\sigma_c^4} \moy{\sigma_c}^2}}$
that squeezes this regime in the vicinity of ${\alpha_c}$.
As for the third limit ${\alpha \ll \tilde{\alpha}_c}$, we recover that if we remove  the diffusion term, we eventually destroy the Herschel-Bulkley behavior, and the macroscopic yield stress tends to ${\alpha_c/\moy{\sigma_c}}$, which is slightly different from the arithmetic average between the mean local yield stress $\moy{\sigma_c}$ and the local stress after a plastic event ($0$ in the full relaxation assumption).

\begin{center}
\begin{figure}[!htb]
\subfigure{\includegraphics[width=\columnwidth]{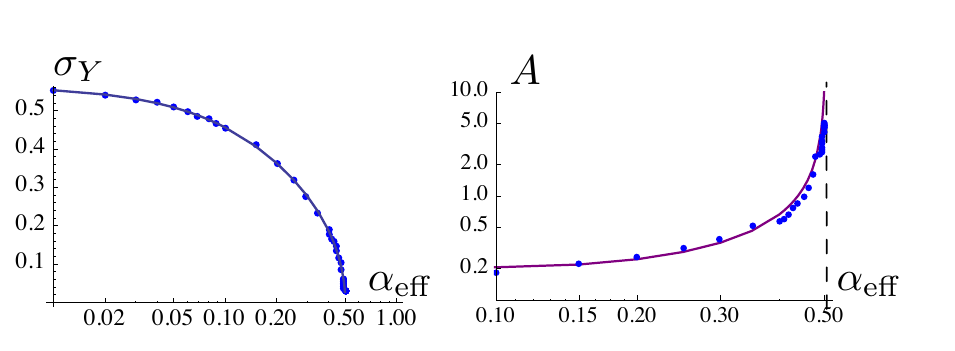}}
\subfigure{\includegraphics[width=\columnwidth]{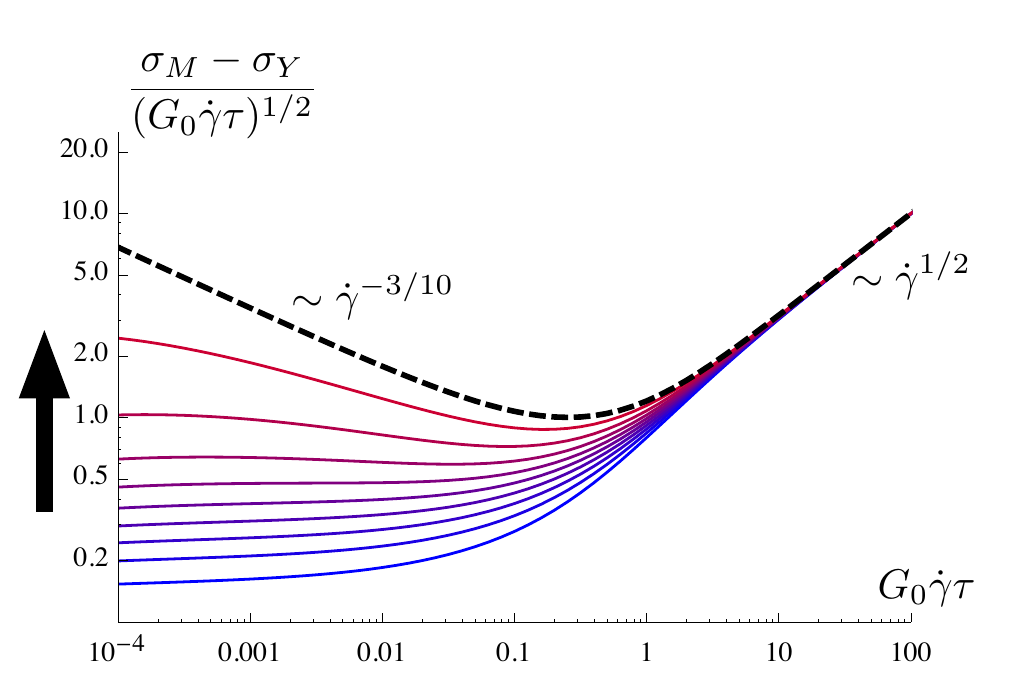}}
 \caption{
 \textit{Top:}~Macroscopic yield stress $\sigma_Y$ and prefactor $A$ of the Herschel-Bulkley behavior predicted at ${\alpha_{\text{eff}}< \alpha_c}$, in the limit of ${\dot{\gamma} \to 0}$ (for ${\rho_0 (\sigma_c)}$ given by eq.~\eqref{eq-def-rho-SGR}). Their behavior is in agreement with the predictions in eq.~\eqref{eq-sigmaM-withshear-smallgammadot-disord-2}.
 On the left, $\sigma_Y$ is evaluated by computing the average stress ${\sigma_M}$ at a very low shear rate ${G_0 \dot{\gamma} \tau =10^{-5}}$ (dots), showing a good agreement with the theoretical prediction eq.~\eqref{eq-HL-law-sigmaY-disord-1} obtained by using the diffusion coefficient computed numerically (continuous line).
 On the right, the prefactor $A$ is evaluated similarly, using again the diffusion coefficient computed numerically and the corresponding prediction for ${\sigma_Y}$.
 \textit{Bottom:}~Test of the validity range of the Herschel-Bulkley behavior ${\sigma_M \approx \sigma_Y + A \, (G_0\dot{\gamma}\tau)^{1/2}}$, for increasing values of ${\alpha_{\text{eff}} \in \arga{0.01, 0.05, 0.1, 0.15, 0.2, 0.25, 0.3, 0.35, 0.4, 0.45, 0.49}}$ (blue to red, as indicated by the black arrow) and ${\alpha_{\text{eff}}=0.5}$ (dashed black).
 The plateaux at low shear rate define the prefactor $A$, and their validity range actually shrinks when ${\alpha_{\text{eff}}}$ tends to~${\alpha_c}$, in agreement with eq.~\eqref{eq-validity-HB-gammadot-max}.}
 \label{fig:sigmaY-A-totale}
\end{figure}
\end{center}

However, we can comment furthermore on the validity range in ${\dot{\gamma}}$ of the Herschel-Bulkley behavior, and hence on the definition of the prefactor $A$.
The consistency of the perturbative expansions of the diffusion coefficient~\eqref{eq-DHL-withshear-smallgammadot-disord-1} at a fixed low ${\dot{\gamma}}$ requires that ${D_{\text{HL}}(\alpha_{\text{eff}}<\alpha_c) < D_{\text{HL}}(\alpha_{\text{eff}}=\alpha_c)}$, which implies in general that
${ G_0 \dot{\gamma} \tau < (\widetilde{C}/C)^5}$, and when ${\alpha_{\text{eff}} \lesssim \alpha_c}$ that:
\begin{equation}
\label{eq-validity-HB-gammadot-max}
 G_0 \dot{\gamma} \tau
 < \argp{\frac{\widetilde{C}}{C}}^5
 \stackrel{\eqref{eq-DHL-withshear-smallgammadot-disord-2}}{\approx}
 \argp{\frac{24}{\moy{\sigma_c^4}}}^{1/2} \argp{\frac{(\alpha_c  - \alpha_{\text{eff}})^{5/2}}{\moy{\sigma_c}^2}}
\end{equation}
This condition defines a much more restrictive upper bound for ${\dot{\gamma}}$ than the condition ${ D_{\text{HL}}\tau \leq \alpha_{\text{eff}}}$, and it explains in fig.~\ref{fig:sigmaY-A-totale} (\textit{bottom}) the shrinking of the plateaux when ${\alpha_{\text{eff}} \to \alpha_c^-}$.
%
%
Given the fact that the scalings of the diffusion coefficient with respect to the shear rate are the same with or without disorder, this argument is valid both for the standard HL model and for our disordered version of it.

So we have shown in this whole section that, when we include a distribution of threshold stress values ${\rho (\sigma_c)}$ in the HL model,
the three scaling regimes at low constant shear rates for the macroscopic stress ${\sigma_M (\dot{\gamma})}$ are the same as in eq.~\eqref{eq-recap-HL-predictions}.
Provided that we generalize the definition of the `critical' coupling parameter
${\alpha_c \equiv \moy{\tilde{\alpha}_c (\sigma_c)}=\moy{\sigma_c^2}/2}$,
the expressions of the prefactors are almost identical to eq.~\eqref{eq-sigmaM-withshear-smallgammadot-2} in the vicinity of ${\alpha_c}$, the terms ${\sigma_c^k}$ being replaced by combinations of the first moments 
of the distribution ${\rho (\sigma_c)}$.
%
%
We emphasize that it is the scaling behavior of the averaged stress
${\sigma_M \equiv \int_{-\infty}^{\infty} d \sigma \, \sigma \, \mathcal{P}_{\text{st}}(\sigma)}$
which is qualitatively robust with respect to a structural disorder;
the fluctuations of the stress characterized by the full PDF ${\mathcal{P}_{\text{st}}(\sigma)}$ are of course modified, as illustrated in fig.~\ref{fig:PDF-3D}.

\subsection{Stationary distribution of local yield stress ${\tilde{\rho}_{\text{st}} (\sigma_c)}$}
\label{section-HL-model-stationary-rho-tilde-disord}

\begin{center}
\begin{figure}[!htb]
\includegraphics[width=\columnwidth]{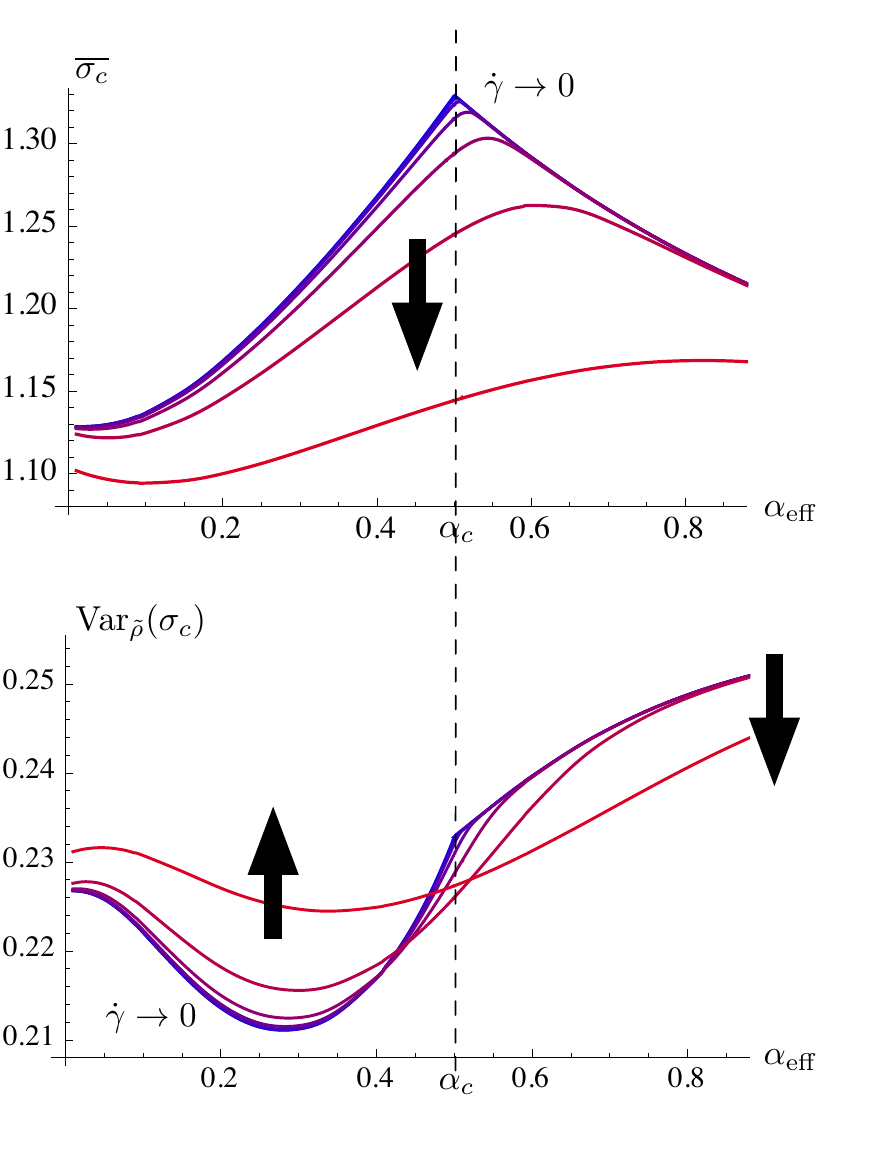}
 \caption{
 (Color online) Mean and variance of the dynamical distribution of local yield stress ${\tilde{\rho}(\sigma_c)}$ as a function of $\alpha_{\text{eff}}$, for increasing values of ${G_0 \dot{\gamma} \tau \in \arga{10^{-7},10^{-6},10^{-5},10^{-4},10^{-3},10^{-2},10^{-1}}}$ (blue to red), as indicated by the black arrow.
The mean and variance are computed from eq.~\eqref{eq-rhotilde-stationary-disord-final-HL}.
The expected limits ${\alpha_{\text{eff}} \to 0}$ and ${\alpha_{\text{eff}} \to \alpha_c}$, computed at ${\dot{\gamma} \to 0}$ for ${\rho_0(\sigma_c)}$ of eq.~\eqref{eq-def-rho-SGR}, are recovered.
Results for larger shear rates are shown in fig.~\ref{fig:rhotilde-meanvariance-at-fixed-alpha}, for fixed values of ${\alpha}$.
}
 \label{fig:rhotilde-meanvariance-WRTG0gammadottau}
\end{figure}
\end{center}

\begin{center}
\begin{figure}[!htb]
\includegraphics[width=\columnwidth]{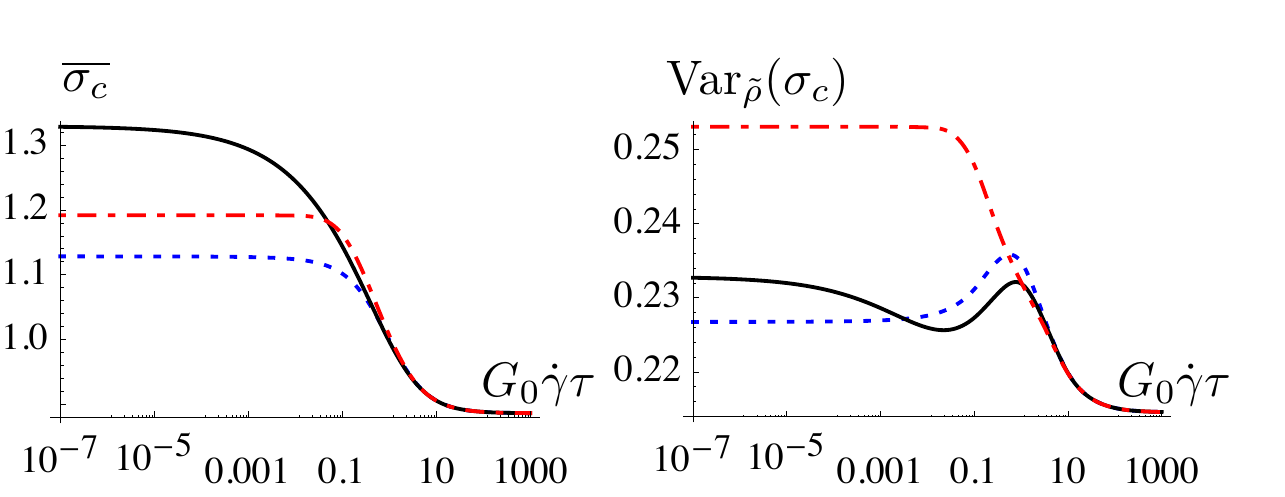}
 \caption{
 (Color online) Mean and variance of the dynamical distribution of local yield stress ${\tilde{\rho}(\sigma_c)}$, as a function of ${G_0 \dot{\gamma} \tau}$,
for three  values of $\alpha_{\text{eff}}$:  ${\alpha_{\text{eff}}=0.3}$ (dotted blue),  ${\alpha_{\text{eff}}=0.5}$ (continuous black) and ${\alpha_{\text{eff}}=1}$ (dot-dashed red).
 The corresponding curves are sections of fig.~\ref{fig:rhotilde-meanvariance-WRTG0gammadottau} at fixed $\alpha_{\text{eff}}$.
  In the limit of large shear rates, we recover as expected from eq.~\eqref{eq-rhotilde-stationary-disord-final-HL-largeshearrate}, for ${\rho_0 (\sigma_c)}$, that ${\overline{\sigma_c} \approx \moy{\sigma_c} \approx 0.886}$ 
  and ${\text{Var}_{\tilde{\rho}}(\sigma_c) \approx \text{Var}_{\rho}(\sigma_c) \approx 0.215}$. 
   }
 \label{fig:rhotilde-meanvariance-at-fixed-alpha}
\end{figure}
\end{center}

Combining eqs.~\eqref{eq-rhotilde-stationary-disord-final-generic} and~\eqref{eq-implicit-for-D-disord}, we have access to the stationary distribution of local yield stress:
\begin{equation}
\label{eq-rhotilde-stationary-disord-final-HL}
 \tilde{\rho}_{\text{st}} (\sigma_c)
 = \rho(\sigma_c) \, \frac{\tilde{f}_{\sigma_c} \argp{\sqrt{D \tau}, \frac{G_0 \dot{\gamma} \tau}{D \tau}}}{\alpha_{\text{eff}}}
\end{equation}
with ${\tilde{f}_{\sigma_c}}$ given by eq.~\eqref{eq-factorf-HL-withshear} and ${D=D_{\text{HL}}(\dot{\gamma},\alpha_{\text{eff}})}$ determined by solving eq.~\eqref{eq-implicit-for-D-disord}.
In what follows, we denote by $\overline{\mathcal{O}}$ the average of an observable $\mathcal{O}$ over the distribution $\tilde{\rho}_{\text{st}} (\sigma_c)$, while $\langle \mathcal{O} \rangle$ still denotes the average of $\mathcal{O}$ over the distribution $\rho(\sigma_c)$.
The main features of this distribution can then be characterized by its mean value and its second moment, respectively:
\begin{eqnarray}
\label{eq-rhotilde-stationary-disord-final-HL-meanvalue}
 \overline{\sigma_c} \,\,
 & \equiv & \int_0^{\infty} d \sigma_c \, \sigma_c \,  \tilde{\rho}_{\text{st}} (\sigma_c) \\
\label{eq-rhotilde-stationary-disord-final-HL-secondmoment}
 \overline{\sigma_c^2} \,\,
 & \equiv & \int_0^{\infty} d \sigma_c \, \sigma_c^2 \,  \tilde{\rho}_{\text{st}} (\sigma_c) 
\end{eqnarray}
We emphasize again that the stationary distribution ${\tilde{\rho}_{\text{st}} (\sigma_c)}$ is \emph{not} equivalent to the a priori distribution ${\rho(\sigma_c)}$, as we can show explicitly in the limit of vanishing shear rate (see the last remark in Appendix~\ref{A-appendix-perturb-diff-coeff}).
The set of predictions given in this section are furthermore illustrated in figs.~\ref{fig:rhotilde-meanvariance-WRTG0gammadottau} and \ref{fig:rhotilde-meanvariance-at-fixed-alpha} for the specific choice of ${\rho_0 (\sigma_c)=2 \sigma_c \exp \argp{-\sigma_c^2}}$.

We start with the case ${\alpha_{\text{eff}} < \alpha_c}$, whose strictly vanishing shear rate limit predicts:
\begin{equation}
\label{eq-rhotilde-stationary-disord-final-HL-belowalphac}
\lim_{\dot{\gamma}\to 0} \frac{\tilde{\rho}_{\text{st}}(\sigma_c)}{\rho(\sigma_c)}
= \frac{\sigma_c^2}{2 \alpha_{\text{eff}}} \frac{\tanh \argp{\sigma_c / \sigma_c^*}}{\sigma_c/\sigma_c^*},
\, \quad \sigma_c^* \equiv 2 C
\end{equation}
with $C$ the prefactor of the diffusion coefficient determined by eq.~\eqref{eq-DHL-belowalphac-smallgammadot-disord-1}.
Using the relations~\eqref{eq-DHL-withshear-smallgammadot-disord-2}, we can write explicitly the usual specific limits, first at ${\alpha_{\text{eff}} \lesssim \alpha_c = \moy{\sigma_c^2}/2}$:
\begin{eqnarray}
\label{eq-rhotilde-stationary-disord-final-HL-justbelowalphac}
\lim_{\dot{\gamma}\to 0} \frac{\tilde{\rho}_{\text{st}}(\sigma_c)}{\rho(\sigma_c)}
	& \approx & \frac{\sigma_c^2}{2 \alpha_c} - \argp{1- \frac{\alpha_{\text{eff}}}{\alpha_c}} \, \frac{\sigma_c^4}{\moy{\sigma_c^4}} \\
\label{eq-rhotilde-stationary-disord-final-HL-justbelowalphac-meanvalue}
\lim_{\dot{\gamma}\to 0} \overline{\sigma_c}
	& \approx & \frac{\moy{\sigma_c^3}}{\moy{\sigma_c^2}} - \argp{1- \frac{2 \alpha_{\text{eff}}}{\moy{\sigma_c^2}}} \, \frac{\moy{\sigma_c^5}}{\moy{\sigma_c^4}} \\
\label{eq-rhotilde-stationary-disord-final-HL-justbelowalphac-secondmoment}
\lim_{\dot{\gamma}\to 0} \overline{\sigma_c^2}
	& \approx & \frac{\moy{\sigma_c^4}}{\moy{\sigma_c^2}} - \argp{1- \frac{2 \alpha_{\text{eff}}}{\moy{\sigma_c^2}}} \, \frac{\moy{\sigma_c^6}}{\moy{\sigma_c^4}}
\end{eqnarray}
and secondly the case at ${\alpha_{\text{eff}} \ll \alpha_c}$:
\begin{eqnarray}
\label{eq-rhotilde-stationary-disord-final-HL-wellbelowalphac}
\lim_{\dot{\gamma}\to 0} \frac{\tilde{\rho}_{\text{st}}(\sigma_c)}{\rho(\sigma_c)}
	& \stackrel{(\sigma_c^* \to 0)}{\approx} & \frac{\sigma_c^2}{2 \alpha_{\text{eff}}} \, \frac{\sigma_c^*}{\sigma_c}
	\stackrel{(\alpha_{\text{eff}} \ll \alpha_c)}{\approx} \frac{\sigma_c}{\moy{\sigma_c}} \\
\label{eq-rhotilde-stationary-disord-final-HL-wellbelowalphac-meanvalue}
\lim_{\dot{\gamma}\to 0} \overline{\sigma_c}
	& \approx & \frac{\moy{\sigma_c^2}}{\moy{\sigma_c}}
	= \moy{\sigma_c} + \frac{\text{Var}(\sigma_c)}{{\moy{\sigma_c}}}
	 \\
\label{eq-rhotilde-stationary-disord-final-HL-wellbelowalphac-secondmoment}
\lim_{\dot{\gamma}\to 0} \overline{\sigma_c^2}
	& \approx & \frac{\moy{\sigma_c^3}}{\moy{\sigma_c}}
\end{eqnarray}
with ${\text{Var}(\sigma_c) \equiv \moy{\sigma_c^2}-\moy{\sigma_c}^2}$.
eq.~\eqref{eq-rhotilde-stationary-disord-final-HL-wellbelowalphac-meanvalue} shows that,
when the diffusion is suppressed (${\alpha \to 0}$),
we have ${\overline{\sigma_c} \neq \moy{\sigma_c}}$ and the distribution ${\tilde{\rho}_{\text{st}}(\sigma_c)}$ does not coincide with ${\rho(\sigma_c)}$, except if ${\rho(\sigma_c)}$ has a zero variance, in other words if $\sigma_c$ can take only one value.

We consider now the case at ${\alpha_{\text{eff}} = \alpha_c}$, for which we can write down the two lowest orders at low shear rate:
\begin{equation}
\label{eq-rhotilde-stationary-disord-final-HL-atalphac}
\frac{\tilde{\rho}_{\text{st}}(\sigma_c)}{\rho(\sigma_c)}
	\stackrel{(\dot{\gamma} \to 0)}{\approx}
	\frac{\sigma_c^2}{2 \alpha_c}
	+ \frac{1}{\alpha_c} \argp{\sigma_c \, \widetilde{C}^{1/2} - \sigma_c^4 \, \frac{1}{\widetilde{C}^2}} (G_0 \dot{\gamma} \tau)^{2/5}
\end{equation}
with $\widetilde{C} = \argp{\frac{\moy{\sigma_c^4}}{24 \moy{\sigma_c}}}^{2/5}$ the prefactor of the diffusion coefficient, as defined in eq.~\eqref{eq-DHL-withshear-smallgammadot-disord-1}, leading to:
\begin{equation}
\label{eq-rhotilde-stationary-disord-final-HL-atalphac-meanvalue}
	\overline{\sigma_c}
	\stackrel{(\dot{\gamma} \to 0)}{\approx}
	\frac{\moy{\sigma_c^3}}{\moy{\sigma_c^2}}
	+ \argp{2 \widetilde{C}^{1/2} - \frac{\moy{\sigma_c^5}}{12 \moy{\sigma_c^2}} \, \frac{1}{\widetilde{C}^2}} (G_0 \dot{\gamma} \tau)^{2/5}
\end{equation}
\begin{equation}
\label{eq-rhotilde-stationary-disord-final-HL-atalphac-secondmoment}
	\overline{\sigma_c^2}
	\stackrel{(\dot{\gamma} \to 0)}{\approx}
	\frac{\moy{\sigma_c^4}}{\moy{\sigma_c^2}}
	+ \argp{2 \frac{\moy{\sigma_c^3}}{\moy{\sigma_c^2}} \widetilde{C}^{1/2} - \frac{\moy{\sigma_c^6}}{12 \moy{\sigma_c^2}} \, \frac{1}{\widetilde{C}^2}} (G_0 \dot{\gamma} \tau)^{2/5}
\end{equation}

We consider at last the case at ${\alpha_{\text{eff}} > \alpha_c}$, for which we can also write down the two lowest orders at low shear rate:
\begin{equation}
\label{eq-rhotilde-stationary-disord-final-HL-abovealphac}
\frac{\tilde{\rho}_{\text{st}}(\sigma_c)}{\rho(\sigma_c)}
	\stackrel{(\dot{\gamma} \to 0)}{\approx}
	\frac{ \argp{x_0^2+\sigma_c x_0 + \sigma_c^2/2}}{\alpha_{\text{eff}}}
	+ \mathcal{O}\argp{ \dot{\gamma}^{2}}
\end{equation}
with ${x_0=\sqrt{D_{\text{HL}} \tau}}$ given by eq.~\eqref{eq-solution-D-without-gammadot-disord},
${D_{\text{HL}}}$ being the diffusion coefficient at zero shear rate.
This leads to:
\begin{equation}
\label{eq-rhotilde-stationary-disord-final-HL-abovealphac-meanvalue}
	\overline{\sigma_c}
	\stackrel{(\dot{\gamma} \to 0)}{\approx}
	\frac{\moy{\sigma_c} x_0^2+ \moy{\sigma_c^2} x_0 + \moy{\sigma_c^3}/2}{\alpha_{\text{eff}}}
	+\mathcal{O}\argp{ \dot{\gamma}^{2}}
\end{equation}
\begin{equation}
\label{eq-rhotilde-stationary-disord-final-HL-abovealphac-secondmoment}
	\overline{\sigma_c^2}
	\stackrel{(\dot{\gamma} \to 0)}{\approx}
	\frac{\moy{\sigma_c^2} x_0^2+ \moy{\sigma_c^3} x_0 + \moy{\sigma_c^4}/2}{\alpha_{\text{eff}}}
	+\mathcal{O}\argp{ \dot{\gamma}^{2}}
\end{equation}
In the case ${\alpha_{\text{eff}} \gtrsim \alpha_c}$, the expansion \eqref{eq-rhotilde-stationary-disord-final-HL-abovealphac} is actually valid only for shear rates such that
${G_0 \dot{\gamma} \tau < \frac{12 (\alpha_c - \alpha_{\text{eff}})^2 \moy{\sigma_c^2}}{G_0 \tau \, \moy{\sigma_c^4} \moy{\sigma_c}^2}}$
as already noticed after eq.~\eqref{eq-sigmaM-withshear-smallgammadot-disord-2} for the mean stress ${\sigma_M}$ expansion above ${\alpha_c}$.
Furthermore, reading directly the limit ${\dot{\gamma} \to 0}$ of eqs.~\eqref{eq-rhotilde-stationary-disord-final-HL-abovealphac}-\eqref{eq-rhotilde-stationary-disord-final-HL-abovealphac-secondmoment}, we can write explicitly the expansions at ${\alpha_{\text{eff}} \gtrsim \alpha_c}$ using that ${x_0 \approx \frac{\alpha_{\text{eff}}-\alpha_c}{\moy{\sigma_c}}}$:
\begin{eqnarray}
\label{eq-rhotilde-stationary-disord-final-HL-abovealphac-strictlyzerogammadot-1}
 \lim_{\dot{\gamma}\to 0} \frac{\tilde{\rho}_{\text{st}}(\sigma_c)}{\rho(\sigma_c)}
 & \approx & \frac{\sigma_c^2}{2 \alpha_c} + \frac{\sigma_c}{\moy{\sigma_c}} \frac{\alpha_{\text{eff}}-\alpha_c}{\alpha_c}
 \\
\label{eq-rhotilde-stationary-disord-final-HL-abovealphac-strictlyzerogammadot-2}
 \lim_{\dot{\gamma}\to 0} \overline{\sigma_c}
 & \approx & \frac{\moy{\sigma_c^3}}{\moy{\sigma_c^2}} + \frac{\moy{\sigma_c^2}}{\moy{\sigma_c}} \argp{\frac{\alpha_{\text{eff}}}{\alpha_c} -1}
 \\
\label{eq-rhotilde-stationary-disord-final-HL-abovealphac-strictlyzerogammadot-3}
 \lim_{\dot{\gamma}\to 0} \overline{\sigma_c^2}
 & \approx & \frac{\moy{\sigma_c^4}}{\moy{\sigma_c^2}} + \frac{\moy{\sigma_c^3}}{\moy{\sigma_c}} \argp{\frac{\alpha_{\text{eff}}}{\alpha_c} -1}
\end{eqnarray}

The complete behaviors of the mean value ${\overline{\sigma_c}}$ and the variance ${\text{Var}_{\tilde{\rho}}(\sigma_c) = \overline{\sigma_c^2}-\overline{\sigma_c}^2}$ are illustrated in figs.~\ref{fig:rhotilde-meanvariance-WRTG0gammadottau} and~\ref{fig:rhotilde-meanvariance-at-fixed-alpha}, respectively at fixed ${\dot{\gamma}}$ and at fixed $\alpha_{\text{eff}}$, for the exponentially decaying distribution ${\rho_0 (\sigma_c)}$ of eq.~\eqref{eq-def-rho-SGR}.
We can immediately see that ${\alpha_{\text{eff}}=\alpha_c}$ plays a special role in the limit ${\dot{\gamma} \to 0}$, as it corresponds to the maximum value of ${\overline{\sigma_c}}$.
Physically, the dynamical distribution ${\tilde{\rho}_{\text{st}} (\sigma_c)}$ is an interplay between the a priori distribution ${\rho(\sigma_c)}$ and how fast the local yield stress values are refreshed by the plastic events triggered by the shear rate. The larger the shear rate, the faster the local yield stress is refreshed and the more small energy barriers $\propto \sigma_c^2$ can be present in the system. On the contrary, the lower the shear rate, the less plastic events we have, so the largest barriers survive longer and the distribution ${\tilde{\rho}_{\text{st}} (\sigma_c)}$ allows larger values of $\sigma_c$ on average.

%
These different sets of predictions for the stationary distribution ${\tilde{\rho}(\sigma_c)}$ provide new features to test numerically or experimentally, in addition to the characterization of the mean stress ${\sigma_M (\dot{\gamma})}$ and more generally of the complete stress distribution ${\mathcal{P}_{\text{st}}(\sigma)}$.
The complete mapping between the dynamical distribution ${\tilde{\rho}(\sigma_c)}$ and the a priori distribution ${\rho (\sigma_c)}$ is provided by eq.~\eqref{eq-rhotilde-stationary-disord-final-HL}.
In the case of a low shear rate, the strict limit ${\dot{\gamma} \to 0}$ can be computed exactly for the different regimes in $\alpha_{\text{eff}}$;
in the specific cases of ${\alpha_{\text{eff}} \ll \alpha_c}$, ${\alpha_{\text{eff}} \lesssim \alpha_c}$ and ${\alpha_{\text{eff}} \gtrsim \alpha_c}$ it turns out to depend solely on the first moments of the distribution ${\rho (\sigma_c)}$.
Moreover, the next order in ${G_0 \dot{\gamma} \tau}$ is available at ${\alpha_{\text{eff}}=\alpha_c}$ and at ${\alpha_{\text{eff}}>\alpha_c}$, and the discrepancy between these scalings
--~respectively ${\sim (G_0 \dot{\gamma} \tau)^{2/5}}$ and ${\sim (G_0 \dot{\gamma} \tau)^{2}}$~-- provides an additional way to probe in which regime in ${\alpha_{\text{eff}}}$ we might be.

%
In a numerical or experimental test of these predictions, if we have access simultaneously to the complete dynamical distribution ${\tilde{\rho}_{\text{st}}(\sigma_c)}$ and to the parameters $\arga{G_0,D,\dot{\gamma},\tau,\alpha_{\text{eff}}}$, then we can use eqs.~\eqref{eq-rhotilde-stationary-disord-final-HL}, \eqref{eq-rhotilde-stationary-disord-final-HL-belowalphac}, \eqref{eq-rhotilde-stationary-disord-final-HL-atalphac} or \eqref{eq-rhotilde-stationary-disord-final-HL-abovealphac} in order to determine the a priori distribution ${\rho (\sigma_c)}$.
If we do not have access to the complete ${\tilde{\rho}_{\text{st}}(\sigma_c)}$, but only to its mean value and variance for instance, then the connexion is still possible but we need to assume a given shape for ${\rho(\sigma_c)}$~\cite{Arxiv-2014-Puosi_Olivier_Martens}.

%
We can also mention the opposite limit of large shear rate, in which
${\Gamma_{\text{st}}\tau \approx 1}$ (all the sites are brought above the local yield stress in the time interval $\tau$, in the stationary state), so ${D_{\text{HL}} \tau \approx \alpha_{\text{eff}}}$ and we have at lowest order in~$1/\dot{\gamma}$:
\begin{equation}
\label{eq-rhotilde-stationary-disord-final-HL-largeshearrate}
 \frac{\tilde{\rho}_{\text{st}}(\sigma_c)}{\rho(\sigma_c)}
 \stackrel{(\dot{\gamma}\to \infty)}{\approx}
 \frac{\tilde{f}_{\sigma_c} \argp{\sqrt{\alpha_{\text{eff}}}, \frac{G_0 \dot{\gamma} \tau}{\alpha_{\text{eff}}}}}{\alpha_{\text{eff}}}
 \stackrel{\eqref{eq-factorf-HL-withshear}}{\approx}
 1+\frac{\sigma_c}{G_0 \dot{\gamma} \tau}
\end{equation}
Consequently, we predict at large shear rates that ${\tilde{\rho}_{\text{st}}(\sigma_c) \approx \rho (\sigma_c)}$, with the same mean value and variance, as is indeed the case in fig.~\ref{fig:rhotilde-meanvariance-at-fixed-alpha}.
Physically, at large shear rates the system is essentially overstressed and on the verge of yielding everywhere, so
first the diffusion coefficient tends to a constant,
secondly the mean stress ${\sigma_M \sim \dot{\gamma}}$ has a Newtonian behavior,
and thirdly the distinction between the dynamical and the a priori distributions of local yield stress $\sigma_c$ has been washed out.

%
It is interesting at this stage to try to provide a physical interpretation of the results obtained for $\tilde{\rho}_{\text{st}}(\sigma_c)$, or in other words, to try to understand how the a priori distribution $\rho(\sigma_c)$ is reweighted to give $\tilde{\rho}_{\text{st}}(\sigma_c)$. A naive reasoning suggests that if the dynamics of $\sigma$ is dominated by the drift $\dot{\gamma}$, the time needed to go from $\sigma=0$ to $\sigma_c$ should be proportional to $\sigma_c$, so that one would expect $\tilde{\rho}_{\text{st}}(\sigma_c) \propto \sigma_c \, \rho(\sigma_c)$. Similarly, if the dynamics is dominated by the diffusion, the time to go from $0$ to $\sigma_c$ should scale as $\sigma_c^2$, so that one expects $\tilde{\rho}_{\text{st}}(\sigma_c) \propto \sigma_c^2 \, \rho(\sigma_c)$.

The first situation (dynamics dominated by the drift) should occur for large $\dot{\gamma}$. However, eq.~(\ref{eq-rhotilde-stationary-disord-final-HL-largeshearrate}) shows that for $\dot{\gamma} \to \infty$, ${\tilde{\rho}_{\text{st}}(\sigma_c)=\rho(\sigma_c)}$, so that the a priori distribution is not reweighted by $\sigma_c$ as expected from the above naive argument. This is actually due to the fact that $\sigma$ does not jump to $0$ exactly at $\sigma_c$, but has only a probability $1/\tau$ per unit time to jump to $0$ when $\sigma > \sigma_c$. For large $\dot{\gamma}$, the local stress $\sigma$ thus reaches values much larger than the local yield stress $\sigma_c$ before jumping to $0$.
The `lifetime' of a state with a given $\sigma_c$ is ${\tau+\sigma_c/(G_0\dot{\gamma})}$, and it leads to the reweighting given in eq.~(\ref{eq-rhotilde-stationary-disord-final-HL-largeshearrate}).

On the other hand, the diffusive regime is expected to occur in the low $\dot{\gamma}$ limit, and for ${\alpha_{\text{eff}}>\alpha_c}$ so that the diffusion coefficient does not vanish. Here again, the result given in eq.~(\ref{eq-rhotilde-stationary-disord-final-HL-abovealphac}) seems to rule out the naive expectation, since the reweighting of $\rho(\sigma_c)$ is not proportional to $\sigma_c^2$. However, a closer look actually shows that intuition was not wrong. Considering a fixed diffusion coefficient $D$ (\textit{i.e.}, discarding the closure relation eq.~(\ref{eq-closure-DHL-disord-stat1})) and taking the limit $\tau \to 0$ to make the stress jump to zero sharply at $\sigma_c$, we indeed get from eq.~(\ref{eq-rhotilde-stationary-disord-final-HL-abovealphac}) that the reweighting of $\rho(\sigma_c)$ is proportional to $\sigma_c^2$. The generic deviation from this scaling in eq.~(\ref{eq-rhotilde-stationary-disord-final-HL-abovealphac}) again results from the fact that $!
 \tau$ is finite, so that $\sigma$ can increase above $\sigma_c$. This can be quantified by comparing $\tau$ to the diffusion time ${\sigma_c^2/D}$ to reach $\sigma_c$ starting from $\sigma=0$: the small $\tau$ regime corresponds to ${\tau \ll \sigma_c^2/D}$, or equivalently to ${D\tau \ll \sigma_c^2}$.
Note that taking into account the closure relation eq.~(\ref{eq-closure-DHL-disord-stat1}), the self-consistent diffusion coefficient $D_{\text{HL}}$ is determined via the product ${D_{\text{HL}} \tau}$ (see eq.~(\ref{eq-DHL-withshear-smallgammadot-disord-1})), so that the limit $\tau \to 0$ cannot be taken at fixed ${D_{\text{HL}}}$. Actually, the control parameter is $\alpha_{\text{eff}}$ rather than $\tau$, and one finds that ${D_{\text{HL}} \tau}$ is small for $\alpha_{\rm eff}$ close to $\alpha_c$ (with ${\alpha_{\rm eff} > \alpha_c}$); in this regime, one recovers a reweighting proportional to $\sigma_c^2$ to leading order, as seen in eq.~(\ref{eq-rhotilde-stationary-disord-final-HL-abovealphac-strictlyzerogammadot-1}).

\subsection{Connection with the disordered KEP model}
\label{section-HL-model-stationary-KEP-disord}

Now that we have generalized the analytical predictions of the original HL model by including the distribution ${\rho (\sigma_c)}$, we can conclude this study by examining the disordered extension of the KEP construction \cite{bocquet_PhysRevLett103_036001} (recalled in sect.~\ref{section-HL-model-stationary-KEP}), in order to provide a justification for our disordered HL model, and in particular for the choice of a diffusion coefficient independent of $\sigma_c$.

First the evolution equation ${\partial_t \mathcal{P}_i(\sigma ,t)}$ in~\cite{bocquet_PhysRevLett103_036001} can be extended to ${\partial_t \widetilde{\mathcal{P}}_i(\sigma_c,\sigma ,t)}$, allowing for stress propagation between regions with different yield stress values, replacing for that purpose the operator `${\mathcal{L}_{i,\sigma} (\mathcal{P},\mathcal{P})}$' by:
\begin{equation}
\begin{split}
 \mathcal{L}_{i,\sigma_c , \sigma} (\mathcal{P},\mathcal{P})
 = & \sum_{j (\neq i)} \int_0^{\infty} \!\!\!\!\! d \sigma_c' \, \frac{1}{\tau} \int_{\valabs{\sigma '}>\sigma_c'} \!\!\!\!\!\!\!\!\!\!\!\!\!\! d\sigma'
 \, \widetilde{P}_j \argp{\sigma_c',\sigma',t} \\
 & \times \argc{ \widetilde{P}_i \argp{\sigma_c,\sigma + \delta \sigma_i^{(j)},t} - \widetilde{P}_i \argp{\sigma_c,\sigma,t}}
 \end{split}
\end{equation}
Note that this operator implicitly assumes that the local distributions  ${\widetilde{\mathcal{P}}_i(\sigma ,t)}$ are independent, allowing for the factorization of their joint distributions (\textit{i.e.}, ${\widetilde{\mathcal{P}}_{ij}=\widetilde{\mathcal{P}}_i \, \widetilde{\mathcal{P}}_j}$).
This evolution equation then transforms into an effective HL equation, with an effective local shear rate ${\dot{\gamma}_i (t)}$ and a diffusion coefficient ${D_i(t)}$, with the following slightly modified set of assumptions:
\textit{(i)}~A plastic rearrangement occurs at a site $j$ soon after its local stress ${\sigma'}$ has exceeded the local yield stress, so ${\valabs{\sigma'} \approx \sigma_c^{(j)}}$.
\textit{(ii)}~The stress is fully relaxed at the site $j$ (${\sigma' \to 0}$) hence the stress that propagates from the site $j$ to the site $i$ is ${\delta \sigma_i^{(j)} \approx - G_{ij} \sigma' \approx - G_{ij} \sigma_c^{(j)}}$, with ${G_{ij}}$ the microscopic elastic propagator.
\textit{(iii)}~This stress ${\delta \sigma_i^{(j)}}$ acts only as a perturbation that can trigger a plastic event only if the site $i$ is already on the verge of yielding, \textit{i.e.}, ${\valabs{\frac{\delta \sigma_i^{(j)}}{\sigma_c^{(i)}}} \approx \valabs{G_{ij} \, \frac{\sigma_c^{(j)}}{\sigma_c^{(i)}}} \ll 1}$.
So this construction remains valid assuming either that a plastic rearrangement occurs sufficiently far away, or that the amplitude of the elastic propagator is small, but only as long as the ratio ${\frac{\sigma_c^{(j)}}{\sigma_c^{(i)}}}$ remains sufficiently small as well, a criterion that might constrain the variance of the distribution ${\rho (\sigma_c)}$.

With the operator ${\mathcal{L}_{i,\sigma_c , \sigma} (\mathcal{P},\mathcal{P})}$ being a linear combination of the contributions of the different values of $\sigma_c$,
the KEP relation between the diffusion coefficient ${D(\mathbf{r},t)}$ and the local plastic activity ${\Gamma (\mathbf{r},t)}$ of eq.~\eqref{eq-closure-DHL-versionKEP-1}-\eqref{eq-closure-DHL-versionKEP-2} becomes:
\begin{equation}
\label{eq-closure-DHL-versionKEP-disord-1}
 D(\mathbf{r},t)
  = \int_0^{\infty} \!\!\!\!\!\! d \sigma_c \argc{\tilde{m}(\sigma_c) \, \partial_{\mathbf{r}}^2 \widetilde{\Gamma} (\sigma_c,\mathbf{r},t) + \tilde{\alpha}(\sigma_c) \, \widetilde{\Gamma} (\sigma_c,\mathbf{r},t)}
\end{equation}
and in the stationary case, using eq.~\eqref{eq-rhotilde-stationary}, we finally obtain:
\begin{eqnarray}
\label{eq-closure-DHL-versionKEP-disord-3}
 && D_{\text{KEP}}(\mathbf{r})
 = m_{\text{eff}}\left[ \rho \right] \, \partial_{\mathbf{r}}^2 \Gamma_{\text{st}} (\mathbf{r}) + \alpha_{\text{eff}}\left[ \rho \right] \, \Gamma_{\text{st}} (\mathbf{r}) \\
\label{eq-closure-DHL-versionKEP-disord-4}
 && \left\lbrace \begin{array}{ccccl}
 m_{\text{eff}}\left[ \rho \right]
 &=& \int_0^{\infty} d \sigma_c \, \tilde{m}(\sigma_c) \, \rho(\sigma_c)
 & \equiv & \langle \tilde{m}(\sigma_c) \rangle_{\rho} \\
 \alpha_{\text{eff}}\left[ \rho \right]
 &=& \int_0^{\infty} d \sigma_c \, \tilde{\alpha}(\sigma_c) \, \rho(\sigma_c)
 & \equiv & \langle \tilde{\alpha}(\sigma_c) \rangle_{\rho}
\end{array} \right.
\end{eqnarray}
This last result justifies, as anticipated, our choice of the time-dependent closure relation \eqref{eq-closure-DHL-disord} in general, the linear closure relation \eqref{eq-closure-DHL-disord-stat1} and the definition of the coupling parameter \eqref{eq-closure-DHL-disord-stat2} in the stationary case.

\section{Discussion and outlook}
\label{section-discussion}

\subsection{Discussion on the HL model assumptions}
\label{section-interplay-timescales}

%
Regarding the definition of the original HL model and its disordered counterpart, one pending issue is the interpretation of the physical mechanism underlying the choice of the plastic rate
${\nu_{\text{HL}} (\sigma , \sigma_c) \equiv \frac{1}{\tau} \theta (\valabs{\sigma} - \sigma_c)}$
in eqs.~\eqref{eq-nu-HL} and \eqref{eq-nu-HL-disord},
\textit{i.e.}, the assumption of a typical fixed rate ${1/\tau}$ of having a plastic event when the local stress exceeds the local threshold $\sigma_c$.
It has been highlighted in an earlier work, that one other key ingredient in order to obtain a shear-rate dependent flow-curve in athermal systems is the existence of at least one additional intrinsic timescale, that will be a material-dependent property~\cite{nicolas_martens_barrat_2014_EurPhysLett107_44003}.
This timescale has been identified as the dissipative time, which describes roughly the typical duration of the local relaxation process. The original definition of the fixed rate  ${1/\tau}$ in the HL model is not equivalent to this dissipative time, but it rather introduces a `local overshoot' regarding the local yield stress value. 
The physical interpretation of this process remains somehow unclear.
However, it is possible to interpret the time $\tau$ in the HL model as the dissipation time defined in ref.~\cite{nicolas_martens_barrat_2014_EurPhysLett107_44003} by introducing a small correction in the definition of the macroscopic stress, namely
\begin{equation}
\label{eq-sigmaM-corr-over-sigmac}
\begin{split}
 \sigma_M^\mathrm{corr}
 & \equiv \int^{\infty}_{0} \!\! d \sigma_c  \int_{\mathbb{R}} d \sigma \, \text{min}(\sigma,\sigma_c) \, {\widetilde{\mathcal{P}}_{\text{st}}(\sigma_c,\sigma)} \\
 & = \sigma_M^{\text{(under)}} + \frac{ \moy{\sigma_c} D \tau}{\moy{\tilde{f}_{\sigma_c} \argp{\sqrt{D \tau},\frac{G_0 \dot{\gamma} \tau}{D\tau}}}}
\end{split}
\end{equation}
in other words assuming that in the `overstressed' regions the local stress does not exceed its local yield value.
Such a modification of the macroscopic stress definition does not alter the rheological behavior of the HL model at low shear rate, at least for the regimes ${\alpha_{\text{eff}} \leq \alpha_c}$ (at ${\alpha_{\text{eff}} > \alpha_c}$ it induces a finite macroscopic yield stress).
%
For small driving shear, this average is in fact dominated by the contributions of local stresses $\sigma$ below the local yield stress values~${\sigma_c}$, and the Herschel-Bulkley exponent is thus robust with respect to this subtle change.
Note however that the large shear rate regime will be influenced by such a correction, so we think it is important to eliminate the rather unphysical existence of locally overstressed regions for future considerations.
Also the existence of a finite dissipation time allows
for the appearance of shear localization \cite{martens_bocquet_barrat_2012_SoftMatter8_4197,nicolas_2014_SoftMatter10_4648}, a feature which is however out of the scope of the present study.

%
Another very strong assumption of the HL model is the full relaxation of the local stress after yielding, as in eq.~\eqref{eq-continuous-evol-sigma} of our toy model. This scenario is not always --~actually, rather rarely~-- satisfied \cite{Arxiv-2014-Puosi_Olivier_Martens}.
Partial relaxation is expected to modify the prefactors in the predicted scalings, for which we are still missing explicit analytical expressions in this case. Since in this study we have focused on the mechanism that leads to the onset of non-linearity and on its corresponding exponent, we are not concerned with the effect of partial relaxation of stresses.
For a quantitative comparison, we would need either to compute numerically the stationary joint PDF ${\widetilde{\mathcal{P}}_{\text{st}}(\sigma_c,\sigma)}$ at fixed $\dot{\gamma}$ 
replacing in eq.~\eqref{eq-dist-Psigma-HL-disord} the $\delta(\sigma)$ with a more general distribution $\Delta (\sigma)$.
Or one can alternatively study numerically the whole set of physical quantities we have defined, starting from the hybrid dynamics of eqs.~\eqref{eq-continuous-evol-sigma}-\eqref{eq-variance-xiplast} on a set of independent sites with local stress $\arga{\sigma_i(t)}$.

%
Finally, we can comment on the assumptions regarding the underlying distribution of local yield stress needed for our results to hold.
In the definition of our disordered HL model, we have assumed a generic ${\rho (\sigma_c)}$ whose moments are finite.
For instance, the perturbation expansion of the diffusion coefficient at low shear rate already involves its fourth moment ${\moy{\sigma_c^4}}$.
However, although the mean value and the variance are reasonably robust features of a distribution, in practice its higher moments can be strongly sensitive to the system size and to the limited available statistics.
So, although the power-law expansions of the HL mean stress (hence the qualitative behavior of the rheological law at low shear rate) are predicted to be robust to the addition of structural disorder, their corresponding prefactors depend on a combination of moments ${\moy{\sigma_c^k}}$ that might display a dependence on the system size.
A direct comparison between atomistic simulations and our analytical predictions should thus take into account such a possible dependence, and would require a careful characterization of the distributions of local yield stress ${\tilde{\rho} (\sigma_c)}$ and ${\rho (\sigma_c)}$.

\subsection{Summary and outlook}
\label{section-conclusion}

In this study we have identified on a mean-field level the necessary ingredients for the modeling of yield stress materials in the case of athermally activated yielding events.
Within the HL model, which we argue to better represent the underlying physical picture in athermally driven systems, we have studied analytically the robustness of the predictions with respect to an additional and usually important physical ingredient, namely the disorder in the yield energy barriers (or equivalently, in the local yield stress).
We find that, although a key ingredient in the SGR model, a distribution of threshold stresses does not modify qualitatively the HL predictions at low constant shear rate, thus predicting a universal critical behavior at the flow transition~${\dot{\gamma} \to 0}$.

The generality of the different analytical expressions in this paper, distinguishing the specific cases or limits taken, allows us not only to recover all known results on the HL model but also to go beyond them.
It enables us on the one hand to estimate numerically all the relevant physical quantities for a given distribution ${\rho (\sigma_c)}$, and on the other hand to consider alternative closure relations for the diffusion coefficient.

In future work we would like to address further questions to render the mean field equations even more consistent with the underlying physical dynamics.
It is known that, close to the flow transitions, complex dynamical heterogeneities in form of avalanches of yielding events play an important role in the plastic response to shear \cite{lemaitre_caroli_2009_PhysRevLett_103_065501,martens_bocquet_barrat_2011_PhysRevLett106_156001,nicolas_2014_SoftMatter10_4648}. 
So an important issue that remains is to better understand how the spatio-temporal correlations of the yielding events can be captured, within the mean field modeling approach, notably within the formulation of the effective noise term.
We emphasize that in the HL-like models, the effective noise is approximated by the Gaussian white noise assumption, whose variance (\textit{i.e.} the diffusion coefficient) is coupled to the plastic activity in the system.
In this work we have studied exclusively the stationary case at fixed low shear rate,
but this assumption might be questioned even more in the further study of transient regimes or for an oscillating shear rate.

Further we will be interested in combining the present picture and model with the notion of thermal noise and activation, in order to mimic a sheared material that is additionally subject to thermal activation of yielding events, in the spirit of our toy model picture.
It would be very interesting to understand in detail the role of disorder in this combined picture, already at the mean-field level, and later on in a complete statistical field theory of the stress field.

\begin{acknowledgments}

We acknowledge financial support from ERC grant ADG20110209.
JLB is supported by IUF.
E.A. acknowledges financial support by a Fellowship for Prospective Researchers Grant No P2GEP2-15586 from the Swiss National Science Foundation. 
KM acknowledges financial support of the French Agence Nationale de la Recherche,
under grant ANR-14-CE32-0005 (project FAPRES).
This research was supported in part by the National Science Foundation under Grant No. NSF PHY11-25915.
Furthermore, we would like to thank Ezequiel Ferrero, Alexandre Nicolas, Julien Olivier, and Francesco Puosi for fruitful discussions.

\end{acknowledgments}


\appendix

\section{Stationary plastic activity at fixed diffusion coefficient}
\label{A-appendix-factorf-sigmac}

In sect.~\ref{section-HL-model-stationary-solution} we have defined the function 
${\tilde{f}_{\sigma_c}}$
as the ratio between the stationary diffusion coefficient $D$ and the corresponding plastic activity ${\Gamma_{\text{st}}(D)}$, for the standard HL case of a single value for $\sigma_c$.
This definition was motivated by the specific closure relation ${D_{\text{HL}}=\alpha \Gamma_{\text{st}}}$ given in eq.~\eqref{eq-closure-DHL}, implying that either ${D=0}$, or ${D>0}$ according to the equality ${\tilde{f}_{\sigma_c} \argp{\sqrt{D \tau}, \frac{G_0 \dot{\gamma} \tau}{D \tau}}=\alpha}$.

In fact, the function ${\tilde{f}_{\sigma_c}  (x_1,x_2)}$ is completely fixed by the functional form of the stationary PDF, if we are able to solve the specific equation \eqref{eq-dist-Psigma-HL} ${\partial_t \mathcal{P}(\sigma,t)=0}$.
If this is the case, we can then define the normalized PDF ${p_{\pm}(\sigma)}$
from ${\mathcal{P}_{\text{st}}(\sigma \lessgtr 0)=\Gamma_{\pm} \tau \, p_{\pm} (\sigma)}$
and ${\int_{\pm \sigma_c}^{\pm \infty} d \sigma \, p_{\pm} (\sigma)= \pm 1}$.
This leads generically to:
\begin{equation}
\begin{split}
 \frac{\tilde{f}_{\sigma_c} \argp{\sqrt{D \tau}, \frac{G_0 \dot{\gamma} \tau}{D \tau}}}{D \tau}
 \equiv 1
 		& + \frac{p_+(0)}{p_+(0)+p_-(0)} \int_{-\sigma_c}^0 d \sigma \, p_-(\sigma) \\
 		& + \frac{p_-(0)}{p_+(0)+p_-(0)} \int^{\sigma_c}_0 d \sigma \, p_+(\sigma)
\end{split}
\end{equation}
and is equal to ${1/\Gamma_{\text{st}}(D)}$, see eq.~\eqref{eq-Gamma-HL-D-factorf}.

In sect.~\ref{section-HL-model-stationary-solution-disord}, we have furthermore generalized this definition to the case with an arbitrary distribution of local yield stress values ${\rho (\sigma_c)}$.
The normalization of the PDF implies respectively eq.~\eqref{eq-rhotilde-stationary-disord-1} for a generic ${\widetilde{D}(\sigma_c)}$, and eq.~\eqref{eq-rhotilde-stationary-disord-2} for a diffusion coefficient independent of $\sigma_c$.
In both cases we can use the function ${\tilde{f}_{\sigma_c}}$ previously determined for a fixed value of $\sigma_c$, as explained in Appendix~\ref{A-appendix-explicit-expressions}.

\section{Perturbative expansions of the stationary diffusion coefficient}
\label{A-appendix-perturb-diff-coeff}

In sect.~\ref{section-HL-model-stationary-low-shearrate} and~\ref{section-HL-model-stationary-low-shearrate-disord}, we give the perturbative expansions of the diffusion coefficient ${D_{\text{HL}}}$ at low shear rate, respectively for the standard HL model in eq.~\eqref{eq-DHL-withshear-smallgammadot-1} and for its disordered counterpart in eq.~\eqref{eq-DHL-withshear-smallgammadot-disord-1}.
The perturbative expansion of the mean local stress $\sigma_M$ is then straightforwardly obtained by substituting into its exact expression \eqref{eq-sigmaM-appendix} the expansion of ${D_{\text{HL}}}$, and expanding the resulting expression at small ${G_0 \dot{\gamma} \tau }.$

In this appendix, we first recall exact mathematical results regarding the perturbative expansion of ${D_{\text{HL}}}$ in the standard HL model.
The general structures of the perturbative expansions of ${D_{\text{HL}}}$ and $\sigma_M$ are discussed in Chapter~2 of ref.~\cite{phdthesis_JulienOlivier2011},
and are given explicitly by the theorem~{4.1} of ref.~\cite{olivier_renardy_2011_SIAMJApplMath71_1144}:
%
\begin{equation}
\label{eq-JO-thm41-expansion-DHL}
	D_{\text{HL}} \stackrel{(\dot{\gamma} \to 0)}{\approx}
	\left\lbrace \begin{array}{ll}
	\mathcal{O} (\dot{\gamma}^{0}) + \mathcal{O} (\dot{\gamma}^{1}) + \mathcal{O} (\dot{\gamma}^{2}) + \dots
	& (\alpha > \alpha_c) \\
	\mathcal{O} (\dot{\gamma}^{4/5}) + \mathcal{O} (\dot{\gamma}^{1}) + \dots
	& (\alpha = \alpha_c)  \\
	\mathcal{O} (\dot{\gamma}^{1}) + \mathcal{O} (\dot{\gamma}^{3/2}) + \dots
	& (\alpha < \alpha_c)
	\end{array} \right.
\end{equation}
and its corollary~{4.2}~\cite{olivier_renardy_2011_SIAMJApplMath71_1144}:
\begin{equation}
\label{eq-JO-cor42-expansion-sigmaM}
	\sigma_M \stackrel{(\dot{\gamma} \to 0)}{\approx}
	\left\lbrace \begin{array}{ll}
	\mathcal{O} (\dot{\gamma}^{1}) + \mathcal{O} (\dot{\gamma}^{2}) + \dots
	& (\alpha > \alpha_c) \\
	\mathcal{O} (\dot{\gamma}^{1/5}) + \mathcal{O} (\dot{\gamma}^{2/5}) + \dots
	& (\alpha = \alpha_c)  \\
	\mathcal{O} (\dot{\gamma}^{0}) + \mathcal{O} (\dot{\gamma}^{1/2}) + \dots
	& (\alpha < \alpha_c)
	\end{array} \right.
\end{equation}
Note that in these references, the quantities of interest are made dimensionless with respect to the single value $\sigma_c$, with the definitions
${\mu = \alpha/\sigma_c^2}$, ${\phi = D \tau / \sigma_c^2}$ and hence ${\Gamma_{\text{st}} \tau = D_{\text{HL}} \tau / \alpha = \phi / \mu}$.
These perturbative expansions are systematically constructed and well-controlled, without any a priori knowledge of the corresponding convergent series, using `asymptotic expansions' of the stationary solution of the PDF~\cite{olivier_renardy_2011_SIAMJApplMath71_1144}.
Although such a procedure can in principle be generalized to the disordered HL case, it is not straightforward.

We have thus taken a shortcut for the disordered HL case, in order to obtain the lowest order of the expansion given in eq.~\eqref{eq-recap-HL-predictions}, using the exponents in eq.~\eqref{eq-JO-thm41-expansion-DHL} as guides in the standard Taylor expansions.
This shortcut consists in assuming the following ansatz at low $\dot{\gamma}$:
\begin{equation}
\label{eq-ansatz-diff-coeff-small-gammadot}
\begin{split}
 & x^2 = D_{\text{HL}}\tau
 \stackrel{(\dot{\gamma} \to 0)}{\approx}
 C_1 \, (G_0 \dot{\gamma} \tau)^{\delta_1}
 \\
 & y = \frac{G_0 \dot{\gamma} \tau}{x^2}
 \stackrel{(\dot{\gamma} \to 0)}{\approx}
 \frac{(G_0 \dot{\gamma} \tau)^{1-\delta_1}}{C_1}
\end{split}
\end{equation}
with ${0 \leq \delta_{1} \leq 1}$.
In the limit ${\dot{\gamma} \to 0}$, we have three possible cases:
\begin{equation}
\label{eq-ansatz-diff-coeff-small-gammadot-3cases}
 \begin{array}{cll}
	\delta_1=1 \, : \quad & x \to 0 \, , \quad & y \to 1/C_1 \\
	0< \delta_1 <1 \, : \quad & x \to 0 \, , \quad & y \to 0 \\
	\delta_1=0 \, : \quad & x \to x_0 \, , \quad & y \to 0
 \end{array}
\end{equation}
for which we expand the function ${\tilde{f}_{\sigma_c} \argp{x, y}}$ given in eq.~\eqref{eq-factorf-HL-withshear} at low shear rate.
If ${\delta_1=1}$, we have:
\begin{equation}
\label{eq-expansion-fidtilde-casedelta1}
  \lim_{x \to 0} \tilde{f}_{\sigma_c} (x, 1/C_1)
 = C_1 \sigma_c \tanh \argp{\frac{\sigma_c}{2 C_1}}
\end{equation}
If ${\delta_1=0}$, we have:
\begin{equation}
\label{eq-expansion-fidtilde-casedelta0}
 \tilde{f}_{\sigma_c} \argp{x_0,G_0 \dot{\gamma}\tau /x_0^2}
 = x_0^2 + \sigma_c x_0 + \frac{\sigma_c^2}{2} + \mathcal{O} \argp{\dot{\gamma}^2}
\end{equation}
And if ${0<\delta_1<1}$, we have:
\begin{equation}
\label{eq-expansion-fidtilde-casedelta01}
 \tilde{f}_{\sigma_c} \argp{x, y}
 = \frac{\sigma_c^2}{2} + \sigma_c x -\frac{\sigma_c^4}{24}y^2+ x^2 + \mathcal{O} \argp{x y^2} + \mathcal{O} \argp{y^4}
\end{equation}
We can identify which value of $\delta_1$ is associated to each regime of ${\alpha_{\text{eff}}}$ and determine the corresponding prefactor $C_1$, by solving at lowest order the equation deduced from the closure relation~\eqref{eq-closure-DHL-disord}:
$${\moy{ \tilde{f}_{\sigma_c} \argp{x, G_0 \dot{\gamma} \tau / x^2}} - \alpha_{\text{eff}} =0}$$
We start from the case ${\delta_1=0}$, that yields:
\begin{equation}
\label{eq-solutionC1-casedelta0}
 x_0^2 + \moy{\sigma_c} x_0 + \frac12 \moy{\sigma_c^2} = \alpha_{\text{eff}}
\end{equation}
which admits a positive solution of ${x_0 = \sqrt{D_{\text{HL}}\tau}}$ only if ${\alpha_{\text{eff}} > \alpha_c= \frac12 \moy{\sigma_c^2}}$, as given by eq.~\eqref{eq-solution-D-without-gammadot-disord}.
We then turn to the case ${0<\delta_1<1}$, where ${x^2 \sim \dot{\gamma}^{\delta_1}}$ is small compared to ${x \sim \dot{\gamma}^{\delta_1/2}}$, so we can cancel the two lowest orders provided that:
\begin{equation}
\label{eq-solutionC1-casedelta01}
\begin{split}
 & \alpha_{\text{eff}}=\frac12 \moy{\sigma_c^2}=\alpha_c \\
 & \moy{\sigma_c} C_1^{1/2} \, (G_0 \dot{\gamma} \tau)^{\delta_1/2} -\frac{ \moy{\sigma_c^4}}{24 C_1^2} \, (G_0 \dot{\gamma} \tau)^{2-2\delta_1} =0
\end{split}
\end{equation}
implying that ${\delta_1=4/5}$ and $C_1=\widetilde{C}$ as given by eq.~\eqref{eq-DHL-withshear-smallgammadot-disord-1}.
The last case ${\delta_1=1}$ should thus correspond to ${\alpha_{\text{eff}}<\alpha_c}$, and it actually yields:
\begin{equation}
\label{eq-solutionC1-casedelta1}
 \moy{C_1 \sigma_c \tanh \argp{\frac{\sigma_c}{2 C_1}}}=\alpha_{\text{eff}}
\end{equation}
as given in eq.~\eqref{eq-DHL-withshear-smallgammadot-disord-1}.
So it is the specific function ${\tilde{f}_{\sigma_c}(x,y)}$ of eq.~\eqref{eq-factorf-HL-withshear} that allows us to order the lowest orders in the perturbation, on the sole assumption that ${0 \leq \delta_1 \leq 1}$, and then identifying which value of $\delta_1$ correspond to each regime in $\alpha_{\text{eff}}$.
The predictions for the disordered HL model are gathered in eq.~\eqref{eq-DHL-withshear-smallgammadot-disord-1}, and we can recover their counterparts for the standard HL model by replacing all the moments ${\moy{\sigma_c^k}}$ by $\sigma_c^k$, as listed in eq.~\eqref{eq-DHL-withshear-smallgammadot-1}.

Actually, in order to obtain the derivation of the Herschel-Bulkley behavior of $\sigma_M$ at ${\alpha_{\text{eff}}< \alpha_c}$, we need to compute the second lowest order of $D_{\text{HL}}$. We thus start from the ansatz:
\begin{equation}
\label{eq-ansatz-diff-coeff-small-gammadot-bis}
 D_{\text{HL}}\tau = x^2
 \stackrel{(\dot{\gamma} \to 0)}{\approx}
 C_1 \, G_0 \dot{\gamma} \tau \argc{1 + C_2 \, (G_0 \dot{\gamma} \tau)^{1/2}} 
\end{equation}
as suggested by eq.~\eqref{eq-JO-thm41-expansion-DHL} for the standard HL model,
and the same procedure as before leads to the following relation between $C_2$ and $C_1$:
\begin{equation}
\label{eq-HL-law-diffcoeff-C2-equa}
 C_2 = \sqrt{C_1} \frac{\frac{\moy{\sigma_c}}{2} + C_1 \moy{\tanh \argp{\frac{\sigma_c}{2 C_1}}}  -\frac12 \moy{\sigma_c \tanh^2 \argp{\frac{\sigma_c}{2 C_1}}} }{\frac{\moy{\sigma_c^2}}{2} - \moy{C_1 \sigma_c \tanh \argp{ \frac{\sigma_c}{2 C_1}}} - \frac12 \moy{\sigma_c^2 \tanh^2 \argp{ \frac{\sigma_c}{2 C_1}}}} 
\end{equation}
The resulting predictions for the mean stress $\sigma_M$, and specifically for the prefactor $A$ of the stress contribution in ${(G_0 \dot{\gamma} \tau)^{1/2}}$, are given explicitly in Appendix~\ref{A-appendix-prefactorA-disordHL}.

The expression for ${C_1=C_1(\alpha_{\text{eff}})}$ and consequently for ${C_2=C_2(C_1)}$ can be considerably simplified in the two limiting cases ${\alpha_{\text{eff}} \lesssim \alpha_c}$ and ${\alpha_{\text{eff}} \ll \alpha_c}$, and they lead to eq.~\eqref{eq-DHL-withshear-smallgammadot-disord-2}.
The argument is the following:
first, for each coupling parameter ${\alpha_{\text{eff}}}$ below ${\alpha_c}$, we can define a typical value ${\sigma_c^*=2 C_1}$.
Then, on the one hand, close to $\alpha_c$ we have ${C_1 \to \infty}$ and ${\sigma_c^* \to \infty}$, so we can safely neglect the contributions of ${\sigma_c > \sigma_c^*}$.
For the contributions of ${\sigma_c < \sigma_c^*}$, we can approximate the hyperbolic tangent with its Taylor expansion at ${2 \sigma_c/C_1 \ll 1}$.
On the other hand, with ${\alpha_{\text{eff}}}$ close to zero, we have ${C_1 \to 0}$ and ${\sigma_c^* \to 0}$, so we can neglect the contributions of ${\sigma_c < \sigma_c^*}$ and use for ${\sigma_c > \sigma_c^*}$ the approximation ${\tanh (2 \sigma_c/C_1) \approx 1}$.
In practice, we can decompose the average ${\moy{\mathcal{O}}}$ in eq.~\eqref{eq-DHL-belowalphac-smallgammadot-disord-1} into two separate averages, restricted on the contributions from ${\sigma_c \lessgtr \sigma_c^*}$.
%
So for ${\alpha_{\text{eff}} \lesssim \alpha_c}$ we have
${\moy{\mathcal{O}}_{\sigma_c > \sigma_c^*} \approx 0}$
and
${\moy{\mathcal{O}} \approx \moy{\mathcal{O}}_{\sigma_c < \sigma_c^*}}$,
whereas at ${\alpha_{\text{eff}} \ll \alpha_c}$ we have
${\moy{\mathcal{O}}_{\sigma_c < \sigma_c^*} \approx 0}$
and
${\moy{\mathcal{O}} \approx \moy{\mathcal{O}}_{\sigma_c > \sigma_c^*}}$.
These approximations eventually lead to the following expressions,
on the one hand at ${\alpha_{\text{eff}} \lesssim \alpha_c}$:
\begin{equation}
 \label{eq-HL-law-diffcoeff-C12-justbelowalphac}
\begin{split}
 C_1 & \approx \argc{\frac{\moy{\sigma_c^4}}{24 \argp{\alpha_c -\alpha_{\text{eff}}}}}^{1/2} \\
 C_2 & 
 		\approx - \frac{\moy{\sigma_c^4}^{1/4} \moy{\sigma_c} }{2^{7/4} \times 3^{1/4} \argp{\alpha_c - \alpha_{\text{eff}}}^{5/4}}
\end{split}
\end{equation}
and at ${\alpha_{\text{eff}} \ll \alpha_c}$:
\begin{equation}
 \label{eq-HL-law-diffcoeff-C12-wellbelowalphac}
\begin{split}
 C_1 \approx \frac{\alpha_{\text{eff}}}{\moy{\sigma_c}}
 \, , \quad
 C_2 
 		\approx - \argp{\frac{\alpha_\text{eff}}{\moy{\sigma_c}^3}}^{1/2}
\end{split}
\end{equation}

Note finally that the expansions of ${\tilde{f}_{\sigma_c} \argp{x, G_0 \dot{\gamma} \tau / x^2}}$ given in eqs.~\eqref{eq-expansion-fidtilde-casedelta1}-\eqref{eq-expansion-fidtilde-casedelta01}-\eqref{eq-expansion-fidtilde-casedelta0}, \emph{before} averaging over the values of ${\sigma_c}$, allow us to obtain the predictions for ${\tilde{\rho}_{\text{st}}(\sigma_c)}$ discussed in sect.~\ref{section-HL-model-stationary-rho-tilde-disord}.
Indeed, we have derived the expressions listed in eqs.~\eqref{eq-rhotilde-stationary-disord-final-HL-belowalphac}-\eqref{eq-rhotilde-stationary-disord-final-HL-atalphac}-\eqref{eq-rhotilde-stationary-disord-final-HL-abovealphac} by substituting into these expansions of ${\tilde{f}_{\sigma_c}}$ the low-shear-rate diffusion coefficient.

\section{Normalization condition for a generic diffusion coefficient ${\widetilde{D}(\sigma_c)}$}
\label{A-appendix-diffusion-coeff-depending-on-sigmac}

In sect.~\eqref{section-HL-model-defmodel-disord}, we have derived the normalization condition for the stationary PDF ${\widetilde{\mathcal{P}}_{\text{st}}(\sigma_c,\sigma )}$ with the restriction that the diffusion coefficient does not depend on the local yield stress $\sigma_c$, but is rather a global quantity controlling the evolution of the PDF according to eq.~\eqref{eq-dist-Psigma-HL-disord}.

If the diffusion coefficient is more generically of the form ${\widetilde{D}(\sigma_c,t)}$, in the stationary case the normalization condition \eqref{eq-rhotilde-stationary-disord-0} is:
\begin{equation}
 \label{eq-rhotilde-stationary-disord-0-appendix}
 \tilde{\rho}_{\text{st}} (\sigma_c) = \widetilde{\Gamma}_{\text{st}} (\sigma_c) \tau \, \frac{\tilde{f}_{\sigma_c} \argp{\sqrt{\widetilde{D}(\sigma_c) \tau}, \frac{G_0 \dot{\gamma} \tau}{\widetilde{D}(\sigma_c) \tau}}}{\widetilde{D}(\sigma_c) \tau}
\end{equation}
where ${\tilde{f}_{\sigma_c}}$ is exactly the same function as in eq.~\eqref{eq-Gamma-HL-D-factorf},
for instance the parabola \eqref{eq-factorf-HL-noshear} in absence of shear rate and the function \eqref{eq-factorf-HL-withshear} in presence of a constant shear rate.
Using again the relation \eqref{eq-rhotilde-stationary}, we obtain the counterpart of eq.~\eqref{eq-Gamma-HL-D-factorf} for the global plastic activity:
\begin{equation}
 \label{eq-rhotilde-stationary-disord-1-appendix}
 \Gamma_{\text{st}} \tau \int_0^{\infty} \!\!\!\! d \sigma_c\, \rho(\sigma_c) \, \frac{\tilde{f}_{\sigma_c} \argp{\sqrt{\widetilde{D}(\sigma_c) \tau}, \frac{G_0 \dot{\gamma} \tau}{\widetilde{D}(\sigma_c) \tau}}}{\widetilde{D}(\sigma_c) \tau} = 1
\end{equation}
but this expression does not simplify into eq.~\eqref{eq-rhotilde-stationary-disord-2}-\eqref{eq-rhotilde-stationary-disord-3}, and thus the closure relation~\eqref{eq-implicit-for-D-disord} is modified by the $\sigma_c$-dependence of ${\widetilde{D}(\sigma_c)}$.
The previous relation~\eqref{eq-rhotilde-stationary-disord-1-appendix} can be used to compute ${\Gamma_{\text{st}}}$, at least numerically if not analytically, for any choice of ${\rho(\sigma_c)}$ and ${\widetilde{D}(\sigma_c)}$.

Nevertheless, for the sake of completeness, we can parametrize the stationary diffusion coefficient according to:
\begin{equation}
 \widetilde{D}(\sigma_c) = D \, \tilde{d}(\sigma_c)
 \, , \quad
 \int_0^{\infty}  \!\!\!\! d \sigma_c \, \tilde{d}(\sigma_c) = 1
\end{equation}
where on the one hand, $D$ is the diffusion coefficient integrated over all the possible values of $\sigma_c$ (on which we could for instance impose a closure relation for ${\Gamma_{\text{st}}(D)}$),
and on the other hand, ${\tilde{d}(\sigma_c)}$ characterizes how the diffusion affects the sites with different values of the local yield stress $\sigma_c$.
If such a parametrization is relevant for a given amorphous system, then eq.~\eqref{eq-rhotilde-stationary-disord-1-appendix} simply becomes:
\begin{equation}
 \label{eq-rhotilde-stationary-disord-2-appendix}
\Gamma_{\text{st}} \tau \int_0^{\infty} \!\!\!\! d \sigma_c\, \rho(\sigma_c) \, \frac{\tilde{f}_{\sigma_c} \argp{\sqrt{D \tau} \sqrt{\tilde{d}(\sigma_c)}, \frac{G_0 \dot{\gamma} \tau}{D \tau} / \tilde{d}(\sigma_c)}}{D \tau \, \tilde{d}(\sigma_c)} = 1.
\end{equation}
So, combined with the closure relation \eqref{eq-closure-DHL-disord-stat1}, this last relation provides us with the generalized counterpart of eq.~\eqref{eq-implicit-for-D}:
\begin{equation}
\label{eq-implicit-for-D-disord-appendix}
 \moy{\frac{\tilde{f}_{\sigma_c} \argp{x \sqrt{\tilde{d}(\sigma_c)}, y / \tilde{d}(\sigma_c)}}{\tilde{d}(\sigma_c)}}
 = \alpha_{\text{eff}}
\end{equation}
with ${x = \sqrt{D \tau}}$ and ${y = G_0 \dot{\gamma} \tau/x^2}$.
This defines an `effective' function ${f_{\text{eff}} \argp{x, y;\tilde{d}(\sigma_c)}}$ similarly to eq.~\eqref{eq-implicit-for-D-disord}.
The diffusion coefficient ${D=D_{\text{HL}}}$ can then be determined uniquely as a function of the shear rate $\dot{\gamma}$ and the effective coupling parameter ${\alpha_{\text{eff}}}$. Note at last that the shape of ${\tilde{d}(\sigma_c)}$ should be justified separately, as it is here introduced as an arbitrary input of the model.

\section{Explicit analytical expressions for the stationary case at fixed diffusion coefficient}
\label{A-appendix-explicit-expressions}

%
In this section we sketch the derivation and give the explicit expressions of the stationary solution of the disordered HL evolution equation~\eqref{eq-dist-Psigma-HL-disord}, on the one hand the complete PDFs and on the other hand the corresponding mean stress, at fixed diffusion coefficient (in other words, before using any specific closure relation for ${\Gamma_{\text{st}}(D)}$).

%
The equation of the stationary joint PDF decomposes into the following structure, respectively on ${\valabs{\sigma} > \sigma_c}$ and ${\valabs{\sigma} \leq \sigma_c}$:
\begin{eqnarray*}
\argc{ \partial_\sigma^2 - \beta_0 \partial_\sigma } \widetilde{P}(\sigma) = 0
 	& \Rightarrow &
 	\widetilde{P}(\sigma) = c_1 \, e^{\beta_0 \sigma} + c_2 \\
\argc{ \partial_\sigma^2 - \beta_0 \partial_\sigma } \widetilde{P}(\sigma) = \frac{\widetilde{P}(\sigma)}{D\tau} 
 	& \Rightarrow &
 	\widetilde{P}(\sigma) = \tilde{c}_1 \, e^{\beta_{(-)} \sigma} + \tilde{c}_2 \, e^{\beta_{(+)} \sigma}
\end{eqnarray*}
with ${\beta_{(\pm)} = \frac{\beta_0}{2} \pm \sqrt{\argp{\frac{\beta_0}{2}}^2+\frac{1}{D\tau}}}$
and the constants ${\arga{c_1,\tilde{c}_1,\tilde{c}_2}}$ fixed by the boundary conditions at ${\sigma \in \arga{-\infty,-\sigma_c,0,\sigma_c,\infty}}$.
Adapting first these solutions to our notations with ${\beta_0 = y = \frac{G_0 \dot{\gamma} \tau}{D \tau}}$,
the joint PDF ${\widetilde{\mathcal{P}}_{\text{st}}(\sigma_c,\sigma)}$ can be decomposed into:
\begin{equation}
\label{eq-jointPDF-explicit-appendix}
\widetilde{\mathcal{P}}_{\text{st}}(\sigma_c,\sigma)
 = \rho(\sigma_c) \frac{\Gamma_{\text{st}} \tau}{D \tau} \, \tilde{\kappa}(\sigma_c) \,  \tilde{p}_{\sigma_c} (\sigma)
\end{equation}
as announced in sect.~\ref{section-HL-model-stationary-solution-disord},
with ${\widetilde{\Gamma}_{\text{st}}(\sigma_c)= \Gamma_{\text{st}} \, \rho (\sigma_c)}$ according to eq.~\eqref{eq-rhotilde-stationary}, and
\begin{equation}
\label{eq-jointPDF-explicit-appendix-bis}
 \tilde{p}_{\sigma_c} (\sigma)
 = \left\lbrace \begin{array}{ll}
 e^{\beta_{(-)} \sigma}
 	& \!\!\!\! : \sigma > \sigma_c
 \\ \\
 \frac{\beta_{(-)}}{y} e^{+\beta_{(-)} \sigma_c} \argc{e^{y (\sigma - \sigma_c)} +  \frac{\beta_{(+)}}{\beta_{(-)}}} & \!\!\!\! : 0 \leq \sigma \leq \sigma_c
 \\ \\
 \frac{\beta_{(+)}}{y} e^{-\beta_{(+)} \sigma_c} \argc{e^{y (\sigma + \sigma_c)} +  \frac{\beta_{(-)}}{\beta_{(+)}}} & \!\!\!\! : -\sigma_c \leq \sigma \leq 0
 \\ \\
 e^{\beta_{(+)} \sigma}
 	& \!\!\!\! : \sigma < -\sigma_c
 \end{array} \right.
\end{equation}
The definition of the partial plastic activity ${\widetilde{\Gamma}_{\text{st}}(\sigma_c)}$ in eq.~\eqref{eq-Gamma-nu-HL-disord-partial} allows one to determine the normalization factor:
\begin{equation}
\begin{split}
 \tilde{\kappa}(\sigma_c)
 &= D \tau \, \argc{\int_{\valabs{\sigma}>\sigma_c} \!\!\!\!\!\!\!\! d\sigma \, \tilde{p}_{\sigma_c} (\sigma)}^{-1} \\
 &= \argc{\beta_{(+)} e^{+\beta_{(-)} \sigma_c } - \beta_{(-)} e^{-\beta_{(+)} \sigma_c}}^{-1}
\end{split}
\end{equation}
The function ${\tilde{f}_{\sigma_c}(x,y)}$ is then defined with respect to the dynamical distribution of local yield stress in the stationary case,
${\tilde{\rho}(\sigma_c) \equiv \rho(\sigma_c) \frac{\Gamma_{\text{st}} \tau}{D \tau} \tilde{f}_{\sigma_c} (x,y)}$, whose definition~\eqref{eq-PDF-sigmac-HLdisord} implies that
\begin{equation}
 \tilde{f}_{\sigma_c} (x,y) = \tilde{\kappa}(\sigma_c) \int_{\mathbb{R}} d\sigma \, \tilde{p}_{\sigma_c} (\sigma)
\end{equation}
which is thus exactly the same expression~\eqref{eq-factorf-HL-withshear} as for the standard HL model.
From the global normalization of the PDF, we conclude that the global plastic activity at fixed $D$ is given by ${\Gamma_{\text{st}}\tau = D\tau / \moy{\tilde{f}_{\sigma_c} (x,y)}}$, as stated in eqs.~\eqref{eq-rhotilde-stationary-disord-2}-\eqref{eq-rhotilde-stationary-disord-3}.
Once ${\tilde{f}_{\sigma_c} (x,y)}$ is known, the dynamical distribution of local yield stress ${\tilde{\rho}(\sigma_c)}$ can be fully determined according to eq.~\eqref{eq-rhotilde-stationary-disord-final-generic}.

The main novelty in eq.~\eqref{eq-jointPDF-explicit-appendix}, compared to previous references on the \emph{standard} HL model \cite{cances_catto_gati_2006_SIAMJMathAnal37_60,olivier_2010_ZAngewMathPhys61_445}, is that the global plastic activity at fixed diffusion coefficient, ${\Gamma_{\text{st}} (D)}$, is replaced by its partial counterpart ${\widetilde{\Gamma}_{\text{st}}(\sigma_c)= \Gamma_{\text{st}} \, \rho (\sigma_c)}$.
Moreover, we have explicitly kept the ratio ${\Gamma_{\text{st}}/D}$, with the global plastic activity fixed by eqs.~\eqref{eq-rhotilde-stationary-disord-2}-\eqref{eq-rhotilde-stationary-disord-3}-\eqref{eq-Gamma-HL-D-factorf}; so the solution~\eqref{eq-jointPDF-explicit-appendix} remains valid for any closure relation, and in particular for the HL closure relation~\eqref{eq-closure-DHL-disord}. In the latter case, that we have studied throughout this paper, the ratio ${\Gamma_{\text{st}}/D}$ can simply be replaced by ${1/\alpha_{\text{eff}}}$.

%
Since all the dependences on the local yield stress $\sigma_c$ have been made explicit, the stress PDF ${\mathcal{P}_{\text{st}}(\sigma)}$ can be computed by integrating ${\widetilde{\mathcal{P}}_{\text{st}}(\sigma_c,\sigma)}$ over the possible values of local yield stress.
%
Nevertheless, for an arbitrary a priori distribution ${\rho (\sigma_c)}$, no explicit expression can be written down, because of the $\sigma_c$-dependence of the stress division itself (${\valabs{\sigma} \lessgtr \sigma_c}$).

%
We come at last to the prediction for the mean stress ${\sigma_M(x,y)}$, with ${x^2=D\tau}$ and ${y=G_0 \dot{\gamma} \tau/x^2}$ fixed, using ${D/\Gamma_{\text{st}}(D)=\moy{\tilde{f}_{\sigma_c}(x,y)}}$ according to eqs.~\eqref{eq-rhotilde-stationary-disord-2}-\eqref{eq-rhotilde-stationary-disord-3}.
We distinguish the contributions at fixed local yield stress of overstressed and understressed regions:
\begin{equation}
\label{eq-sigmaM-properaverages}
\left\lbrace \begin{array}{ll}
 \sigma_M^{\text{(over)}}
 &=	 \frac{1}{\moy{\tilde{f}_{\sigma_c}(x,y)}}
 		\moy{\tilde{\kappa}(\sigma_c) \, \int_{\valabs{\sigma}>\sigma_c} \!\! d \sigma \,  \sigma \, \tilde{p}_{\sigma_c} (\sigma)}
 \\ \\
 \sigma_M^{\text{(under)}}
 &=	 \frac{1}{\moy{\tilde{f}_{\sigma_c}(x,y)}}
 		\moy{\tilde{\kappa}(\sigma_c) \, \int_{\valabs{\sigma}<\sigma_c} \!\! d \sigma \,  \sigma \, \tilde{p}_{\sigma_c} (\sigma)}
\end{array} \right.
\end{equation}
We start with the contribution of the overstressed regions:
\begin{equation}
 \tilde{\kappa}(\sigma_c) \, \int_{\valabs{\sigma}>\sigma_c} \!\!\!\!\!\!\!\! d \sigma \,  \sigma \, \tilde{p}_{\sigma_c} (\sigma)
 = \tilde{f}_{\sigma_c}(x,y) \, x^2 y
\end{equation}
which, combined to eq.~\eqref{eq-rhotilde-stationary-disord-final-generic}, leads to:
\begin{equation}
\label{eq-sigmaM-over-appendix}
 \sigma_M^{\text{(over)}}
 = \int_0^{\infty} \!\! d\sigma_c \, \tilde{\rho}(\sigma_c) \, G_0 \dot{\gamma} \tau
 = G_0 \dot{\gamma} \tau
\end{equation}
We emphasize that this result does not depend on a specific choice for the closure relation ${\Gamma_{\text{st}}(D)}$, it stems solely from the specific functional of the stationary joint PDF ${\widetilde{P}(\sigma_c,\sigma)}$.
We turn now to the contribution of the understressed regions:
\begin{equation}
\begin{split}
 & \tilde{\kappa}(\sigma_c) \, \int_{\valabs{\sigma}<\sigma_c} \!\!\!\!\!\!\!\! d \sigma \,  \sigma \, \tilde{p}_{\sigma_c} (\sigma) \\
 & = \frac{\sigma_c^2}{2y} + \frac{\sigma_c}{y^2} \frac{1- \argp{\frac{2}{\sigma_c y} + \sqrt{1+ \frac{4}{x^2 y^2}}} \tanh \argp{\frac{\sigma_c y}{2}}}{\tanh \argp{\frac{\sigma_c y}{2}}+ \sqrt{1+ \frac{4}{x^2 y^2}}} \\
 & = \frac{\argp{\frac{\sigma_c^2}{2}-\tilde{f}_{\sigma_c}(x,y) + x^2}}{y} +\frac{1}{y^2} \frac{2 \sigma_c}{\sqrt{1+\frac{4}{x^2 y^2}}+\tanh \argp{\frac{\sigma_c y}{2}}}
 \end{split}
\end{equation}
which leads to
\begin{equation}
\label{eq-sigmaM-under-appendix}
\begin{split}
 \sigma_M^{\text{(under)}}
 =	& \frac{1}{y} \argc{\frac{\moy{\sigma_c^2}/2 - \moy{\tilde{f}_{\sigma_c}(x,y)} + x^2}{\moy{\tilde{f}_{\sigma_c}(x,y)}}} \\
	& + \frac{1}{y^2 \, \moy{\tilde{f}_{\sigma_c}(x,y)}} \moy{\frac{2 \sigma_c}{\sqrt{1+\frac{4}{x^2 y^2}}+\tanh \argp{\frac{\sigma_c y}{2}}}}
\end{split}
\end{equation}
where we can recognize ${\alpha_c=\moy{\sigma_c^2}/2}$ in the first term.
The total mean stress can eventually be computed by combining eqs.~\eqref{eq-sigmaM-over-appendix} and~\eqref{eq-sigmaM-under-appendix} into
\begin{equation}
\label{eq-sigmaM-appendix}
 \sigma_M = \sigma_M^{\text{(over)}} + \sigma_M^{\text{(under)}}
\end{equation}
Moreover, while discussing the assumption of a typical fixed rate ${1/\tau}$ in sect.~\ref{section-interplay-timescales}, we have suggested the alternative definition of the `macroscopic' stress given in eq.~\eqref{eq-sigmaM-corr-over-sigmac}.
It simply consists in the replacement of ${\sigma_M^{\text{(over)}}}$ by ${\int_0^{\infty} \!\! d \sigma_c \, \sigma_c \, \widetilde{\Gamma}_{\text{st}}(\sigma_c) \tau }$,
and hence at fixed ${(x,y)}$:
\begin{equation}
 \sigma_M^\mathrm{corr}
 = \frac{ \moy{\sigma_c} x^2}{\moy{\tilde{f}_{\sigma_c}(x,y)}}
 	+ \sigma_M^{\text{(under)}}
\end{equation}

The limit of low shear rate ${\dot{\gamma} \to 0}$ of $\sigma_M$, with the HL closure relation~\eqref{eq-closure-DHL-disord-stat1}, is discussed in the main text in sect.~\ref{section-HL-model-stationary-low-shearrate-disord}.
Note finally that, before performing any Taylor expansion of $\sigma_M$ at small ${\dot{\gamma}}$, it is crucial \emph{not} to replace ${\moy{\tilde{f}_{\sigma_c}(x,y)}}$ by $\alpha_{\text{eff}}$, in order to capture correctly the lowest orders in the perturbation; the HL closure relation will in fact already be encoded in the diffusion coefficient ${D_{\text{HL}}}$ itself.

\section{Herschel-Bulkley behavior in the disordered HL model at ${\alpha_{\text{eff}}<\alpha_c}$}
\label{A-appendix-prefactorA-disordHL}

As discussed in sect.~\ref{section-HL-model-stationary-low-shearrate-disord},
for ${\alpha_{\text{eff}} < \alpha_c}$ the mean stress displays a Herschel-Bulkley behavior a low shear rate:
$${\sigma_M \stackrel{(\dot{\gamma} \to 0)}{\approx} \sigma_Y+A \, \argp{G_0 \dot{\gamma} \tau}^{1/2}}$$
The macroscopic yield stress $\sigma_Y$ is simply obtained using the lowest-order expansion of the diffusion coefficient (using the minimal ansatz of eq.~\eqref{eq-ansatz-diff-coeff-small-gammadot}), the prefactor $A$ involves its second-order expansion (using the ansatz of eq.~\eqref{eq-ansatz-diff-coeff-small-gammadot-bis}).
In this appendix, we give explicitly the expressions of these two parameters of the Herschel-Bulkley behavior of exponent $1/2$ predicted by the disordered HL model for ${\alpha_{\text{eff}} < \alpha_c}$.
%

We first substitute the ansatz for ${D_{\text{HL}}}$ given in eq.~\eqref{eq-ansatz-diff-coeff-small-gammadot-bis} into the exact expression for the mean stress $\sigma_M$ of eq.~\eqref{eq-sigmaM-appendix}, and expand the result at small ${G_0 \dot{\gamma} \tau}$.
We obtain at ${\mathcal{O}\argp{\dot{\gamma}^{0}}}$:
\begin{equation}
 \label{eq-HL-law-sigmaY-disord-1-appendix}
 \sigma_Y(C_1)
 = C_1 \argc{\frac{\moy{\sigma_c^2} /2}{C_1 \moy{\sigma_c \, \tanh \argp{\frac{\sigma_c}{2 C_1}}}} -1 }
\end{equation}
and at ${\mathcal{O}\argp{\dot{\gamma}^{1/2}}}$:
\begin{equation}
 \label{eq-HL-law-prefactorA-disord-1}
\begin{split}
 & A =
 C_1^{1/2} \frac{\frac{\moy{\sigma_c}}{2} - C1 \moy{\tanh \argp{\frac{\sigma_c}{2 C_1}}} + \frac12 \moy{\sigma_c \tanh^2 \argp{\frac{\sigma_c}{2 C_1}}}}{\moy{\sigma_c \tanh \argp{\frac{\sigma_c}{2 C_1}} }} \\
 & + 2 C_2 \frac{\frac{\moy{\sigma_c^2}}{2} - C1 \moy{\sigma_c \tanh \argp{\frac{\sigma_c}{2 C_1}}} - \frac14 \moy{\sigma_c^2 \tanh^2 \argp{\frac{\sigma_c}{2 C_1}}}}{\moy{\sigma_c \tanh \argp{\frac{\sigma_c}{2 C_1}} }}
\end{split}
\end{equation}
with ${C_2 (C_1)}$ given by eq.~\eqref{eq-HL-law-diffcoeff-C2-equa}, and ${C_1(\alpha_{\text{eff}})}$ by eq.~\eqref{eq-solutionC1-casedelta1}.

Secondly, we can simplify this expression in the usual limiting cases of ${\alpha_{\text{eff}} \lesssim \alpha_c}$ and ${\alpha_{\text{eff}} \ll \alpha_c}$, as announced in eq.~\eqref{eq-sigmaM-withshear-smallgammadot-disord-2}.
On the one hand, we use eq.~\eqref{eq-HL-law-diffcoeff-C12-justbelowalphac} at ${\alpha_{\text{eff}} \lesssim \alpha_c}$:
\begin{equation}
\label{eq-HL-law-prefactorA-disord-1-justbelowalphac}
\begin{split}
\sigma_Y
 &	\stackrel{\eqref{eq-HL-law-sigmaY-disord-1}}{\approx}
 		\frac{\moy{\sigma_c^4}/\moy{\sigma_c^2}}{12 C}
 	\stackrel{\eqref{eq-DHL-withshear-smallgammadot-disord-2}}{\approx}
 		\frac{\argp{\alpha_c - \alpha_{\text{eff}}}^{1/2}}{\sqrt{6}} \frac{\moy{\sigma_c^4}^{1/2}}{\moy{\sigma_c^2}}
 	\\
 A
 &	\approx
		\frac{\moy{\sigma_c}}{\moy{\sigma_c^2}}  C^{3/2}
	\approx \frac{\argp{\alpha_c - \alpha_{\text{eff}}}^{-3/4}}{2^{3/2} \times 6^{3/4}} \frac{\moy{\sigma_c^4}^{3/4} \!\! \moy{\sigma_c} }{\moy{\sigma_c^2}}
\end{split}
\end{equation}
and on the other hand, we use eq.~\eqref{eq-HL-law-diffcoeff-C12-wellbelowalphac} at ${\alpha_{\text{eff}} \ll \alpha_c}$:
\begin{equation}
\label{eq-HL-law-prefactorA-disord-1-wellbelowalphac}
\begin{split}
 \sigma_Y
 &	\stackrel{\eqref{eq-HL-law-sigmaY-disord-1}}{\approx}
  		\frac{\moy{\sigma_c^2}/\moy{\sigma_c}}{2} - C
	\stackrel{\eqref{eq-DHL-withshear-smallgammadot-disord-2}}{\approx}
 		\frac{\alpha_c - \alpha_{\text{eff}}}{\moy{\sigma_c}}
 	\\
 A
 &	\approx
		\argp{1- \frac{\moy{\sigma_c^2}}{2 \moy{\sigma_c}^2}} \sqrt{C}		
	\approx \argp{1- \frac{\moy{\sigma_c^2}}{2 \moy{\sigma_c}^2}} \argp{\frac{\alpha_{\text{eff}}}{\moy{\sigma_c}}}^{1/2}
\end{split}
\end{equation}

The predictions of the standard HL model, presented in sect.~\ref{section-HL-model}, respectively eqs.~\eqref{eq-sigmaM-withshear-smallgammadot-1}-\eqref{eq-HL-law-prefactorA-1}-\eqref{eq-sigmaM-withshear-smallgammadot-2}, are of course recovered by removing all the averages over ${\rho (\sigma_c)}$ from the three last equations.
We emphasize that the absence of averages over ${\rho (\sigma_c)}$ allows one to simplify considerably these expressions, removing in particular the non-trivial combinations of moments ${\moy{\sigma_c^k}}$.



\end{document}